\documentclass[twocolumn,showpacs,aps,floatfix,prd]{revtex4}

\usepackage{graphicx}
\usepackage{dcolumn}
\usepackage{amsmath}
\usepackage{epsfig}

\input{pubboard/babarsym}

\newcommand{\BABARPubNumber}  {08/045}

\newcommand{\SLACPubNumber} {13437}
\newcommand{\LANLNumber} {xxxx}
\newcommand{\modeTwo}{$B^-\rightarrow\psi(2S)\pi^-K_S^0$}
\newcommand{\modeFour}{$B^-\rightarrow J/\psi\pi^-K_S^0$}
\newcommand{\modeSix}{$B^0\rightarrow \psi(2S)\pi^-K^+$}
\newcommand{\modeEight}{$B^0\rightarrow J/\psi\pi^-K^+$}
\newcommand{\modeTwoSix}{$B^{0,-}\rightarrow \psi(2S)\pi^-K^{+,0}$}
\newcommand{\modeFourEight}{$B^{0,-}\rightarrow J/\psi\pi^-K^{+,0}$}
\newcommand{\kpi}{$K\pi^-$}
\newcommand{\mkpi}{$m_{K\pi^-}$}
\newcommand{\psipi}{$\psi\pi^-$}
\newcommand{\jpsipi}{$J/\psi\pi^-$}
\newcommand{\psitwospi}{$\psi(2S)\pi^-$}
\newcommand{\mpsitwospi}{$m_{\psi(2S)\pi^-}$}
\newcommand{\mjpsipi}{$m_{J/\psi\pi^-}$}
\newcommand{\Ksone}{$K^{\ast}(892)$}
\newcommand{\Kstwo}{$K^{\ast}_2(1430)$}
\newcommand{\Ksthree}{$K^{\ast}_3(1780)$}
\newcommand{\z}{$Z(4430)^-$}
\newcommand{\costhk}{$\cos\theta_K$}

\def\figurebox#1#2#3{
    \def\arg{#3}
    \ifx\arg\empty
    {\hfill\vbox{\hsize#2\hrule\hbox to #2{\vrule\hfill\vbox to #1{\hsize#2\vfill}\vrule}\hrule}\hfill}
    \else
    {\hfill\epsfbox{#3}\hfill}
    \fi}

\long\def\inst#1{\par\nobreak\kern 4pt\nobreak
    {\it #1}\par\vskip 10pt plus 3pt minus 3pt}

\begin{document}

\preprint{\babar-PUB-\BABARPubNumber} 
\preprint{SLAC-PUB-\SLACPubNumber} 

\begin{flushleft}
\babar-PUB-\BABARPubNumber\\
SLAC-PUB-\SLACPubNumber\\
hep-ex/\LANLNumber\\ [10mm]
\end{flushleft}

\title{ {\large {\bf \boldmath Search for the \z\ at \babar\ } } }

%
\author{B.~Aubert}
\author{M.~Bona}
\author{Y.~Karyotakis}
\author{J.~P.~Lees}
\author{V.~Poireau}
\author{E.~Prencipe}
\author{X.~Prudent}
\author{V.~Tisserand}
\affiliation{Laboratoire de Physique des Particules, IN2P3/CNRS et Universit\'e de Savoie, F-74941 Annecy-Le-Vieux, France }
\author{J.~Garra~Tico}
\author{E.~Grauges}
\affiliation{Universitat de Barcelona, Facultat de Fisica, Departament ECM, E-08028 Barcelona, Spain }
% ...
\collaboration{The \babar\ Collaboration}
\noaffiliation

\date{\today}
\begin{abstract}
We report the results of a search for \z\ decay to \jpsipi\ or
\psitwospi\ in $B^{-,0}\rightarrow \jpsi\pi^- K^{0,+}$ and
$B^{-,0}\rightarrow \psitwos \pi^- K^{0,+}$ decays. The data were
collected with the \babar\ detector at the SLAC PEP-II
asymmetric-energy $e^+e^-$ collider operating at center of mass energy
10.58 \gev\/, and the sample corresponds to an integrated luminosity
of 413 fb$^{-1}$. Each \kpi\ mass distribution exhibits clear
$K^{\ast}(892)$ and $K^{\ast}_2(1430)$ signals, and the
efficiency-corrected spectrum is well-described by a superposition of
the associated Breit-Wigner intensity distributions, together with an
$S$-wave contribution obtained from the LASS $I=1/2$ \kpi\ scattering
amplitude measurements.  Each \kpi\ angular distribution varies
significantly in structure with \kpi\ mass, and is represented in
terms of low-order Legendre polynomial moments. We find that each
\jpsipi\ or \psitwospi\ mass distribution is well-described by the
reflection of the measured \kpi\ mass and angular distribution
structures. We see no significant evidence for a \z\ signal for any of
the processes investigated, neither in the total \jpsipi\ or
\psitwospi\ mass distribution, nor in the corresponding distributions
for the regions of \kpi\ mass for which observation of the \z\ signal
was reported. We obtain branching fraction upper limits
${\cal{B}}(B^-\rightarrow Z^-\bar{K^0}, Z^-\rightarrow
J/\psi\pi^-)<1.5\times 10^{-5}$, ${\cal{B}}(B^0\rightarrow Z^-K^+,
Z^-\rightarrow J/\psi\pi^-)<0.4\times 10^{-5}$,
${\cal{B}}(B^-\rightarrow Z^-\bar{K^0}, Z^-\rightarrow
\psi(2S)\pi^-)<4.7\times 10^{-5}$, and ${\cal{B}}(B^0\rightarrow
Z^-K^+, Z^-\rightarrow \psi(2S)\pi^-)<3.1\times 10^{-5}$ at $95\%$
confidence level, where the \z\ mass and width have been fixed to the
reported central values.
\end{abstract}

\maketitle

\section{Introduction}
In the original paper in which he proposed the Quark
Model~\cite{GellMann:1964nj}, Gell-Mann stated that ``Baryons can now
be constructed from quarks using the combinations ($q q q$), ($q q q q
\bar{q}$), {\it etc}., while mesons are made out of ($q \bar{q}$), ($q q
\bar{q}\bar{q}$), {\it etc}.''. He chose the lowest configurations to
create the representations describing the known meson and baryon
states. However, the higher configurations were not {\it a priori}
excluded, and experimentalists and theorists have been seeking
evidence supporting the existence of such states ever since.

In the baryon sector, resonant structure in the $KN$ system would be
indicative of five-quark content, and searches for states of this type
have been carried out since the mid-1960's, mainly through
partial-wave analysis of $KN$ elastic and charge-exchange scattering
data. In recent years, there has been a great deal of activity focused
on the search for the conjectured $\Theta(1540)^+$ pentaquark state
decaying to $K^0p$. However, the initial low-statistics signals
claimed in a variety of experimental contexts have not withstood
high-statistics scrutiny, and the existence of this state must be
considered to be in doubt at the present time. The subject is reviewed
in Ref.~\cite{pentaq}, and the status is updated in
Ref.~\cite{Yao:2006px}.

In the meson sector, attention has been focused over the years mainly
on the $a_0(980)$ and the $f_0(980)$ scalar mesons as possible
four-quark states. However, the discovery of the $D_{s0}^*(2317)$ and
the $D_{s1}(2460)$, with their unexpectedly low mass values, and the
observation of many new charmonium-like states above threshold for
decay to open charm, have led to speculation that certain of these may
be four-quark states (see {\it e.g.} Ref.~\cite{maiani3}), although in
no case has this been clearly established. In this regard, it follows
that the recent paper from the Belle Collaboration~\cite{:2007wga}
which reports the observation of a resonance-like structure, the \z\/,
in the $\psi(2S)\pi^-$ system produced in the decays
$B^{-,0}\rightarrow\psi(2S)\pi^-K^{0,+}$~\cite{CC} has generated a
great deal of interest (see {\it e.g.} Ref.~\cite{maianiZ}, and
references therein). Such a state must have a minimum quark content
($c \bar{c} d \bar{u}$), and would represent the unequivocal
manifestation of a four-quark meson state.

It is clearly important to seek confirmation of the Belle observation,
not only in the \psitwospi\ system, but also for the \jpsipi\
combination, which might also show evidence of a \z\ signal or of a
similar lower mass state~\cite{theory}. Consequently, in this paper we
present a \babar\ analysis of the entire Dalitz plot corresponding to
the decays $B^{-,0}\rightarrow\psi(2S)\pi^-K^{0,+}$ and in parallel
pursue an identical analysis of our $B^{-,0}\rightarrow
J/\psi\pi^-K^{0,+}$ data. Both analyses make use of the complete
\babar\ data sample accrued at the $\Upsilon(4S)$ resonance. In this
regard, we first seek a representation of the \kpi\ mass and angular
distribution structures, which dominate the final states under study,
in terms of their expected low-angular-momentum intensity
contributions. We then investigate the reflection of each \kpi\ system
into its associated \psipi\/~\cite{psi} mass distribution in order to
establish the need for any additional narrow signal.

The \babar\ detector and the data sample are described briefly in
Sec.~\ref{sec:det}, and the event selection procedures are discussed
in Sec.~\ref{sec:select}. In Sec.~\ref{sec:dalitz}, the Dalitz plots
and their uncorrected invariant mass projections are shown for the $B$
meson signal regions. Since the analysis emphasizes this search for
narrow structure in the $J/\psi\pi^-$ and $\psi(2S)\pi^-$ mass
distributions, the mass resolution dependence on invariant mass for
these systems is analyzed in Sec.~\ref{sec:resolution}.  Similarly, it
is important to understand the behavior of the event reconstruction
efficiency over each final state Dalitz plot and to correct for it
before assessing the significance of any observed mass structures. The
procedure followed is described in Appendix A, and the results are
summarized in Sec.~\ref{sec:efficiency}. Fits to the corrected \kpi\
mass distributions are discussed in Sec.~\ref{sec:kpi}, and the \kpi\
angular distribution structure as a function of \kpi\ mass is
represented in terms of Legendre polynomial moments as described in
Sec.~\ref{sec:legndre}. In Sec.~\ref{sec:reflection}, the reflections
of the observed \kpi\ mass and angular structures onto the
$J/\psi\pi^-$ and $\psi(2S)\pi^-$ mass distributions are compared to
the corresponding efficiency-corrected distributions, and in
Sec.~\ref{sec:comparison} our results are discussed in relation to
those in the Belle publication. The \babar\ \psipi\ mass distributions
are fitted in Sec.~\ref{sec:babar_fits}, and we present a summary and
our conclusions in Sec.~\ref{sec:conclusions}. Finally,
acknowledgments are expressed in Sec.~\ref{sec:acknow}.

\section{The \babar\ detector and data sample}
\label{sec:det}
The data used in this analysis were collected with the \babar\/
detector at the PEP-II asymmetric-energy $e^+e^-$ collider operating
at a center-of-mass (c.m.) energy of $10.58$ \gev\/.

A detailed description of the \babar\/ detector can be found in
Ref.~\cite{Aubert:2001tu}. Charged particle tracks are detected with a
five-layer, double-sided silicon vertex tracker (SVT) and a 40-layer
drift chamber (DCH), filled with a helium-isobutane gas mixture, and
coaxial with the cryostat of a superconducting solenoidal magnet,
which produces a magnetic field of approximately 1.5 T. The
charged-particle momentum resolution is given by $(\delta
P_T/P_T)^2=(0.0013P_T)^2+(0.0045)^2$, where $P_T$ is the transverse
momentum measured in \gevc\/. The SVT, with a typical coordinate
resolution of 10 $\mum$, measures the impact parameters of charged
particle tracks in both the plane transverse to the beam direction and
along the collision axis; it also supports stand-alone reconstruction
of low-$P_T$ charged particle tracks.

Charged particle types are identified from specific ionization energy
loss ($dE/dx$) measured in the DCH and SVT, and from Cherenkov
radiation detected in a ring-imaging Cherenkov device. Electrons are
identified by means of a CsI(Tl) electromagnetic calorimeter (EMC).

The return yoke of the superconducting coil is instrumented with
resistive plate chambers for the identification of muons and the
detection of clusters produced by $K_L$ and neutron interactions. For
the latter part of the experiment these chambers were replaced by
limited streamer tubes in the barrel region of the
detector~\cite{LST}.

In this analysis, we use a data sample corresponding to an integrated
luminosity of 413 fb$^{-1}$, which is equivalent to the production of
approximately 455 million $B\bar{B}$ pairs.

\section{Event selection} 
\label{sec:select}
We reconstruct events in four decay modes~\cite{CC}:
\begin{eqnarray}
\label{eq:modefour}
B^-&\rightarrow& J/\psi\pi^-K_S^0  \,  , \\
\label{eq:modeeight}
B^0&\rightarrow& J/\psi\pi^-K^+    \, ,  \\
\label{eq:modetwo}
B^-&\rightarrow&\psi(2S)\pi^-K_S^0 \, ,  \\
\label{eq:modesix}
B^0&\rightarrow& \psi(2S)\pi^-K^+ \, .
\end{eqnarray}

The event selection criteria were established by optimizing
signal-to-background ratio using Monte Carlo (MC) simulated signal
events, $B^{-,0}\rightarrow\psi\pi^- K^{0,+}$, and background,
$B\bar{B}$ and $e^+e^-\rightarrow q\bar{q}$ ($q=u,d,s,c$), events.

For the data sample, a \jpsi\ candidate is formed by geometrically
constraining an identified $e^+e^-$ or $\mu^+\mu^-$ pair of tracks to
a common vertex point and requiring a fit probability $>0.001$. For
$\mu^+\mu^-$, the invariant mass of the pair must in addition satisfy
$3.06<m_{\mu^+\mu^-}<3.14$ \gevcc\/, while for $e^+e^-$ the
requirement is $2.95<m_{e^+e^-}<3.14$ \gevcc\/. In the latter case,
the mass interval extends to lower values in order to allow for
electron bremsstrahlung energy loss; if an electron-associated photon
cluster of this type is found in the EMC, its four-momentum vector is
included in the calculation of $m_{e^+e^-}$. The surviving \jpsi\
candidates were fitted to impose a constraint to the nominal mass
value~\cite{Yao:2006px}.

For \psitwos\ decay to $\mu^+\mu^-$ or $e^+e^-$ the same selection
procedures are followed, but with invariant mass requirements
$3.640<m_{\mu^+\mu^-}<3.740$ \gevcc\ or $3.440<m_{e^+e^-}<3.740$
\gevcc\/. For \psitwos\ decay to $J/\psi\pi^+\pi^-$, the \jpsi\
candidate is selected as previously described, and is fit again to
incorporate a constraint to its nominal mass
value~\cite{Yao:2006px}. This \jpsi\ and an identified $\pi^+\pi^-$
pair are geometrically constrained to a common vertex (fit probability
$>0.001$), and required to have an invariant mass in the range
$3.655<m_{J/\psi\pi^+\pi^-}<3.715$ \gevcc\/. In the same manner as for
the \jpsi\/, surviving candidates were then constrained to the nominal
\psitwos\ mass value~\cite{Yao:2006px}.

A $K^0_S$ candidate is formed by geometrically constraining a pair of
oppositely charged tracks to a common vertex (fit probability
$>0.001$); the tracks are treated as pions, but without
particle-identification requirements, and the invariant mass of the
pair must satisfy $0.472<m_{\pi^+\pi^-}<0.522$ \gevcc\/. A charged
kaon candidate from the $B$ meson decay must be identified as a kaon,
but no particle identification is required of corresponding charged
pion candidates.

The $\psi$, $K$ and $\pi$ candidates forming a $B$ meson decay
candidate are geometrically constrained to a common vertex, with fit
probability $>0.001$ required. For decay modes involving a $K^0_S$,
the $K^0_S$ flight length with respect to this vertex must have $>+3$
standard deviation significance in order to reduce combinatoric
background. The $K^0_S$ candidate is not mass constrained, since this
was found to have a negligible effect on resolution.

We further define $B$ meson decay candidates using the energy
difference $\DeltaE=E^{\ast}_B-\sqrt{s}/2$ in the center of mass
(c.m.) frame, and the beam-energy substituted mass
$\mes=\sqrt{((s/2+\vec{p}_i\cdot\vec{p}_B)/E_i)^2-\vec{p}_B^{\,2}}$,
where ($E_i,\vec{p}_i$) is the initial state four-momentum vector in
the laboratory frame and $\sqrt{s}$ is the c.m. energy; $E^{\ast}_B$
is the $B$ meson energy in the c.m. and $\vec{p}_B$ is its laboratory
frame momentum.

We require that $B$ decay signal events satisfy $5.272<\mes<5.286$
\gevcc\ and $|\DeltaE|<0.020$ \gev\/. In order to correct for
background events in the signal region, we define a \DeltaE\ sideband
region by $0.030<\left |\DeltaE\right |<0.050$ \gev\/; we have
verified through MC studies that sideband events in the \mes\ signal
range correctly represent background in the $B$ meson signal
region. We refer to the procedure by which we correct for background
in the signal region by subtracting the \DeltaE\ sideband events in
the \mes\ signal range by the term ``sideband subtraction''.

\begin{figure*}[!htbp]
\begin{center}
\includegraphics[width=8.5cm]{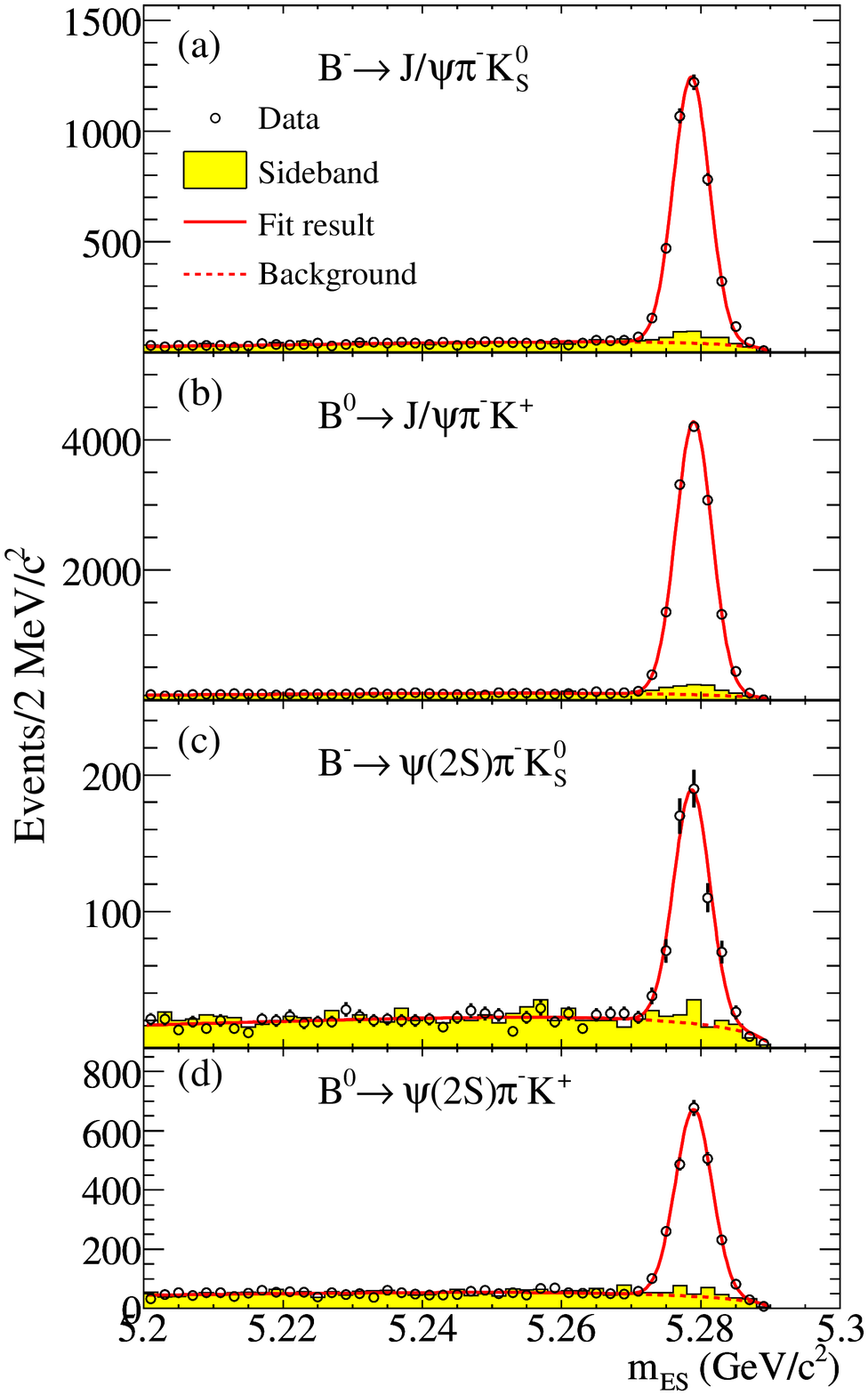}
\includegraphics[width=8.5cm]{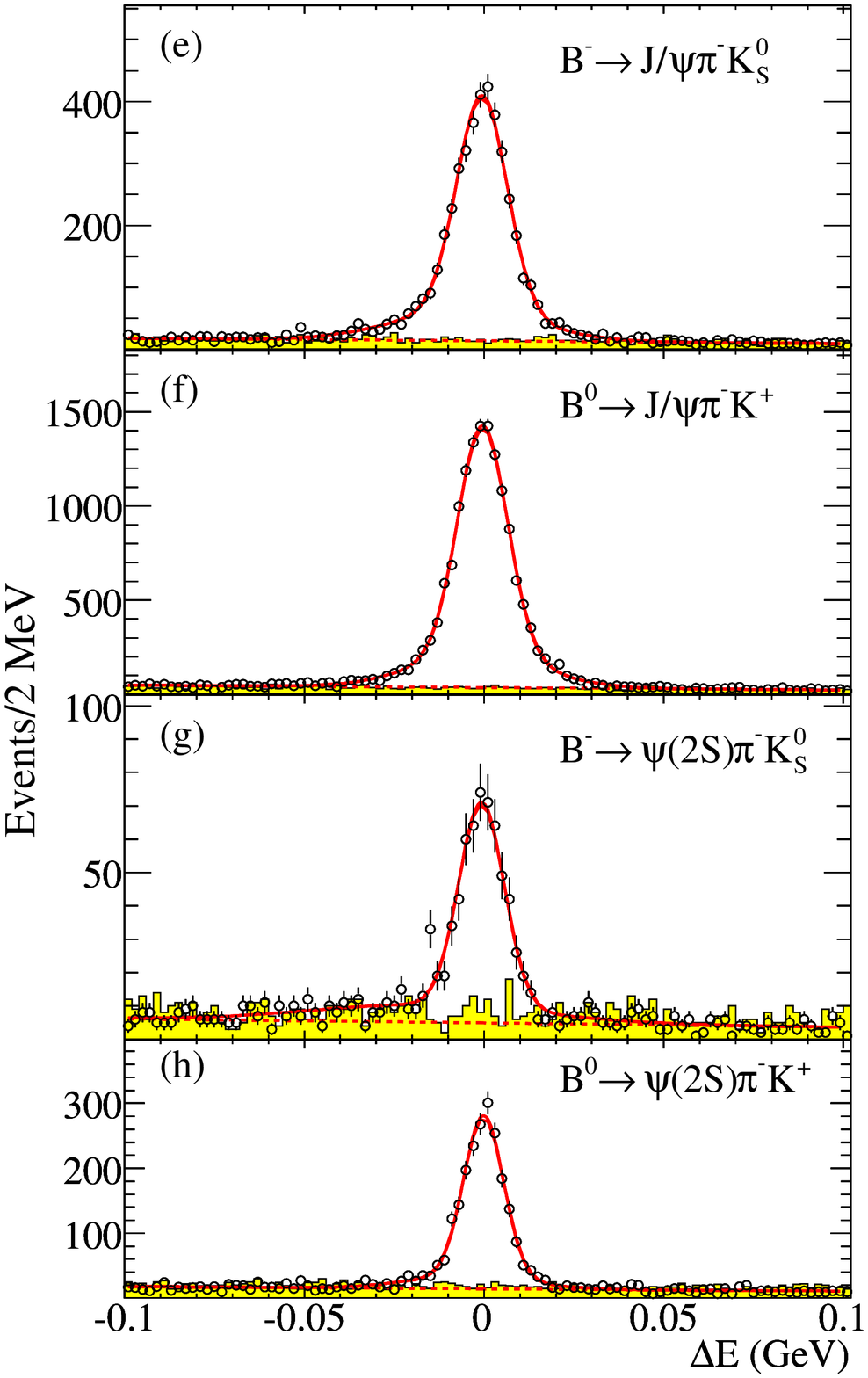}
\caption{The \mes\ distributions, (a)-(d), and (\DeltaE\/)
distributions, (e)-(h), for the decay modes \modeFour\/, \modeEight\/,
\modeTwo\/, and \modeSix\/. The points show the data, and the solid
curves represent the fit functions. The dashed curves indicate the
background contributions, and the filled histograms show the
corresponding distributions for the sideband regions.}
\label{fig:mes:de}
\end{center}
\end{figure*}

In Figs.~\ref{fig:mes:de}(a)-(d) we show the \mes distributions in the
\DeltaE\ signal region for the decay processes of
Eqs.~(\ref{eq:modefour})-~(\ref{eq:modesix}), where the filled
histograms show the sideband distributions. We fit each distribution
with a signal Gaussian function with mass and width as free
parameters, and an ARGUS background function~\cite{Albrecht:1990cs}
with a free exponential slope parameter. In each figure, the solid
curve represents the total function and the dashed curve shows the
background contribution. Clear \mes\ signals are observed in
Figs.~\ref{fig:mes:de}(a)-(d), and in each figure the sideband
distribution is consistent with the fitted background.

The \DeltaE\ distributions for the \mes\ signal region
(Figs.~\ref{fig:mes:de}(e)-(h)) exhibit clear signal peaks. We fit
each distribution with a linear background function and a signal
function consisting of two Gaussian functions with a common center;
all parameters are free in the fits. In each case, the filled
histogram is from the \mes\ sideband region defined by
$5.250<m_{ES}<5.264$ \gevcc\/, and is in good agreement with the
fitted background. For the decay modes of
Eqs.~(\ref{eq:modefour})-(\ref{eq:modesix}), the fraction of events
with more than one $B$ meson signal candidate ranges from 0.5 to 1.1
$\%$. For such events, the candidate with the smallest value of
$|\Delta E|$ is selected.

We summarize the principal selection criteria in
Table~\ref{selections}, and in Table~\ref{table_sig_bkg} provide an
overview of the data samples in the $B$-meson signal region used in
the analysis described in this paper.
\begin{table}
  \begin{center}
    \caption{Summary of the principal criteria used to select $B$
    candidates.}
    \begin{tabular}{lc}\hline\hline \\
    Selection category  & criterion \\ \\ \hline \hline \\
    $\jpsi\rightarrow e^+e^-$      & $2.95<m_{ee}<3.14$ \gevcc\/ \\ \\ 
    $\jpsi\rightarrow \mu^+\mu^-$  & $3.06<m_{\mu\mu}<3.14$ \gevcc\/  \\ \\ \hline \\

    $\psitwos\rightarrow e^+e^-$   & $3.44 < m_{ee}<3.74$ \gevcc\/  \\ \\
    $\psitwos\rightarrow \jpsi\pi^+\pi^-$   & $3.655 < m_{\jpsi\pi\pi}<3.715$ \gevcc\/ \\
    ($\jpsi\rightarrow e^+e^-$)         &                               \\ \\ \hline \\
    $\psitwos\rightarrow \mu^+\mu^-$     & $3.64<m_{\mu\mu}<3.74$ \gevcc\/\\ \\
    $\psitwos\rightarrow \jpsi\pi^+\pi^-$     & $3.655<m_{\jpsi\pi\pi}<3.715$ \gevcc\/ \\
    ($\jpsi\rightarrow \mu^+\mu^-$)      &                              \\ \\ \hline \\

    $K^0_S\rightarrow\pi^+\pi^-$         & $0.472<m_{\pi\pi}<0.522$ \gevcc\ \\ \\
    Flight length significance        & $>+3\sigma$ \\ \\ \hline \\

    \mes\ signal region                 & $5.272<\mes<5.286$  \gevcc\   \\ \\ \hline \\
    \DeltaE\ signal region           & $|\DeltaE|<0.020$ \gev\   \\ \\ \hline\hline
    \end{tabular}
    \label{selections}
  \end{center}
\end{table}

\section{The Dalitz plots and invariant mass projections}
\label{sec:dalitz}
\begin{figure*}[!htbp]
\begin{center}
\includegraphics[width=15.0cm]{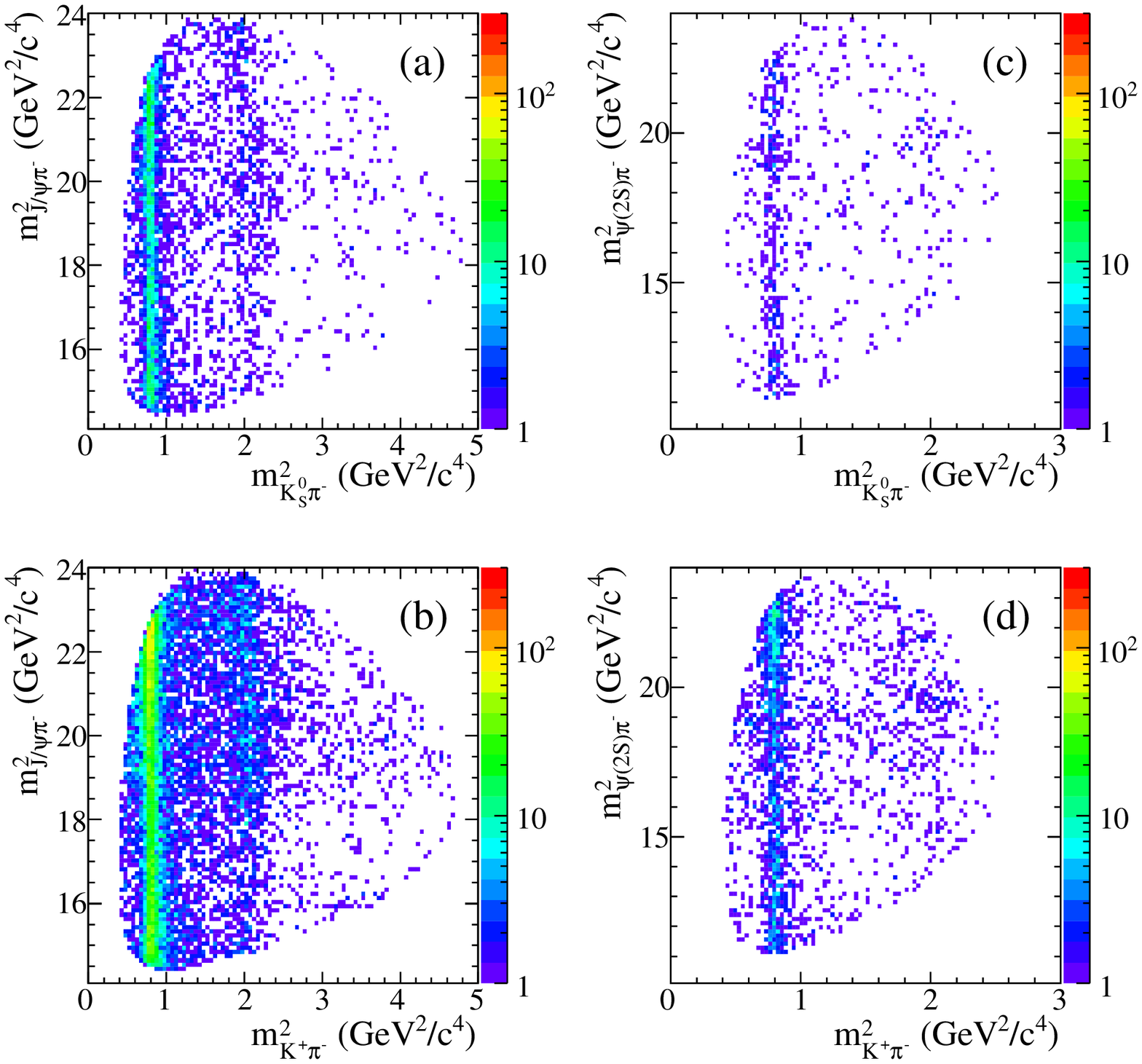}
\caption{The $m_{\psi\pi^-}^2$ versus $m_{K\pi^-}^2$ Dalitz plot
distributions for the signal regions for the decay modes (a)
\modeFour\/, (b) \modeEight\/, (c) \modeTwo\/, (d) \modeSix\/. The
intensity scale is logarithmic.}
\label{fig:dp2}
\end{center}
\end{figure*}

\begin{figure}[!htbp]
\begin{center}
\includegraphics[width=8.5cm]{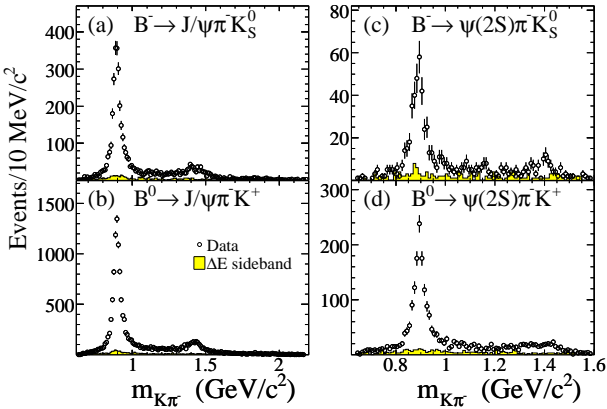}
\caption{The $m_{K\pi^-}$ mass projections for the Dalitz plots of
Fig.~\ref{fig:dp2}. The data points are for the \mes\/-\DeltaE\ signal
regions, and the filled histograms are for the \DeltaE\ sideband
regions.}
\label{fig:proj_kpi}
\end{center}
\end{figure}

\begin{figure}[!htbp]
\begin{center}
\includegraphics[width=8.5cm]{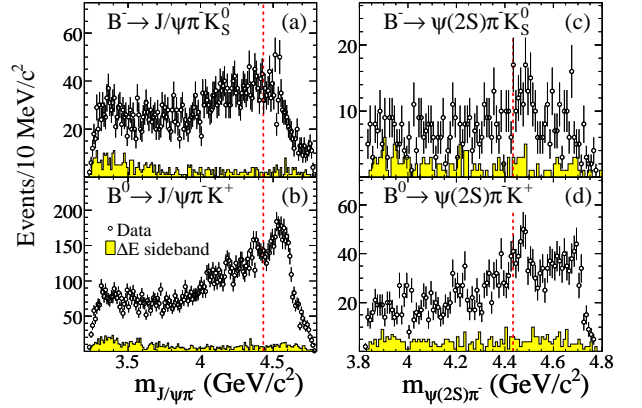}
\caption{The $m_{\psi\pi^-}$ mass projections for the Dalitz plots of
Fig.~\ref{fig:dp2}. The data points are for the \mes\/-\DeltaE\ signal
regions, and the filled histograms are for the \DeltaE\ sideband
regions. The dashed vertical lines indicate $m_{\psi\pi^-}=4.433$
\gevcc\/.}
\label{fig:proj_psipi}
\end{center}
\end{figure}

\begin{figure}[!htbp]
\begin{center}
\includegraphics[width=8.5cm]{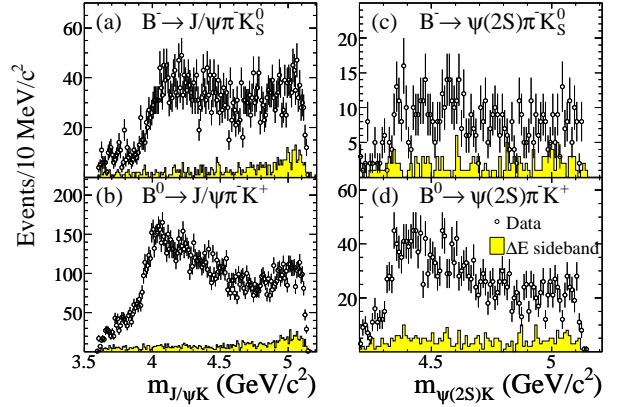}
\caption{The $m_{\psi K}$ mass projections for the Dalitz plots of
Fig.~\ref{fig:dp2}. The data points are for the \mes\/-\DeltaE\ signal
regions, and the filled histograms are for the \DeltaE\ sideband
regions.}
\label{fig:proj_psiK}
\end{center}
\end{figure}

The Dalitz plots of $m^2_{\psi\pi^-}$ versus $m^2_{K\pi^-}$ are shown
in Fig.~\ref{fig:dp2} for the signal regions defined in
Table~\ref{selections} for the $B$ meson decay modes specified in
Eqs.~(\ref{eq:modefour})-(\ref{eq:modesix}). The corresponding
$m_{K\pi^-}$, $m_{\psi\pi^-}$, and $m_{\psi K}$ mass projections are
represented by the data points in Figs.~\ref{fig:proj_kpi},
\ref{fig:proj_psipi}, and \ref{fig:proj_psiK}, respectively. In each
figure the filled histogram is obtained from the relevant \DeltaE\
sideband region.

In Fig.~\ref{fig:proj_kpi}, the contributions due to the \Ksone\
dominate the mass distributions. Small, but clear, \Kstwo\ signals are
evident for the \jpsi\ decay modes, and these seem to be present for
the \psitwos\ modes also. Previous
analyses~\cite{Aubert:2001pe,Aubert:2004cp} have shown that, for the
\jpsi\ modes, the region between the \Ksone\ and \Kstwo\ signals
($\sim 1.1-1.3$ \gevcc\/) contains a significant \kpi\ $S$-wave
contribution. In the \Ksone\ region, the presence of the $S$-wave
amplitude has been demonstrated through its interference with the
\Ksone\ $P$-wave amplitude~\cite{Aubert:2004cp}. This interference
yields a strong forward-backward asymmetry in the \kpi\ angular
distribution, as is seen in the vertical \Ksone\ bands of
Fig.~\ref{fig:dp2}(b) and Fig.~\ref{fig:dp2}(d), and as is shown in
Sec.~\ref{sec:kpi}, Figs.~\ref{fig:moment1}(a) and
Fig.~\ref{fig:moment1}(c). These features of the \kpi\ mass and
angular distributions will be analyzed in detail in
Secs.~\ref{sec:kpi} and \ref{sec:legndre} below.

The $m_{\psi\pi^-}$ distributions of Fig.~\ref{fig:proj_psipi} show no
peaking structure at the mass reported for the \z\/~\cite{:2007wga}
(indicated by the dashed vertical line in each figure). In
Fig.~\ref{fig:proj_psipi}(b) there seems to be a peak at $\sim4.65$
\gevcc\ and perhaps a weaker one just below 4.4 \gevcc\/, while in
Figs.~\ref{fig:proj_psipi}(c) and (d) there seems to be a peak just
below $\sim4.5$ \gevcc\/. These features are discussed in
Secs.~\ref{sec:comparison} and \ref{sec:babar_fits} in conjunction
with reflections resulting from the \kpi\ mass and angular structures.

Similarly, the $m_{\psi K}$ distributions of Fig.~\ref{fig:proj_psiK}
show no evidence of narrow structure. In fact, in overall shape these
distributions approximate mirror images of those in
Fig.~\ref{fig:proj_psipi}. This is not unexpected if both result
primarily from \kpi\ reflection, since then the high-mass region of
one distribution would be correlated strongly with the low mass region
of the other, and {\it vice versa}. Since Fig.~\ref{fig:proj_psiK}
shows no evidence of interesting features, and since our emphasis in
this paper is on the search for the \z\/, we do not discuss the $\psi
K$ systems any further in the present analysis.

We note that in Figs.~\ref{fig:proj_kpi}-\ref{fig:proj_psiK}, and in
other invariant mass distributions to follow, the \jpsi\/-\psitwos\
mass difference causes significant differences in the range spanned in
the respective decay modes. This should be kept in mind when making
\jpsi\/-\psitwos\ comparisons.
\begin{table*}
  \begin{center}
    \caption{The data samples used in the analysis.}
    \begin{tabular*}{1.0\textwidth}{@{\extracolsep{\fill}}lccc} \hline\hline \\
      Decay mode     &   Signal region events     &  \DeltaE\ sideband events           &  Net analysis sample \\ \\ \hline\hline \\
      \modeFour\     & $4229\pm65$         &  $485\pm22$             & $3744\pm68$         \\ \\
      \modeEight\    & $14251\pm119$       &  $1269\pm36$            & $12982\pm124$       \\ \\ 
      \modeTwo\      & $703\pm26$          &  $161\pm13$             & $542\pm29$          \\ \\
      \modeSix\      & $2405\pm49$         &  $384\pm20$             & $2021\pm53$         \\ \\ \hline \hline
    \end{tabular*}
    \label{table_sig_bkg}
\end{center}
\end{table*}

\section{The \psipi\ mass resolution}
\label{sec:resolution}
In Ref.~\cite{:2007wga}, the width of the \z\ is given as
$45^{+35}_{-18}$ \mev\/, where we have combined statistical and
systematic errors in quadrature. This value is very similar to that of
the \Ksone\/~\cite{Yao:2006px}, although with larger uncertainties,
which we have no difficulty observing, as shown in
Fig.~\ref{fig:proj_kpi}. However, mass resolution degrades with
increasing $Q$-value, where $Q$-value is the difference between the
invariant mass value in question and the corresponding threshold mass
value. Since the $Q$-value for the \Ksone\ is only $\sim 260$
\mevcc\/, while that for the \z\ is $\sim 600$ \mevcc\ in the
$\psi(2S)\pi$ mode, and would be $\sim 1200$ \mevcc\ in the
$J/\psi\pi$ mode, the mass resolution should be worse for the \z\ than
for the \Ksone\/, and hence should be systematically investigated.

\begin{figure}[!htbp]
\begin{center}
\includegraphics[width=8.5cm]{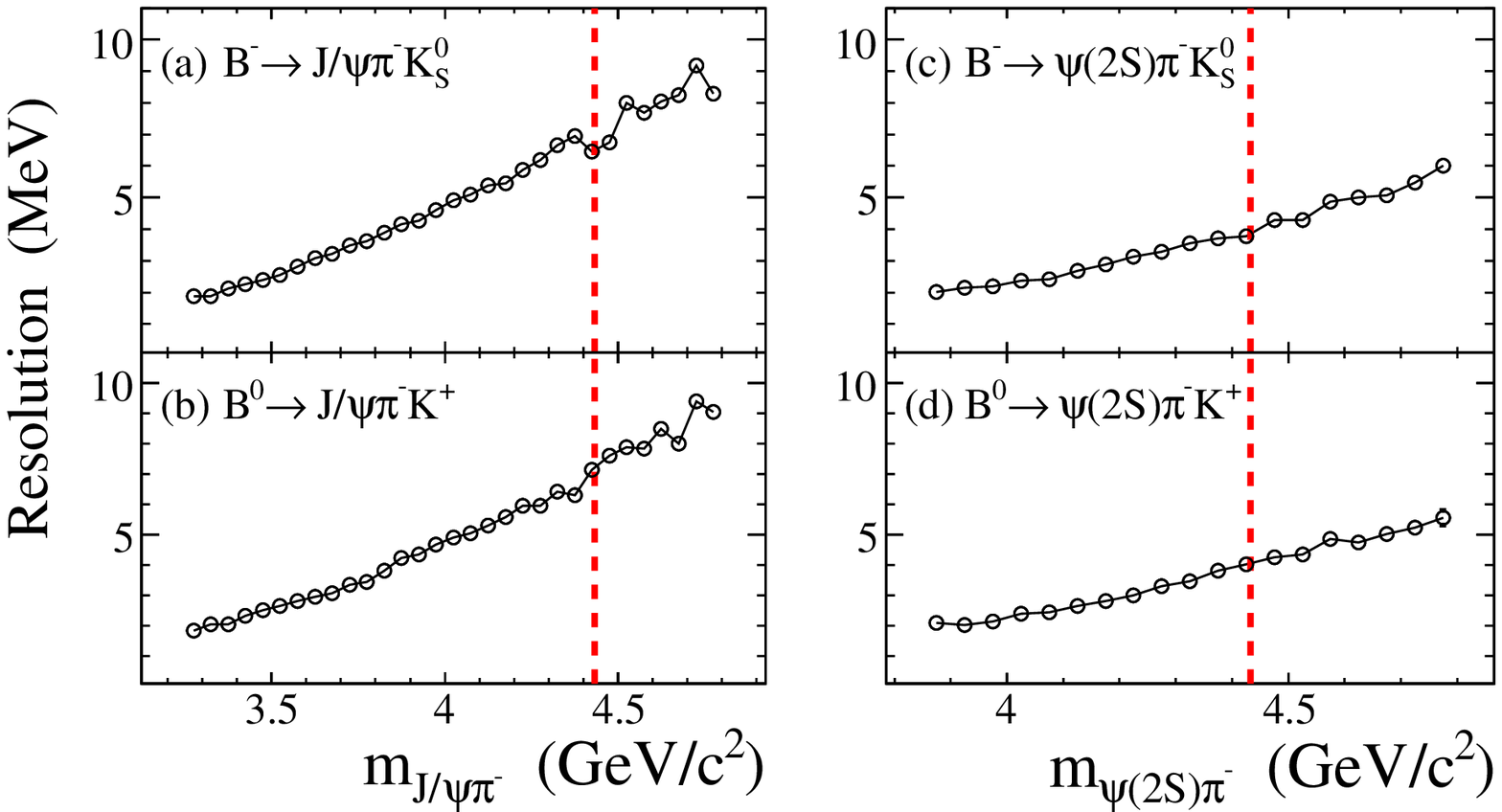}
\caption{The \psipi\ mass resolution in 50 \mevcc\ intervals as a
function of \psipi\ mass for the decay modes (a) \modeFour\/, (b)
\modeEight\/, (c) \modeTwo\/, and (d) \modeSix\/. The dashed vertical
lines indicate $m_{\psi\pi^-}=4.433$ \gevcc\/.}
\label{fig:res}
\end{center}
\end{figure}

We do this by using MC simulated data for the $B$ meson decay modes of
Eqs.~(\ref{eq:modefour})-(\ref{eq:modesix}). For each mode, the
reconstructed events were divided into 50 \mevcc\ intervals of \psipi\
mass, and within each interval it was found that the distribution of
(reconstructed - generated) \psipi\ mass could be described by a
double Gaussian function with a common mean at zero. The local \psipi\
mass resolution is characterized by the half-width-at-half-maximum
(HWHM) value for this line shape, and the dependence of this quantity
on \psipi\ mass is shown in Fig.~\ref{fig:res} for the individual
decay modes. For both \jpsi\ modes, the resolution varies from $\sim
2$ \mevcc\ at threshold to $\sim 9$ \mevcc\ at the maximum mass value,
while for both $\psi(2S)\pi$ modes, the variation is from $\sim 2-6$
\mevcc\/. At the \z\ mass value, indicated by the dashed vertical
lines in Fig.~\ref{fig:res}, the resolution is $\sim 4$ \mevcc\ for
$\psi(2S)\pi$, and $\sim 7$ \mevcc\ for $J/\psi\pi$. We note that, for
$J/\psi\pi$, the resolution at a $Q$-value of $\sim 600$ \mevcc\ is
essentially the same as for $\psi(2S)\pi$ at the \z\/. It follows that
failure to observe the \z\ in its \jpsipi\ or \psitwospi\ decay mode
in the present experiment should not be attributed to inadequate mass
resolution.

\section{Efficiency correction}
\label{sec:efficiency}
In the search for the \z\/, a detailed understanding of event
reconstruction efficiency over the entire final state Dalitz plot for
each of the $B$ meson decay processes of
Eqs.~(\ref{eq:modefour})-(\ref{eq:modesix}) is necessary. This is
because efficiency variation can, in principle, lead to the creation
of spurious signals or to the distortion of real effects such that
their significance is reduced. Even when the process of efficiency
correction leads to no significant change in the interpretation of the
data, it is important to demonstrate clearly that this is in fact the
case.

The efficiency correction procedure which we follow is described in
detail in Appendix~\ref{append1}. For reasons discussed there, we use
a ``rectangular Dalitz plot'' for which the variables are chosen to be
$m_{K\pi^-}$ and $\cos\theta_K$, the normalized dot-product between
the \kpi\ three-momentum vector in the parent-$B$ rest frame and the
kaon three-momentum vector after a Lorentz transformation from the $B$
rest frame to the \kpi\ rest frame. For the Dalitz plots shown in
Fig.~\ref{fig:dp2}, the $y$-axis variable, $m^2_{\psi\pi^-}$, varies
linearly with $\cos\theta_K$.

\begin{figure}[!htbp]
\begin{center}
\includegraphics[width=8.5cm]{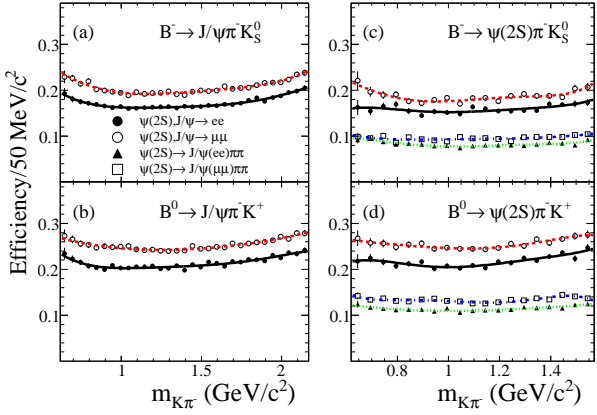}
\caption{The $m_{K\pi^-}$ dependence of the average efficiency, $E_0$,
for the $B$ meson decay processes (a) \modeFour\/, (b) \modeEight\/, (c)
\modeTwo\/, and (d) \modeSix\/. The key in (a) applies to all four
figures, and, as indicated, the individual \jpsi\ and \psitwos\ decay
modes are treated separately. The curves result from fifth-order
polynomial fits to the data points.}
\label{fig:effi_modes}
\end{center}
\end{figure}

The average efficiency $E_0$ depends on $\psi$ decay mode, as shown in
Fig.~\ref{fig:effi_modes}. The individual fitted curves in this figure
are used to calculate the value of $E_0$ for an event at a particular
value of $m_{K\pi^-}$ which has been reconstructed in the relevant
$\psi$ decay mode.

As explained in Appendix~\ref{append1}, $E_0$ is then modulated by a
linear combination of 12 Legendre polynomials in $\cos\theta_K$, whose
multiplicative coefficients $E_1-E_{12}$ are obtained from curves
representing their individual $m_{K\pi^-}$ dependence. In this way, an
efficiency value can be calculated for each reconstructed event in our
data sample. The inverse of this efficiency then provides a
weight-factor which is associated with this event in any distribution
to which it contributes. This enables us to correct any distribution
under study for efficiency-loss effects.

\begin{figure*}[!htbp]
\begin{center}
\includegraphics[width=15.0cm]{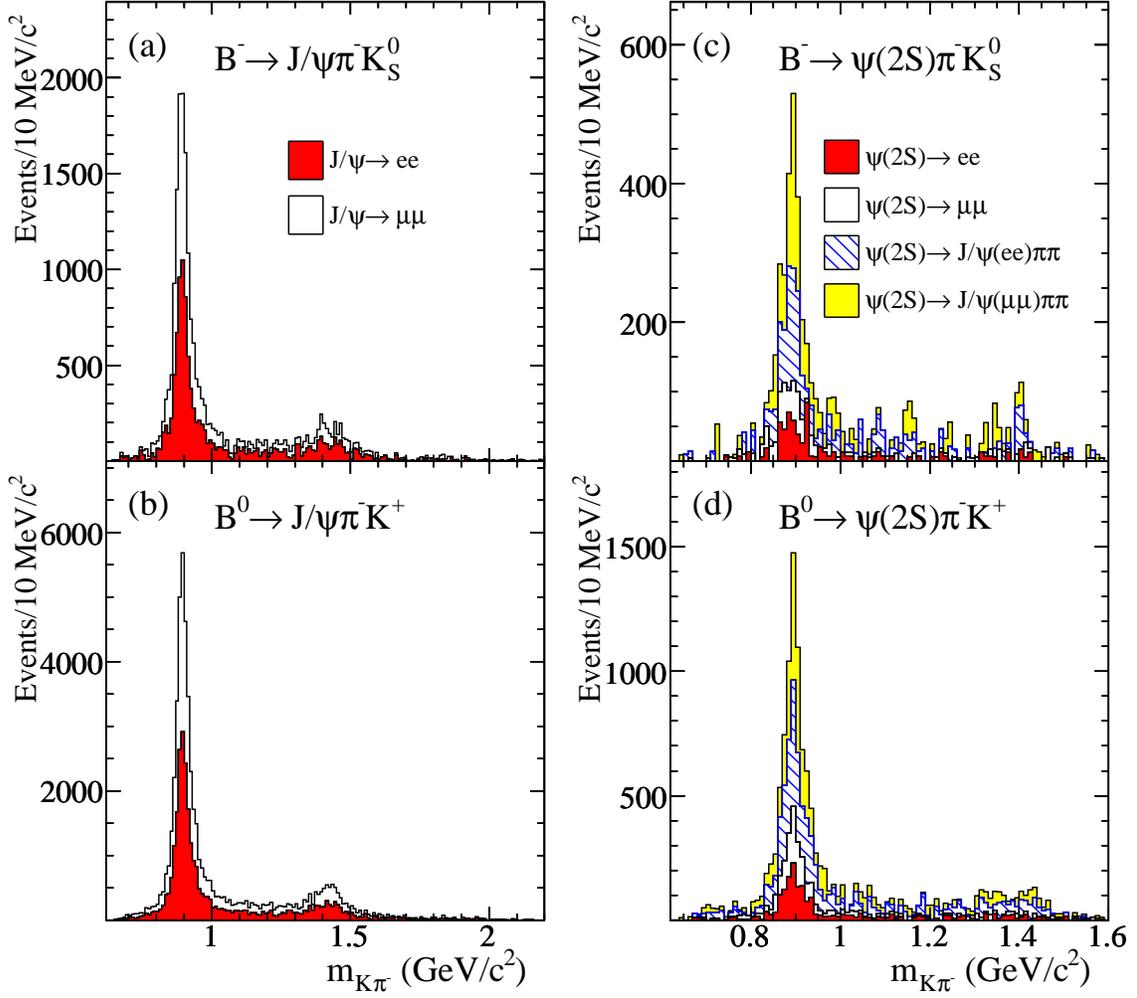}
\caption{The \kpi\ mass distributions, after sideband subtraction and
efficiency correction, for the decay modes (a) \modeFour\/, (b)
\modeEight\/, (c) \modeTwo\/, and (d) \modeSix\/. The contributions
from the individual $\psi$ decay modes are obtained separately, as
described in the text, and are accumulated as indicated by the keys in
(a) and (c), to form the final histograms.}
\label{fig:kpi_psimodes}
\end{center}
\end{figure*}

A specific example is provided by the \kpi\ mass distributions of
Fig.~\ref{fig:kpi_psimodes}. A weight value for each event is
calculated according to its particular $\psi$ decay mode, as described
above. The histograms of Fig.~\ref{fig:kpi_psimodes} are then formed
by summing these weights in each mass interval of each plot. Sideband
subtraction is accomplished by assigning sideband events negative
weight. The contributions from the different $\psi$ decay modes are
distinguished by shading, and the final histograms represent the sum of
these contributions.

\begin{figure}[!htbp]
\begin{center}
\includegraphics[width=8.5cm]{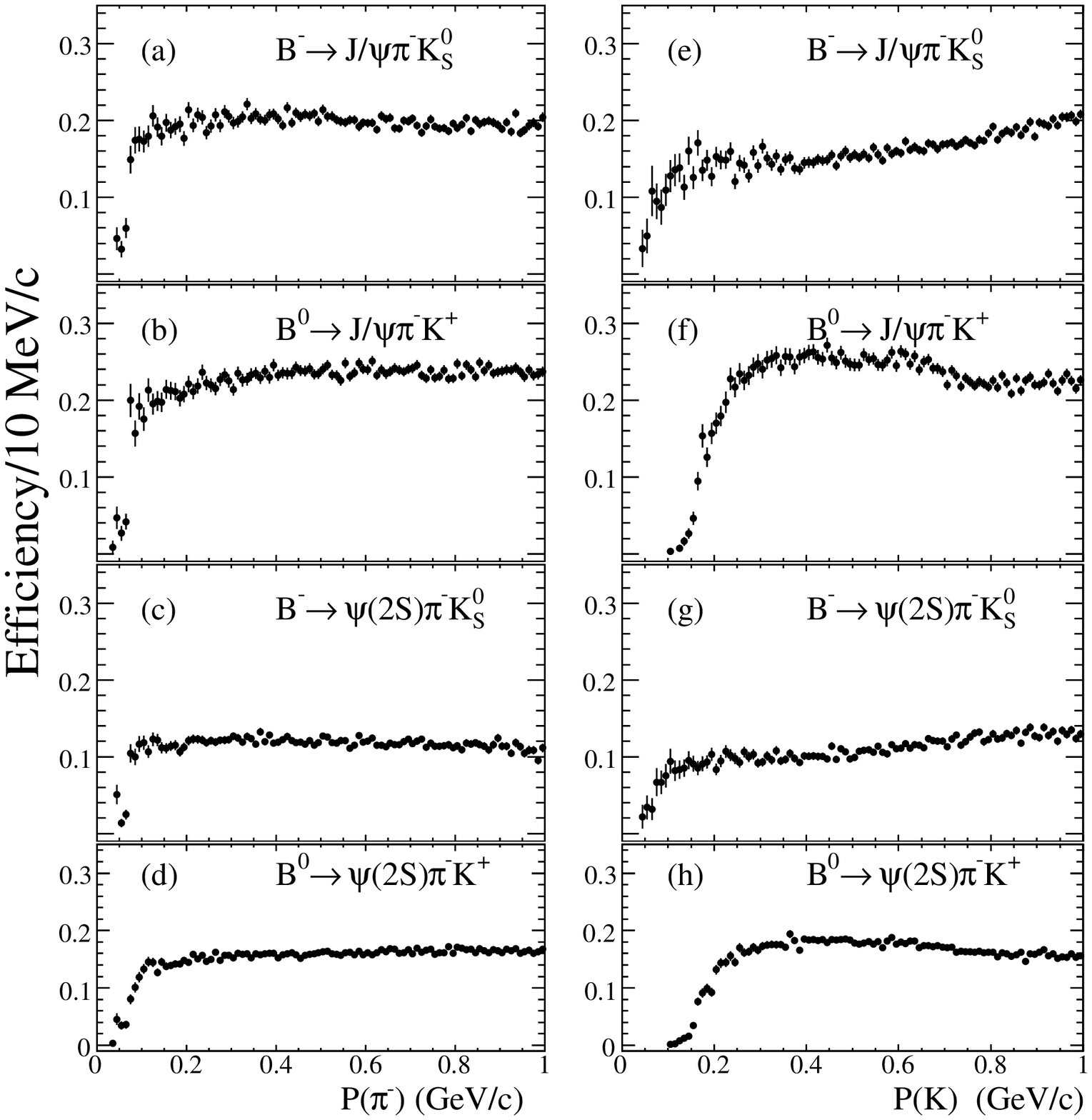}
\caption{(a)-(d): The dependence of efficiency on pion momentum in the
 laboratory frame for the $B$ meson decay modes of
 Eqs.~(\ref{eq:modefour})-(\ref{eq:modesix}). (e)-(h): The
 corresponding efficiency dependence on kaon laboratory frame
 momentum.}
\label{fig:effpik}
\end{center}
\end{figure}

In Appendix~\ref{append1} it is pointed out that the use of high-order
Legendre polynomials is necessary because of significant decrease in
efficiency for $\cos\theta_K\sim +1$ and $0.72<m_{K\pi^-}<0.92$
\gevcc\/, and for $\cos\theta_K\sim -1$ and $0.97<m_{K\pi^-}<1.27$
\gevcc\/. The former loss is due to the failure to reconstruct
low-momentum pions in the laboratory frame, and the latter is due to a similar
failure to reconstruct low-momentum kaons. The dependence of
efficiency on laboratory frame momentum is shown in Fig.~\ref{fig:effpik}.  In
Figs.~\ref{fig:effpik}(a)-(d), there is a significant decrease in
efficiency for pions of momentum $<0.1$ \gevc\/, while in
Figs.~\ref{fig:effpik}(f),(h) there is similar decrease for charged
kaons below $\sim 0.25$ \gevc\/. For $K^0_S$, the effect is similar to
that for charged pions (Figs.~\ref{fig:effpik}(e),(g)).

The Lorentz boost from the laboratory frame to the \kpi\ rest frame translates the
laboratory frame losses into the losses localized in $m_{K\pi^-}$ and $\cos\theta_K$
which our efficiency study reveals. The effect of these regions of low
efficiency on the uncorrected $\psi\pi$ mass distributions is
discussed in Sec.~\ref{sec:reflection}.

\section{Fits to the \kpi\ mass distributions}
\label{sec:kpi}
The most striking aspects of the Dalitz plots of Fig.~\ref{fig:dp2}
pertain to the \kpi\ system, as shown explicitly in
Fig.~\ref{fig:proj_kpi}. In order to investigate the features of the
\psipi\ distributions of Fig.~\ref{fig:proj_psipi}, it is necessary to
understand how structure in the \kpi\ mass and $\cos\theta_K$
distributions reflects into the \psipi\ system. We begin this process
by making detailed fits to the sideband-subtracted and
efficiency-corrected \kpi\ mass distributions of
Fig.~\ref{fig:kpi_psimodes}.

As discussed in Sec.~\ref{sec:dalitz}, we expect that the \kpi\ system
can be described in terms of a superposition of $S$-, $P$-, and
$D$-wave amplitudes. Since we correct for \costhk\ dependence of the
efficiency, it follows that the corrected \kpi\ mass distributions of
Fig.~\ref{fig:kpi_psimodes} can be described by a sum of $S$-, $P$-,
and $D$-wave intensity contributions, since any interference terms
vanish when integrated over \costhk\/. Consequently, we describe the
\mkpi\ mass projections as follows,
\begin{eqnarray}
\lefteqn{\frac{dN}{dm_{K\pi}} = N \times } \\
                               & \left[ f_S\left(\frac{G_S}{\int G_S dm_{K\pi}} \right)
                               +f_P\left(\frac{G_P}{\int G_P dm_{K\pi}} \right)
                               +f_D\left(\frac{G_D}{\int G_D dm_{K\pi}} \right) \right ] \nonumber \, ,
\end{eqnarray}
where the integrals are over the full $m_{K\pi}$ range and the
fractions $f$ are such that
\begin{equation}
 f_S+f_P+f_D=1 \, .
\end{equation}
The $P$- and $D$-wave intensities, $G_P$ and $G_D$, are expressed in
terms of the squared moduli of Breit-Wigner (BW) amplitudes. For the
$P$-wave,
\begin{eqnarray}
G_P(m_{K\pi})=\frac{B_P(m_{K\pi})(p\cdot q) \frac{q^2}{D_P(qR_P)}}{(m_P^2-m^2_{K\pi})^2+m_P^2\Gamma^2_P(q,R_P)} \, , 
\end{eqnarray}
where 
\begin{itemize}
\item $B_P(m_{K\pi})$ describes the $B$-decay vertex;
\item $p$ is the momentum of the $\psi$ in the $B$ rest frame;
\item $q$ is the momentum of the $K$ in the \kpi\ rest frame;
\item $D_P$ is the $P$-wave Blatt-Weisskopf barrier factor with radius
$R_P$~\cite{Blatt};
\item the mass-dependent total width is
\end{itemize}
\begin{eqnarray}
\Gamma_P= \Gamma^0_P\left ( \frac{q^2}{q^2_P}\right )\frac{D_P(q_PR_P)}{D_P(qR_P)}\left ( \frac{q}{q_P}\right )\left ( \frac{m_P}{m_{K\pi}}\right ) \, ,
\end{eqnarray}
with $q_P=q$ evaluated at $m_P$; $m_P$ is the mass, and $\Gamma^0_P$
the width, of the \Ksone\/. We leave the mass and width free in the
fits and choose $R_P=3.0$ GeV$^{-1}$~\cite{Aitala:2005yh}.

Similarly, for the $D$-wave,
\begin{eqnarray}
G_D(m_{K\pi})=\frac{B_D(m_{K\pi})(p\cdot q)\frac{q^4}{D_D(qR_D)}}{(m^2_D-m^2_{K\pi})^2+m^2_D\Gamma^2_D(q,R_D)} \, ,
\end{eqnarray} 
where 
\begin{itemize}
\item $B_D(m_{K\pi})$ describes the $B$-decay vertex;
\item $D_D$ is the $D$-wave Blatt-Weisskopf barrier factor with radius $R_D$~\cite{Blatt};
\item the mass-dependence of the total width is approximated by the
\kpi\ contribution as follows,
\end{itemize}
\begin{eqnarray}
  \Gamma_D=\Gamma^0_D\left(\frac{q^4}{q^4_D}\right)\frac{D_D(q_DR_D)}{D_D(qR_D)}\left(\frac{q}{q_D}\right)\left(\frac{m_D}{m_{K\pi}}\right) \, ,
\end{eqnarray}
with $q_D=q$ evaluated at $m_D$; $m_D$ is the mass, and $\Gamma^0_D$
the width of the \Kstwo\/. We fix $m_D$ and $\Gamma^0_D$ to their
nominal values~\cite{Yao:2006px} and choose $R_D=1.5$
GeV$^{-1}$~\cite{LASS}.

The $S$-wave contribution is described using the $I=1/2$ amplitude for
$S$-wave $K^-\pi^+$ elastic scattering~\cite{Aitala:2005yh}. We write
\begin{eqnarray}
G_S(m_{K\pi})=B_S(m_{K\pi})(p\cdot q) {|T_S|}^2 \, ,
\end{eqnarray}
where $B_S(m_{K\pi})$ describes the $B$-decay vertex; $T_S$ is the
invariant amplitude, which is related to $A_S$, the complex \kpi\
scattering amplitude, by
\begin{eqnarray}
\left |T_S\right |= \left(\frac{m_{K\pi}}{q} \right ) \left| A_S\right | \, .
\end{eqnarray}
For $m_{K\pi^-}>1.5$ \gevcc\/, $\left|A_S\right|$ is obtained by
interpolation from the measured values~\cite{LASS}. For lower mass
values, $A_S$ is a pure-elastic amplitude (within error) and is
parameterized as
\begin{eqnarray}
  \label{eq:elastic}
A_S=\frac{1}{\cot\delta_B -i} +
e^{2i\delta_B}\left(\frac{1}{\cot\delta_R-i}\right) \, ,
\end{eqnarray}
where the first term is non-resonant, and the second is a resonant
term rotated by $2\delta_B$ in order to maintain elastic unitarity. In
Eq.~\ref{eq:elastic},
\begin{eqnarray}
q\cot\delta_B = \frac{1}{a}+\frac{1}{2}rq^2 \, ,
\end{eqnarray}
with $a=1.94$ GeV$^{-1}$ and $r=1.76$ GeV$^{-1}$~\cite{LASS};
\begin{equation}
\cot\delta_R=\frac{m^2_S-m^2_{K\pi}}{m_S\Gamma_S} \, ,
\end{equation}
with 
\begin{equation}
\Gamma_S=\Gamma^0_S\left(\frac{q}{q_S}\right)\frac{m_S}{m_{K\pi}} \, ;
\end{equation}
$m_S$ ($=1.435$ \gevcc\/) is the mass of the $K^{\ast}_0(1430)$
resonance and $\Gamma^0_S$ ($=0.279$ \gev\/) is its
width~\cite{LASS}. We choose to fix the $S$-wave parameters to the
indicated values.

Although the vertex functions $B_S$, $B_P$, and $B_D$ depend, in
principle, on $m_{K\pi}$, we find that our best fits to the mass
distributions are obtained when each vertex function is set to one
($B_S=B_P=B_D=1$). In this regard, we emphasize that our goal is not
to obtain a precise amplitude decomposition of the \kpi\ mass
spectrum. This would require taking account of the angular
correlations between the $\psi$ decay products and the \kpi\
system. For \kpi\ $S$-, $P$-, and $D$-wave this is extremely
complicated~\cite{Stephane}, and is far beyond the scope of the
present analysis. Our aim is to obtain an accurate description of the
\kpi\ mass distributions in terms of the expected angular momentum
contributions so that we can reliably project the observed structures
onto the related \psipi\ mass distributions.

\begin{figure*}[!htbp]
\begin{center}
\includegraphics[width=17.0cm]{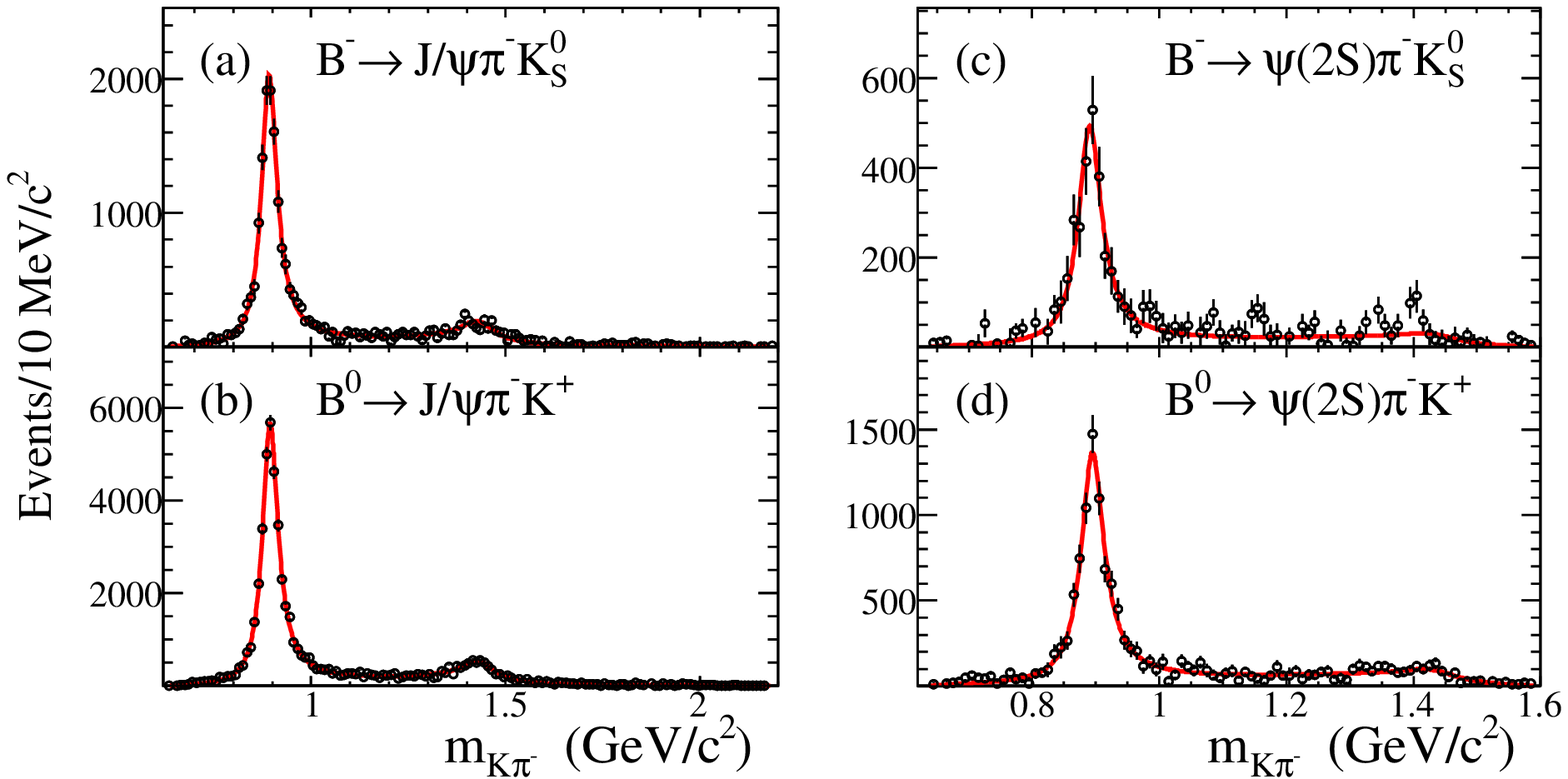}
\caption{The efficiency-corrected and sideband-subtracted \kpi\ mass
distributions of Fig.~\ref{fig:kpi_psimodes} for the decay modes (a)
\modeFour\/, (b) \modeEight\/, (c) \modeTwo\/, and (d) \modeSix\/. The
data are shown as open dots, and the curves correspond to the fits
described in the text.}
\label{fig:4kpi_lin}
\end{center}
\end{figure*}

The results of the fits to the \mkpi\ distributions of
Fig.~\ref{fig:kpi_psimodes} are shown by the curves in
Fig.~\ref{fig:4kpi_lin}. Good descriptions are obtained, even though
each fit has only five free parameters and the fit function has
exactly the same structure in each case.

Table~\ref{table_kpi_numbers} summarizes the output from the fits. The
$\chi^2/NDF$ ($NDF=$ Number of Degrees of Freedom) values are
satisfactory, and the mass and width values for the $K^{\ast}(892)^0$
and $K^{\ast}(892)^-$ modes are internally consistent and agree with
their nominal values~\cite{Yao:2006px}; the width values for the
\psitwos\ modes are slightly low, but the uncertainties are quite
large because of the smaller data sample involved.
\begin{table*}
  \begin{center}
    \caption{Summary of the fit results corresponding to
    Fig.~\ref{fig:4kpi_lin}. For each decay mode, we list the total
    number of $K\pi$ events after efficiency-correction and sideband-subtraction, $\chi^2/NDF$, \Ksone\ mass and width, and
    the percentage $S$-, $P$-, and $D$-wave intensity
    contributions. Only statistical uncertainties are given.}
    \begin{tabular*}{1.0\textwidth}{@{\extracolsep{\fill}}lccccccc} \hline\hline\\
    Mode & Corrected Events & $\chi^2/NDF$ & $m(K^{\ast}(892))$ & $\Gamma(K^{\ast}(892))$ & $S$-wave & $P$-wave & $D$-wave  \\ 
         &        &                      & (\mevcc\/)         & (\mev\/)                & ($\%$)   &  ($\%$)  & ($\%$)    \\ \\ \hline\hline \\
    \modeFour  & 20985$\pm$393 & 117.6/149 & 892.9$\pm$0.8 & 49.0$\pm$1.9& 17.0$\pm$1.6 & 72.5$\pm$1.3 & 10.5$\pm$1.0 \\ \\ 
    \modeEight & 57231$\pm$561 & 171.4/149 & 895.5$\pm$0.4 & 48.9$\pm$1.0& 15.7$\pm$0.8 & 73.5$\pm$0.7 & 10.8$\pm$0.5 \\ \\ 
    \modeTwo   &  5016$\pm$292 &  98.1/90  & 891.6$\pm$2.1 & 44.8$\pm$6.0& 23.4$\pm$4.5 & 71.3$\pm$4.4 &  5.3$\pm$2.7 \\ \\ 
    \modeSix   & 13237$\pm$377 &  81.5/90  & 895.8$\pm$1.0 & 43.8$\pm$3.0& 25.4$\pm$2.2 & 68.2$\pm$2.0 &  6.4$\pm$1.2 \\ \\  \hline \hline 
   \end{tabular*}
\label{table_kpi_numbers}
\end{center}
\end{table*}

The fractional contributions for the two \jpsi\ modes agree well with
each other, and there is similar agreement for the \psitwos\
modes. This demonstrates that the charged and neutral \kpi\
distributions are very similar in shape, and so we combine them for
the remainder of the analysis, unless we explicitly indicate
otherwise.

\begin{figure*}[!htbp]
\begin{center}
\includegraphics[width=17.0cm]{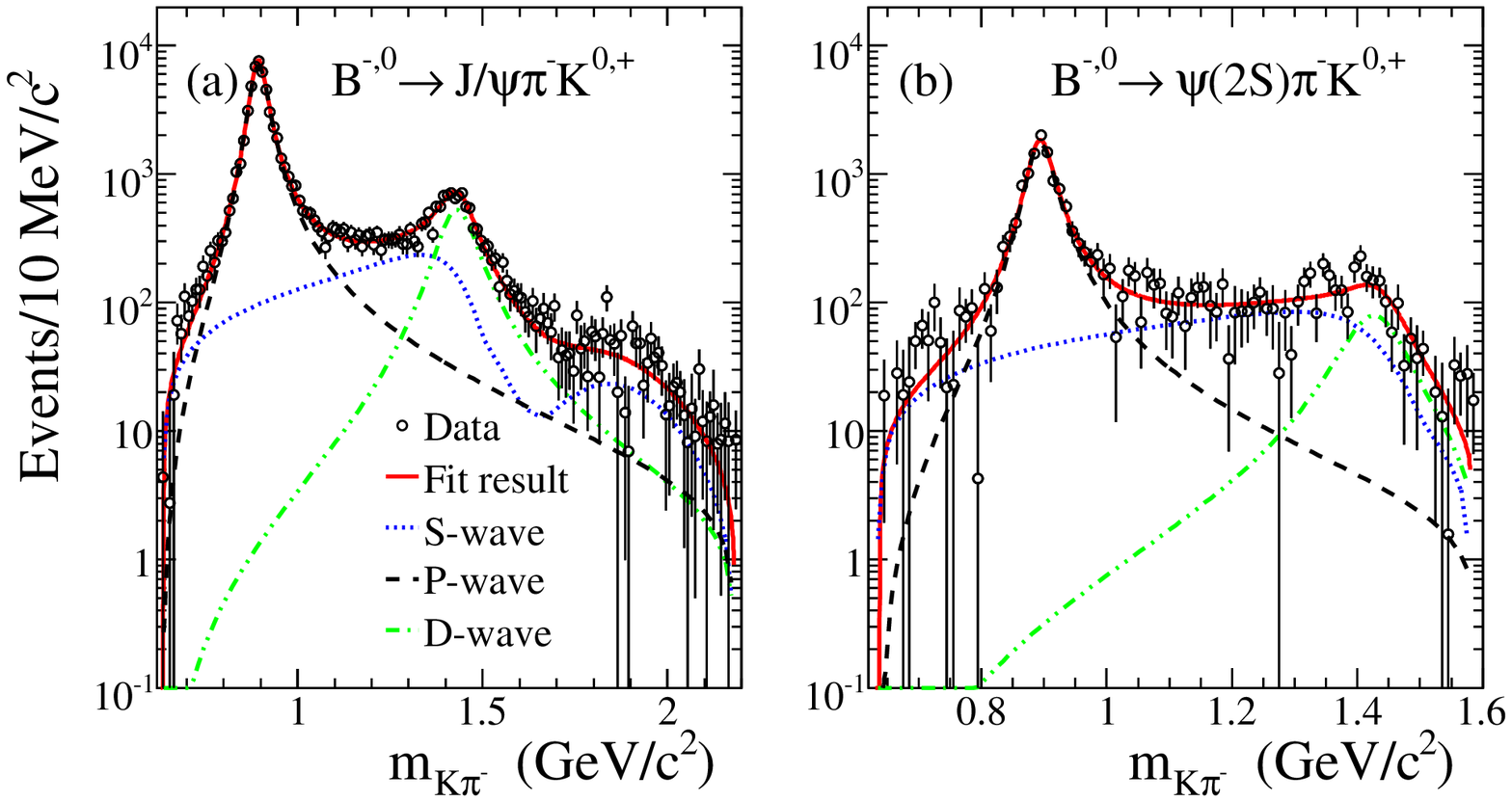}
\caption{The results of the fits to the \kpi\ mass distributions for
the combined \kpi\ charge configurations (a) $B^{-,0}\rightarrow
J/\psi\pi^- K^{0,+}$ and (b) $B^{-,0}\rightarrow \psi(2S)\pi^-
K^{0,+}$. The data are shown as open dots, and the individual fit
contributions are as indicated.}
\label{fig:kpi_combined}
\end{center}
\end{figure*}

The results of repeating the fits for the combined distributions are
shown in Fig.~\ref{fig:kpi_combined}, where we use a logarithmic
$y$-axis scale in order to display the individual contributions more
clearly. The solid curves describe the distributions very well. For
$m_{K\pi}\sim 0.7$ \gevcc\/, the curves in both plots are slightly
below the data, and we believe that this results from the
parameterization of the $S$-wave amplitude at low \kpi\ mass
values. If we normalize the $S$-wave amplitudes of Ref.~\cite{LASS}
and Ref.~\cite{Aitala:2005yh} at 1.2 \gevcc\ and average them from
threshold to 1.2 \gevcc\/, the discrepancy is removed. However, the
fits become slightly worse in the region between the two
$K^{\ast}$'s. Since the latter region is very important to the present
analysis, while the region around 0.7 \gevcc\ is much less so, we do
not make use of the modified $S$-wave amplitude.

Based on Table~\ref{table_kpi_numbers}, the following branching
fractions can be calculated:
\begin{eqnarray}
{\cal B}(B^-\rightarrow J/\psi\pi^- \bar{K}^0)     &=& (1.101\pm0.021)\times 10^{-3} \nonumber \, , \\ \\
\label{eq:brch8}
{\cal B}(B^0\rightarrow J/\psi\pi^- K^+)     &=& (1.079\pm0.011)\times 10^{-3} \nonumber \, ,\\ \\
{\cal B}(B^-\rightarrow \psi(2S)\pi^- \bar{K}^0)   &=& (0.588\pm0.034)\times 10^{-3} \nonumber \, , \\ \\
\label{eq:brch6}
{\cal B}(B^0\rightarrow \psi(2S)\pi^- K^+)   &=& (0.557\pm0.016)\times 10^{-3} \nonumber \, , \\ 
\end{eqnarray}
where we have corrected for the relevant \jpsi\/, \psitwos\/,
$K^0_S\rightarrow\pi^0\pi^0$, and $K^0_L$ branching
fractions~\cite{Yao:2006px}. The quoted errors result from the
statistical uncertainties of Table~\ref{table_kpi_numbers}. The
published value corresponding to Eq.~(\ref{eq:brch8}) is $(1.2\pm
0.6)\times 10^{-3}$~\cite{Yao:2006px}, and for the mode of
Eq.~(\ref{eq:brch6}) an upper limit of $1\times 10^{-3}$ ($90\%$
confidence level (c.l.)) is quoted~\cite{Yao:2006px}. No other
information on these branching fraction values exists; the present
measurements thus represent significant improvements, even in the
absence of systematic error studies.

We note that the charged and neutral $B$ meson decay rates to
$J/\psi\pi^- K$ agree very well, and that this is true also for decay to
$\psi(2S)\pi^- K$; also, the latter decays occur at slightly more than
half the rate of the former.

\section{The \kpi\ Legendre polynomial moments}
\label{sec:legndre}
At the beginning of Sec.~\ref{sec:kpi} we pointed out the need to
understand the \kpi\ mass dependence of the angular structure in the
\kpi\ system. In order to do this, we choose to represent the \kpi\
angular distribution at a given \mkpi\ in terms of a Legendre
polynomial expansion, following much the same procedure as described
in Appendix~\ref{append1} for our efficiency studies. In the notation
of Eq.~(\ref{eq:dndcosthk}), we write
\begin{equation}
\label{eq:legndre}
\frac{dN}{d\cos\theta_K} = N\sum_{i=0}^{L}\langle P_i\rangle P_i(\cos\theta_K) \, ,
\end{equation}
where $N$ is the number of events (after correction) in a small mass
interval centered at \mkpi\/, and $L=2\ell_{max}$, where $\ell_{max}$
is the maximum orbital angular momentum required to describe the \kpi\
system at \mkpi\/. We can re-write Eq.~(\ref{eq:legndre}) as
\begin{eqnarray}
\label{eq:22}
\frac{dN}{d\cos\theta_K}=\frac{N}{2} + \sum_{i=1}^{L}\left(N\langle P_i\rangle\right)P_i(\cos\theta_K) \, ,
\end{eqnarray}
and extract the coefficients from the data using
\begin{eqnarray}
\label{eq:23}
N\langle P_i\rangle\approx\sum_{j=1}^{N}P_i(\cos\theta_{K_{j}}) \, ,
\end{eqnarray}
as in Appendix~\ref{append1}. We refer to this coefficient as ``the
unnormalized $P_i$ moment'' in the course of our discussion. In order
to incorporate the efficiency weighting and sideband subtraction
procedures, we extend Eq.~(\ref{eq:23}) as follows:
\begin{eqnarray}
\label{eq:25}
N\langle P_i\rangle &\approx& \sum_{j=1}^{N}\left(\frac{1}{\epsilon_j}\right)P_i(\cos\theta_{K_j}) \nonumber \\
                  &+&\sum_{k=1}^{N_{SB}}\left(\frac{-1}{\epsilon_k}\right)P_i(\cos\theta_{K_k}) \, ,
\end{eqnarray}
where $N_{SB}$ is the number of sideband events falling in this \mkpi\
interval, and the efficiency values, $\epsilon$, are obtained as
described in Appendix~\ref{append1}. For convenience, we introduce the
notation $\langle P_i^U\rangle =N\langle P_i\rangle $, where the
superscript $U$ indicates that we refer to the unnormalized moment,
and re-write Eq.~(\ref{eq:22}) as
\begin{equation}
\label{eq:26}
\frac{dN}{d\cos\theta_K}=\frac{N}{2}+\sum_{i=1}^{L}\langle  P_i^U\rangle P_i(\cos\theta_K) \, .
\end{equation}

An overview of the $(m_{K\pi^-},\cos\theta_K)$ structure is provided
by the rectangular Dalitz plots (see Appendix~\ref{append1}) of
Fig.~\ref{fig:dp3}. Data for the $B$ decay modes of
Eqs.~(\ref{eq:modefour}) and~(\ref{eq:modeeight}) have been combined
in Fig.~\ref{fig:dp3}(a), and data for those of
Eqs.~(\ref{eq:modetwo}) and~(\ref{eq:modesix}) have been combined in
Fig.~\ref{fig:dp3}(b). In Fig.~\ref{fig:dp3}(a), the intensity for
$\cos\theta_K<0$ is stronger than for $\cos\theta_K>0$ in the region
$m_{K\pi^-}<0.85$ \gevcc\/. A similar asymmetry is present in the
\Ksone\ band, where in addition there is a clear decrease in intensity
around $\cos\theta_K=0$. Above $\sim 1.2$ \gevcc\ the backward region
is enhanced, especially in the increased intensity region at the
\Kstwo\/. For $m_{K\pi^-}>1.5$ \gevcc\ the overall intensity is
significantly decreased, but the backward region of \costhk\ continues
to be favored. Despite the smaller data sample of
Fig.~\ref{fig:dp3}(b), the backward region seems again to be favored
in the \Ksone\ and \Kstwo\ regions, but little more can be said.
\begin{figure*}[!htbp]
\begin{center}
\includegraphics[width=17.0cm]{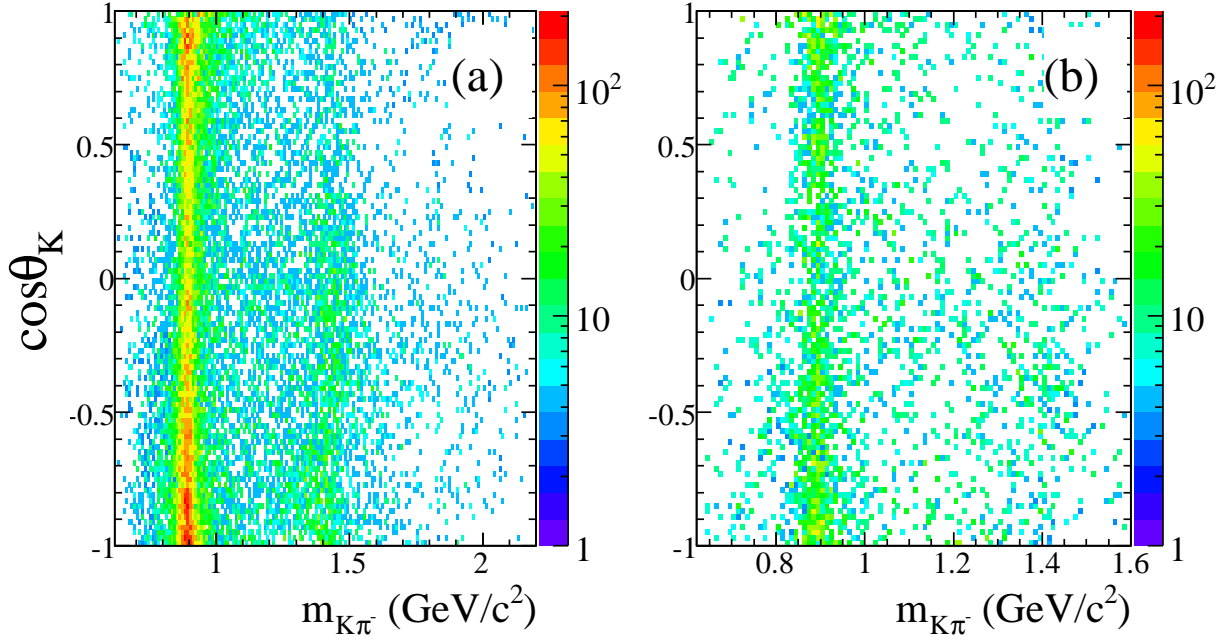}
\caption{The \costhk\ versus \mkpi\ rectangular Dalitz plots for the
combined decay modes (a) $B^{-,0}\rightarrow J/\psi\pi^- K^{0,+}$, (b)
$B^{-,0}\rightarrow \psi(2S)\pi^- K^{0,+}$. The plots are obtained
after efficiency weighting, but without \DeltaE\ sideband
subtraction. The intensity scale is logarithmic, and is the same for
both plots.}
\label{fig:dp3}
\end{center}
\end{figure*}

For both Fig.~\ref{fig:dp3}(a) and~\ref{fig:dp3}(b) it is necessary to
examine the \mkpi\ dependence of the unnormalized moments in order
to quantify these qualitative features.

In order to facilitate discussion of the \mkpi\ dependence of the
unnormalized moments, we first express them in terms of $S$-, $P$- and
$D$-wave \kpi\ amplitudes. These expressions have been obtained from
the $B\rightarrow J/\psi\pi K$ analysis of Ref.~\cite{Stephane}, after
integration over the \jpsi\ decay angles; they apply equally to the
$B$ meson decay to $\psi(2S)\pi K$.

For an interval of \mkpi\ containing $N$ events:
\begin{eqnarray}
\label{eq:heli1}
N=\mbox{\boldmath{$S^2_0+P^2_0+D^2_0$}}+P^2_{+1}+P^2_{-1}+D^2_{+1}+D^2_{-1} \, 
\end{eqnarray}
\begin{eqnarray}
\label{eq:heli2}
\langle P^U_1\rangle&=& \mbox{\boldmath{ $\displaystyle S_0P_0\cos(\delta_{S_0}-\delta_{P_0})$}}  \\
                  &+& \mbox{\boldmath{ $\displaystyle 2\sqrt{\frac{2}{5}}P_0D_0\cos(\delta_{P_0}-\delta_{D_0})$}} \nonumber  \\
                  &+& \sqrt{\frac{6}{5}} [P_{+1}D_{+1}\cos(\delta_{P_{+1}}-\delta_{D_{+1}}) \nonumber \\
                  &+& P_{-1}D_{-1}\cos(\delta_{P_{-1}}-\delta_{D_{-1}}) ] \nonumber  \, ,
\end{eqnarray}
\begin{eqnarray}
\label{eq:heli3}
\langle P_2^U\rangle&=& \mbox{\boldmath{$\displaystyle \sqrt{\frac{2}{5}}P^2_0+\frac{\sqrt{10}}{7}D^2_0$ }}\\ 
                  &+& \mbox{\boldmath{$\displaystyle \sqrt{2}S_0D_0\cos(\delta_{S_0}-\delta_{D_0})$}} \nonumber \, \\
                  &-&\left(\frac{1}{\sqrt{10}}\left(P^2_{+1}+P^2_{-1}\right) +\frac{5\sqrt{10}}{28}\left(D^2_{+1}+D^2_{-1}\right)\right)\nonumber  \, ,
\end{eqnarray}

\begin{eqnarray}
\label{eq:heli4}
\langle P_3^U\rangle&=& \mbox{\boldmath{ $\displaystyle {3\sqrt{\frac{6}{35}}P_0D_0\cos(\delta_{P_0}-\delta_{D_0})}$}} \\
&-& 3\sqrt{\frac{2}{35}}(P_{+1}D_{+1}\cos(\delta_{P_{+1}}-\delta_{D_{+1}}) \nonumber \\
&+&P_{-1}D_{-1}\cos(\delta_{P_{-1}}-\delta_{D_{-1}})) \nonumber \, ,
\end{eqnarray}

\begin{eqnarray}\label{eq:heli5}
\langle P_4^U\rangle=\mbox{\boldmath{$\displaystyle\frac{3\sqrt{2}}{7}D^2_0$}}-\frac{2\sqrt{2}}{7}\left(D^2_{+1}+D^2_{-1}\right) \, .
\end{eqnarray}

The $S_i$, $P_i$, and $D_i$ are amplitude magnitudes and $i$ denotes
the relevant helicity; the corresponding phase angles are denoted by
$\delta$ with the appropriate subscript. The helicity-zero terms,
denoted in bold-face font, provide the corresponding description of
$K^-\pi^+$ elastic scattering.
Equations~(\ref{eq:heli1})-(\ref{eq:heli5}) define the five measurable
quantities accessible to the present analysis if we restrict ourselves
to $S$-, $P$-, and $D$-wave \kpi\ amplitudes. However, the equations
involve seven amplitude magnitudes and six relative phase values, and
so they cannot be solved in each \mkpi\ interval. For this reason, we
can only measure the mass dependence of the moments of the \kpi\
system, and then use the results to understand how the \mkpi\ and
\costhk\ structure reflects into the observed \psipi\ mass
distributions, as will be discussed in Sec.~\ref{sec:reflection}. In
the following, all \kpi\ moments are sideband-subtracted and
efficiency-corrected.

\begin{figure}[!htbp]
\begin{center}
\includegraphics[width=8.5cm]{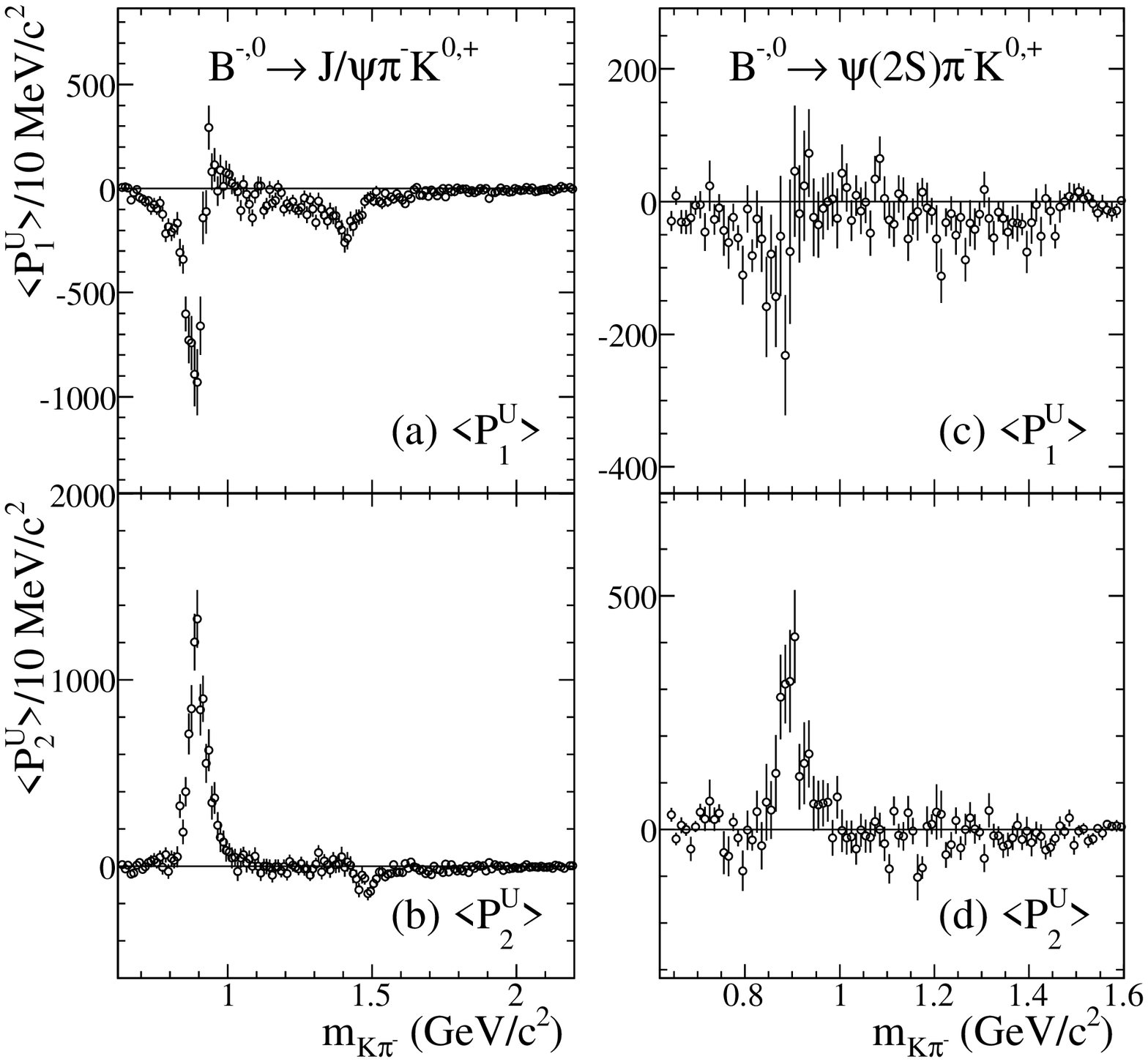}
\caption{The \mkpi\ dependence of (a) $\langle P_1^U\rangle $ and (b)
$\langle P_2^U\rangle $ for \modeFourEight\/; the \mkpi\ dependence of
(c) $\langle P_1^U\rangle $ and (d) $\langle P_2^U\rangle $ for
\modeTwoSix\/.}
\label{fig:moment1}
\end{center}
\end{figure}

The dependence of $\langle P_1^U\rangle $, and of $\langle
P_2^U\rangle $, on \mkpi\ is shown in Fig.~\ref{fig:moment1}. For each
moment, the behavior for \psitwos\ (Fig.~\ref{fig:moment1}(c),(d)) is
very similar to that for \jpsi\ (Fig.~\ref{fig:moment1}(a),(b)), but
the statistical uncertainties are significantly larger because the net
analysis sample is smaller by a factor of approximately six
(Table~\ref{table_sig_bkg}).

\begin{figure}[!htbp]
\begin{center}
\includegraphics[width=8.5cm]{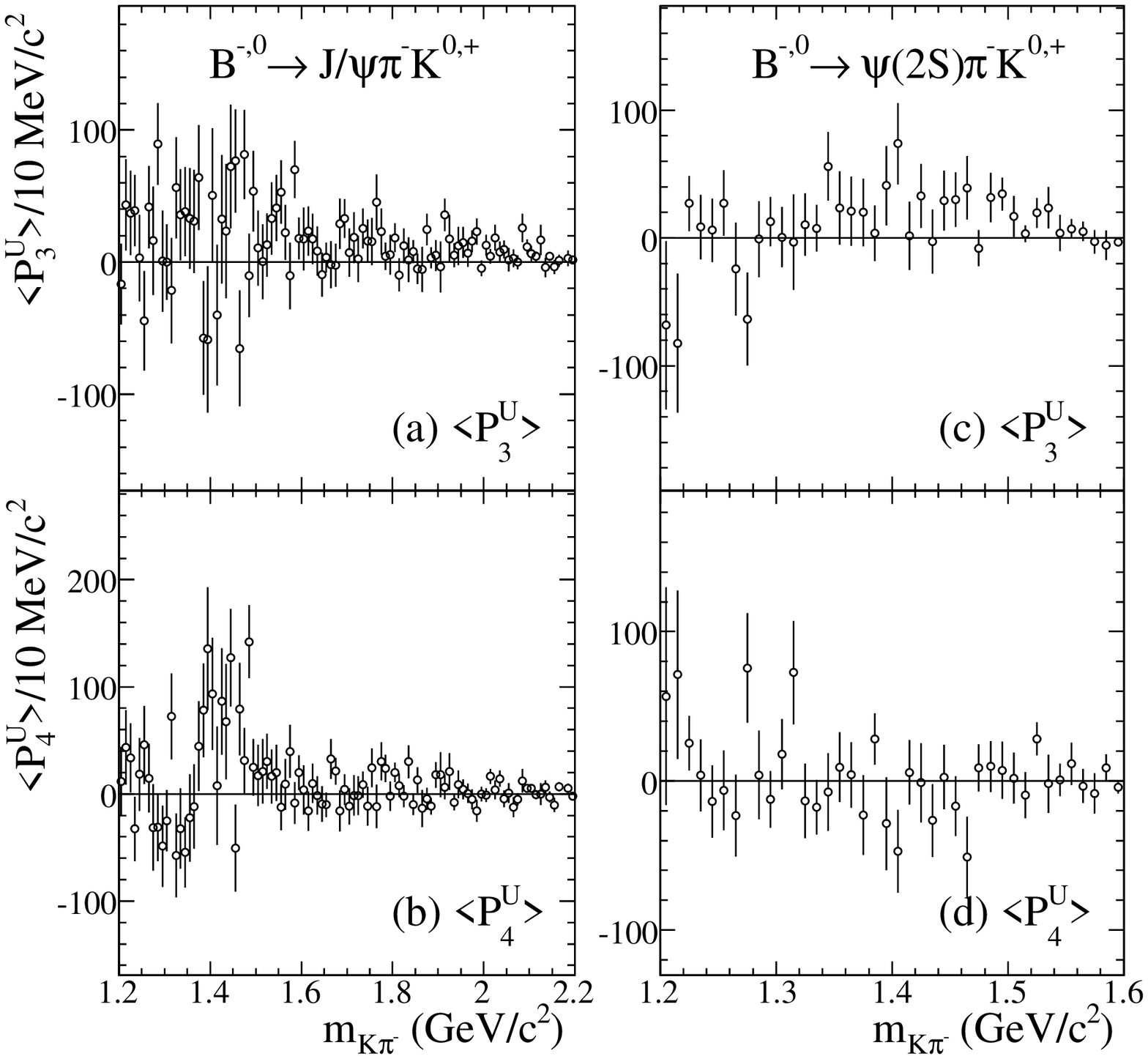}
\caption{The \mkpi\ dependence (for $m_{K\pi^-}>1.2$ \gevcc\/) of (a)
$\langle P_3^U\rangle $ and (b) $\langle P_4^U\rangle $ for
\modeFourEight\/; the \mkpi\ dependence of (c) $\langle P_3^U\rangle $
and (d) $\langle P_4^U\rangle $ for \modeTwoSix\/.}
\label{fig:moment2}
\end{center}
\end{figure}

\begin{figure}[!htbp]
\begin{center}
\includegraphics[width=8.5cm]{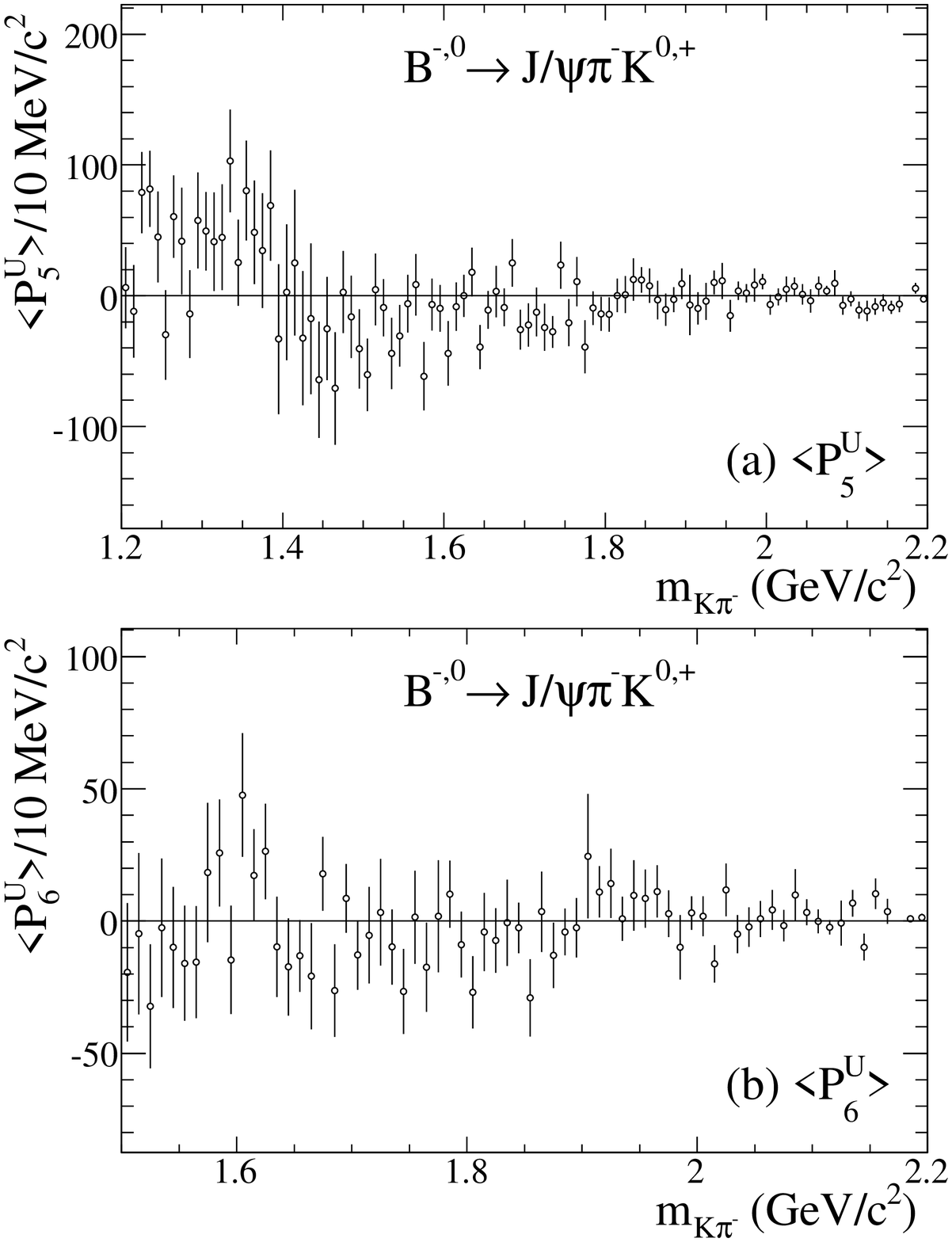}
\caption{(a) The \mkpi\ dependence of $\langle P_5^U\rangle $ for
$m_{K\pi^-}>1.2$ \gevcc\ for \modeFourEight\/; (b) the \mkpi\
dependence of $\langle P_6^U\rangle $ for $m_{K\pi^-}>1.5$ \gevcc\/
for \modeFourEight\/.}
\label{fig:moment3}
\end{center}
\end{figure}

The distributions of Fig.~\ref{fig:4kpi_lin} and
Figs.~\ref{fig:moment1}-\ref{fig:moment3} can be compared to those
observed in Ref.~\cite{LASS}, Fig.~6, for the $K^-\pi^+$ system
produced in the reaction $K^-p\rightarrow K^-\pi^+ n$ at 11 \gevc\
$K^-$ beam momentum; the latter are representative of the moments
structure of $K^-\pi^+$ elastic scattering. A striking overall feature
is the strong suppression of the mass structure for $m_{K\pi^-}>1$
\gevcc\ relative to the \Ksone\ which is observed for the $B$ meson
decay processes of Eqs.~(\ref{eq:modefour})-(\ref{eq:modesix}) in
comparison to $K^-\pi^+$ elastic scattering.

With this in mind, the $\langle P_1^U\rangle $ distribution of
Fig.~\ref{fig:moment1}(a) bears a remarkable similarity to that in
Ref.~\cite{LASS}, except that the sign is reversed. This is entirely
consistent with the analysis of Ref.~\cite{Aubert:2004cp}, performed
with a much smaller \babar\ data sample ($\sim 81$ fb$^{-1}$), which
showed that the $S_0-P_0$ relative phase of Eq.~(\ref{eq:heli2})
differed by $\pi$ from that obtained for $K^-\pi^+$ elastic
scattering. In the \Ksone\ region, the $D$-wave terms in
Eq.~(\ref{eq:heli2}) should be negligible, so that the relative phase
offset should yield the observed sign reversal w.r.t. $K^-\pi^+$
scattering. This sign reversal continues through the \Kstwo\ region,
which suggests that $S_0-P_0$ interference remains the dominant
contribution to $\langle P_1^U\rangle $, especially since $\langle
P_3^U\rangle $ is systematically positive in this region
(Figs.~\ref{fig:moment2}(a),(c)) (Note that the first term in
Eq.~(\ref{eq:heli4}) differs by only $1.8\%$ from the second term in
Eq.~(\ref{eq:heli2})).

The behavior of $\langle P_2^U\rangle $ in the \Ksone\ region is very similar to
that observed in Ref.~\cite{LASS}, and, ignoring $D$-wave
contributions, agrees in magnitude and sign with a calculation using
the values of $P_0$, $P_{+1}$, and $P_{-1}$ from
Ref.~\cite{Aubert:2004cp}. In Ref.~\cite{LASS}, $\langle P_2^U\rangle $ is positive
and much larger at the \Kstwo\ than at the \Ksone\/. Clearly, this is
not the case in Figs.~\ref{fig:moment1}(b) and \ref{fig:moment1}(d),
where $\langle P_2^U\rangle $ is small and negative at the \Kstwo\/. From
Eq.~(\ref{eq:heli3}), this could occur if $S_0$ is also shifted in phase
by $\pi$ relative to $D_0$, while the $D_0^2$ and
($D^2_{+1}+D^2_{-1}$) contributions to Eq.~(\ref{eq:heli3}) essentially
cancel. The latter is suggested by the observation of only a small
$\langle P_4^U\rangle $ signal at the \Kstwo\ in Fig.~\ref{fig:moment2}(b), and the
absence of signal in this region of Fig.~\ref{fig:moment2}(d). In
Eq.~(\ref{eq:heli5}), $D_0^2$ is favored 3:2 over $(D^2_{+1}+D^2_{-1})$,
whereas in Eq.~(\ref{eq:heli3}) the ratio is 4:5, hence the conjecture
that these contributions may in effect be canceling in
Figs.~\ref{fig:moment1}(b) and~\ref{fig:moment1}(d).

The $\langle P_3^U\rangle $ and $\langle P_4^U\rangle $ moments
require that $D$-wave contributions be present. At the statistical
level of the present analysis, we do not expect to observe such
contributions for $m_{K\pi^-}<1.2$ \gevcc\/, and so we begin the
distributions of Fig.~\ref{fig:moment2} at this value. In
Figs.~\ref{fig:moment2}(a) and~\ref{fig:moment2}(c), $\langle
P_3^U\rangle $ is systematically positive for all \mkpi\ mass values
above 1.3 \gevcc\/. This results from the enhancement observed in
Fig.~\ref{fig:dp3} for $\cos\theta_K <0$, and is similar to the
behavior in Ref.~\cite{LASS}, but at a much-reduced intensity level,
as mentioned previously. The $\langle P_4^U\rangle $ moments of
Fig.~\ref{fig:moment2}(b) and~\ref{fig:moment2}(d) have been discussed
already. The dip to negative values at $m_{K\pi^-}\sim 1.3$ \gevcc\ in
Fig.~\ref{fig:moment2}(b) is interesting, since it may indicate that
the relative strength of the $D^2_0$ and $(D^2_{+1}+D^2_{-1})$
contributions to Eq.~(\ref{eq:heli5}) varies with \mkpi\/. The
detailed amplitude analyses of the \Ksone\
region~\cite{Aubert:2004cp,Aubert:2007hz} do not consider such a
possibility for the $P_0$, $P_{+1}$, and $P_{-1}$ amplitudes, but a
mass-independent approach to these analyses would require a much
larger data sample.

Finally, the extended \mkpi\ range available for the combined
$B^{-,0}\rightarrow J/\psi\pi^- K^{0,+}$ data samples allows us to
search for evidence of $F$-wave amplitude contributions associated
with the \Ksthree\/, which has mass $\sim$ 1.78 \gevcc\ and width
$\sim 0.20$ \gev\/~\cite{LASS}. In Fig.~\ref{fig:moment3}(a) we show
the \mkpi\ dependence of $\langle P_5^U\rangle $ for $m_{K\pi^-}>1.2$
\gevcc\/, and in Fig.~\ref{fig:moment3}(b), the dependence of $\langle
P_6^U\rangle $ for $m_{K\pi^-}>1.5$ \gevcc\/. Interference between
$D$- and $F$-wave amplitudes could yield a $\langle P_5^U\rangle $
distribution characterized by the underlying $D$-wave BW
amplitude. For zero relative phase, the \mkpi\ dependence resulting
from the overlap of the leading edge of the $F$-wave BW amplitude with
the entire $D$-wave BW amplitude would resemble the real part of the
$D$-wave BW. Figure~\ref{fig:moment3}(a) exhibits just such behavior;
the intensity increases from near zero to a maximum below the
\Kstwo\/, passes through zero near the nominal mass value, reaches a
minimum just below 1.5 \gevcc\/, returns to zero near 1.6 \gevcc\/,
and has no clear structure thereafter. A $\langle P_6^U\rangle $
moment would involve $F_0^2$ and $(F^2_{+1}+F^2_{-1})$ intensity
contributions of opposite sign, just as for the $P$- and $D$-waves. We
see no clear signal in Fig.\ref{fig:moment3}(b), although the
distribution is systematically negative in the region $1.7-1.9$
\gevcc\/, so it could be that these contributions almost cancel, as
may be the case for the $D$-wave amplitudes. In the overall \mkpi\
distributions, these contributions add, and it is interesting to note
that in Fig.~\ref{fig:kpi_combined}(a) there is a small excess of
events above the fitted curve in the region of the \Ksthree\/. Our fit
to the mass distribution could possibly be improved slightly in this
region by including a \Ksthree\ contribution, but we do not do this at
present.
It seems reasonable to interpret the \mkpi\ dependence of $\langle
P_5^U\rangle $ observed in Fig.~\ref{fig:moment3}(a) as indicating the
presence of a \Ksthree\ amplitude, in which case this would be the
first evidence for $B$ meson decay to a final state including a spin
three resonance.

In summary, the angular structures observed for the \kpi\ systems
produced in $B^{-,0}\rightarrow J/\psi\pi^- K^{0,+}$ and
$B^{-,0}\rightarrow \psi(2S) \pi^- K^{0,+}$ show interesting features,
which can be well understood on the basis of the expected \kpi\
amplitude contributions. Moreover, the main features agree well
between the \jpsi\ and \psitwos\ modes, taking account of the
statistical limitations of the latter data sample.

\section{Reflection of \kpi\ structure into the \psipi\ mass distributions}
\label{sec:reflection}
We now investigate the extent to which reflection of the \kpi\ mass
and angular structures described in Secs.~\ref{sec:kpi} and
\ref{sec:legndre} are able to reproduce the efficiency-corrected and
sideband-subtracted \psipi\ mass distributions.

We do this by using a MC generator which initially creates large
samples of unit weight $B^{-,0}\rightarrow \psi\pi^- K^{0,+}$ events
with the correct $B$ meson production angular distribution in the
overall c.m. frame, and distributed in \kpi\ mass according to the fit
functions obtained as described in Sec.~\ref{sec:kpi}. The
distribution in \costhk\ is uniform at each value of \mkpi\/, and so
is described by
\begin{equation}
\frac{dN}{d\cos\theta_K}=\frac{N}{2} \, ,
\end{equation}
which is just Eq.~(\ref{eq:26}) with no angular structure.
 
Since we are now dealing with efficiency-corrected distributions, and
since mass resolution cannot generate local structure
(Fig.~\ref{fig:res}), we need not subject the generated events to
detector response simulation and subsequent event
reconstruction. Consequently we can create very large MC samples.

\begin{figure}[!htbp]
\begin{center}
\includegraphics[width=8.5cm]{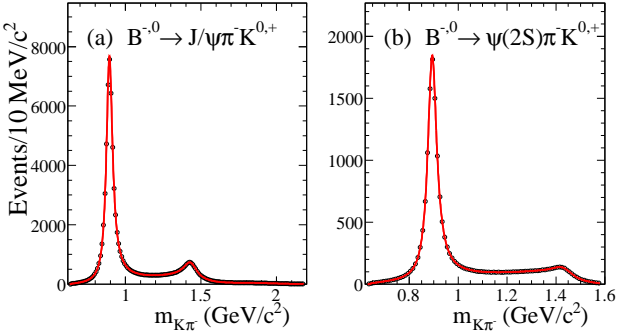}
\caption{The \kpi\ mass distributions generated according to the \kpi\
fit functions of Fig.~\ref{fig:kpi_combined} for (a)
$B^{-,0}\rightarrow J/\psi\pi^- K^{0,+}$ and (b) $B^{-,0}\rightarrow
\psi(2S)\pi^- K^{0,+}$. Each distribution contains ten million events,
normalized to the total number of events observed in the corrected
data sample (Fig.~\ref{fig:kpi_combined}).}
\label{fig:kpigeneration}
\end{center}
\end{figure}

The \kpi\ mass distributions generated in this way for
$B^{-,0}\rightarrow J/\psi\pi^- K^{0,+}$ and $B^{-,0}\rightarrow
\psi(2S)\pi^- K^{0,+}$ are shown in Figs.~\ref{fig:kpigeneration}(a)
and \ref{fig:kpigeneration}(b), respectively. Each distribution
contains ten million events, and has been normalized to the total
number of events in the corresponding corrected data sample; each
represents the relevant fit function of Fig.~\ref{fig:kpi_combined}
very well. We can use these events to create \psipi\ mass
distributions according to any desired selection criteria, and we
provide several examples of this later.

However, in Sec.~\ref{sec:legndre} we showed that there is a great
deal of angular structure in the \kpi\ system, as represented by the
\mkpi\ dependence of the $\langle P_i^U\rangle $ moments, and this affects the
reflection from the \kpi\ system into the \psipi\ mass distribution.

We take this \kpi\ angular structure into account by returning to
Eq.~(\ref{eq:26}), removing a factor of $N/2$ on the right side, and so
obtaining
\begin{eqnarray}
\frac{dN}{d\cos\theta_K}=\frac{N}{2}\left(1+\sum_{i=1}^{L}(\frac{2}{N})\langle P_i^U\rangle P_i(\cos\theta_K)\right) \, ,
\end{eqnarray}
{\it{i.e.}}
\begin{eqnarray}
\frac{dN}{d\cos\theta_K}=\frac{N}{2}\left( 1+\sum_{i=1}^{L}\langle P_i^N\rangle P_i(\cos\theta_K)\right) \, ,
\end{eqnarray}
where 
\begin{equation}
\label{eq:p_in}
\langle P_i^N\rangle=\frac{2}{N}\langle P_i^U\rangle  
\end{equation}
is defined to be the ``the normalized $P_i$ moment''. We can thus
incorporate the measured \kpi\ angular structure into our generator by
giving weight $w_j$ to the $j^{th}$ event generated, where
\begin{equation}
\label{eq:weight}
w_j=1+\sum_{i=1}^{L}\langle P_i^N\rangle P_i(\cos\theta_{K_j}) \, .
\end{equation}
The $\langle P_i^N\rangle $ are evaluated for the \mkpi\ value of the $j^{th}$
event by linear interpolation of the values obtained by normalizing
the $\langle P_i^U\rangle $ of Figs.~\ref{fig:moment1}-\ref{fig:moment3} according
to Eq.~(\ref{eq:p_in}).

\begin{figure}[!htbp]
\begin{center}
\includegraphics[width=8.5cm]{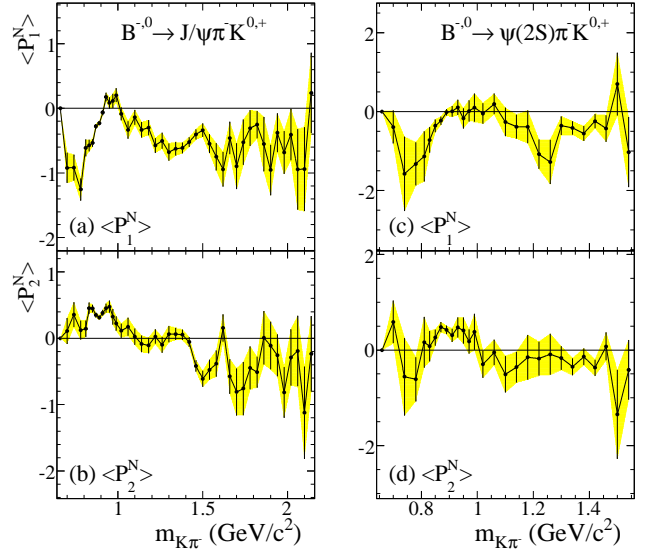}
\caption{The normalized moments corresponding to
Fig.~\ref{fig:moment1} obtained by using Eq.~(\ref{eq:p_in}). Mass
intervals have been combined in order to reduce statistical
fluctuations. The lines indicate the linear interpolations used in
weighting the MC events. The shaded regions indicate the $\pm1\sigma$
variations used to account of statistical uncertainties.}
\label{fig:normoment1}
\end{center}
\end{figure}

\begin{figure}[!htbp]
\begin{center}
\includegraphics[width=8.5cm]{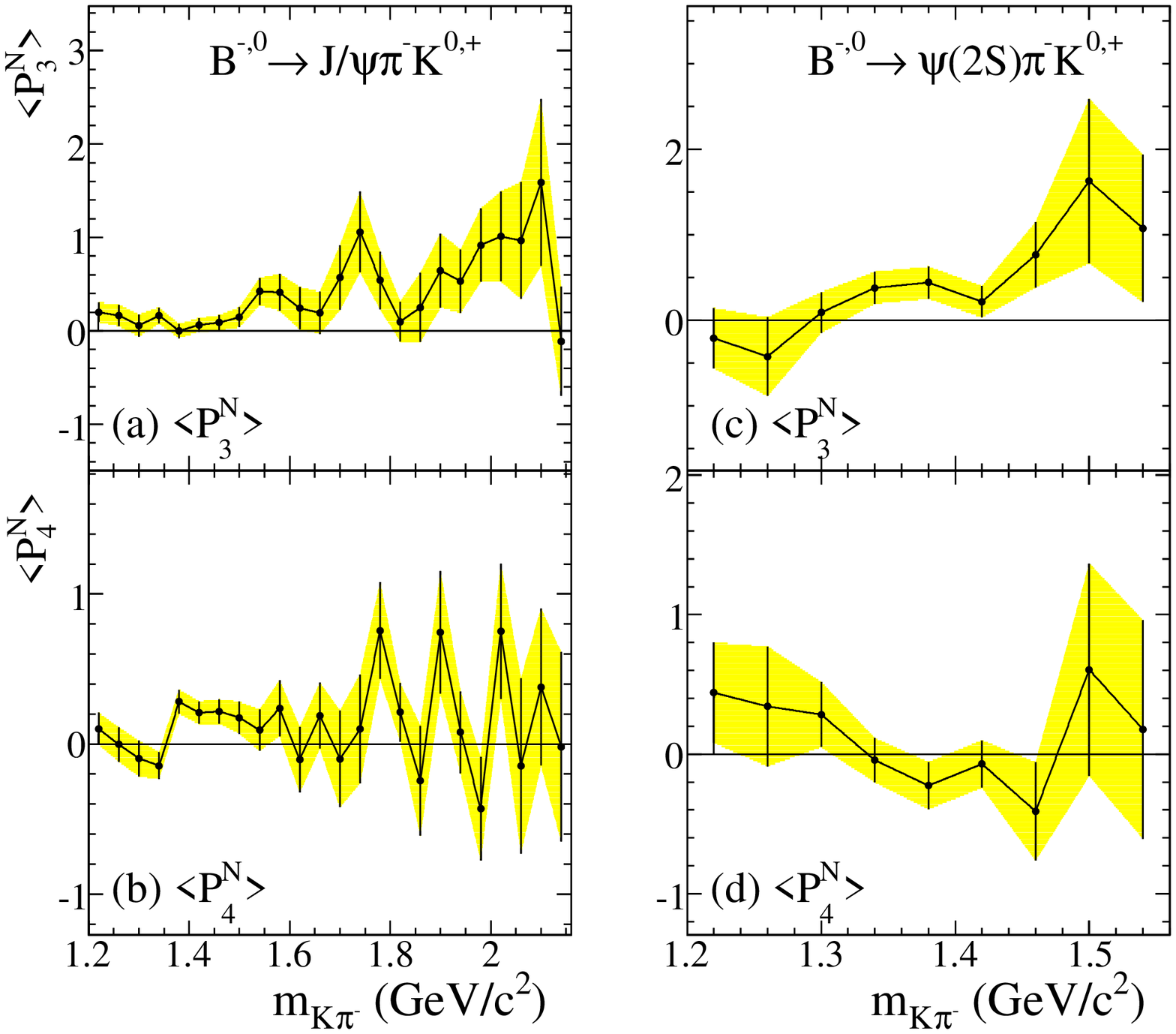}
\caption{The normalized moments corresponding to
Fig.~\ref{fig:moment2} obtained by using Eq.~(\ref{eq:p_in}). Mass
intervals have been combined in order to reduce statistical
fluctuations. The lines indicate the linear interpolations used in
weighting the MC events. The shaded regions indicate the $\pm1\sigma$
variations used to account of statistical uncertainties.}
\label{fig:normoment2}
\end{center}
\end{figure}

\begin{figure}[!htbp]
\begin{center}
\includegraphics[width=8.5cm]{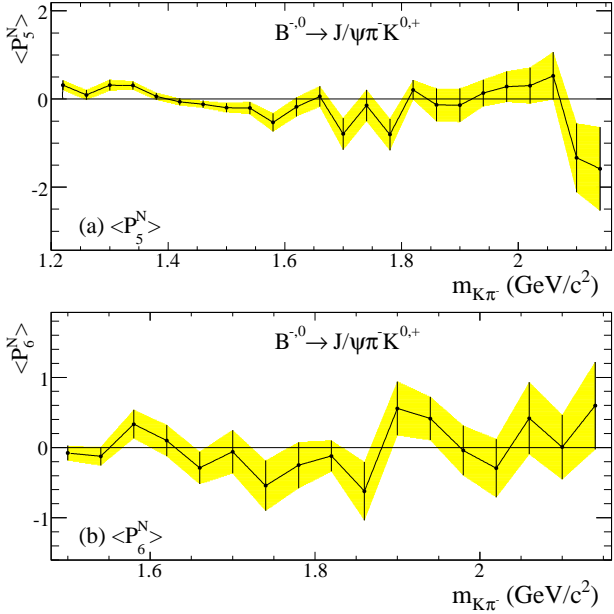}
\caption{The normalized moments corresponding to
Fig.~\ref{fig:moment3} obtained by using Eq.~(\ref{eq:p_in}). Mass
intervals have been combined in order to reduce statistical
fluctuations. The lines indicate the linear interpolations used in
weighting the MC events. The shaded regions indicate the $\pm1\sigma$
variations used to account of statistical uncertainties.}
\label{fig:normoment3}
\end{center}
\end{figure}

The results are shown in
Figs.~\ref{fig:normoment1}-\ref{fig:normoment3}, where we have used
modified mass intervals in order to reduce statistical fluctuation. We
interpolate linearly using the lines connecting the measured values.
In order to take into account the statistical uncertainties, we also
interpolate using lines connecting the $+1\sigma$ error values, and
lines connecting the $-1\sigma$ error values, as shown by the shaded
regions in each plot.

\begin{figure*}[!htbp]
\begin{center}
\includegraphics[width=17.0cm]{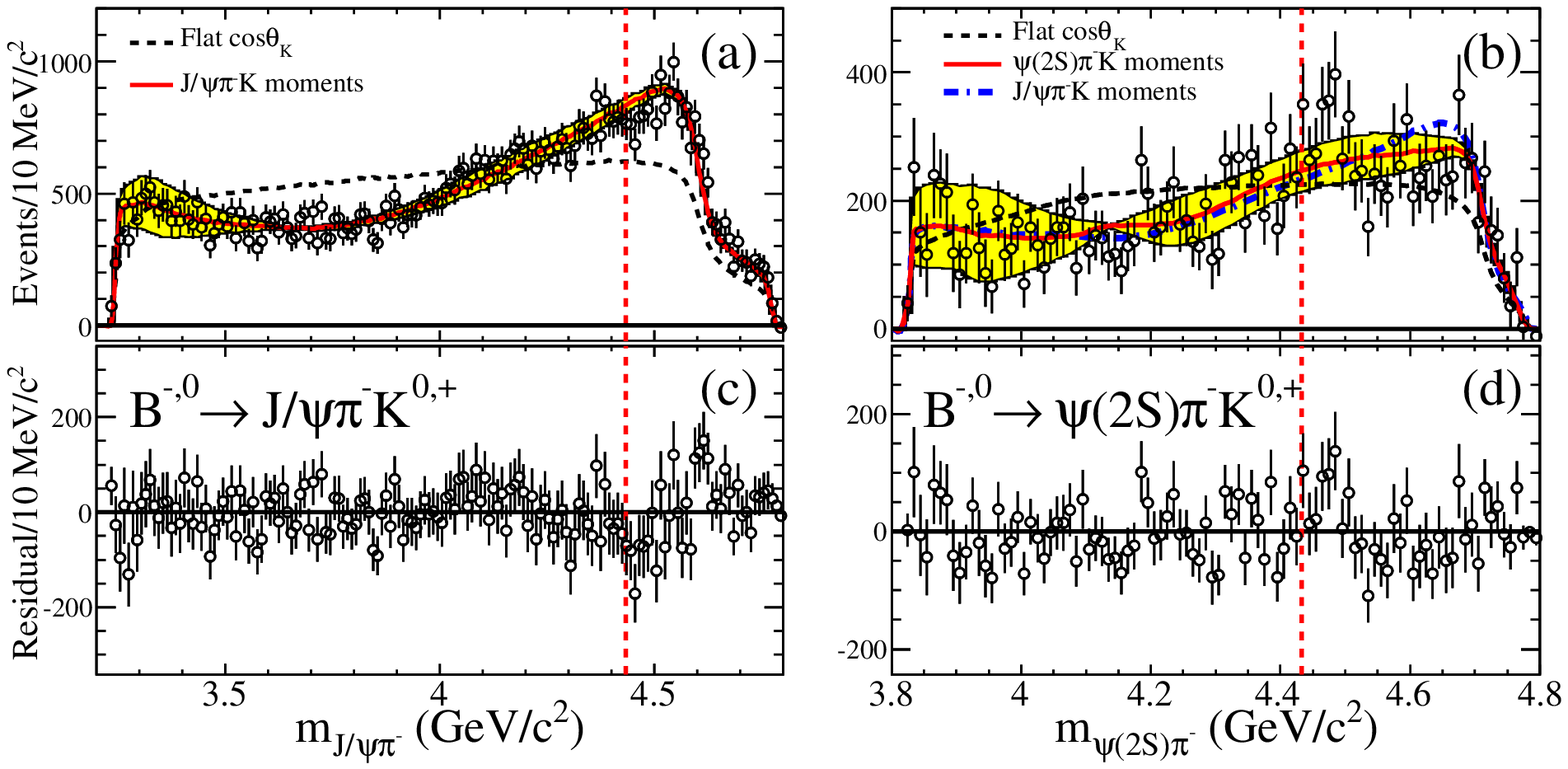}
\caption{The \psipi\ mass distributions for the combined decay modes
(a) $B^{-,0}\rightarrow J/\psi\pi^- K^{0,+}$ and (b)
$B^{-,0}\rightarrow \psi(2S)\pi^- K^{0,+}$. The points show the data
after efficiency correction and \DeltaE\ sideband subtraction. The
dashed curves show the \kpi\ reflection for a flat \costhk\
distribution, while the solid curves show the result of \costhk\
weighting.  The shaded bands represent the effect of statistical
uncertainty on the normalized moments. In (b), the dot-dashed curve
indicates the effect of weighting with the normalized $J/\psi\pi^-K$
moments. The dashed vertical lines indicate the value of
$m_{\psi\pi^-}=4.433$ \gevcc\/. In (c) and (d), we show the residuals
(data-solid curve) for (a) and (b), respectively.}
\label{Z_all_k_corr}
\end{center}
\end{figure*}

The \psipi\ mass distributions for the entire \kpi\ mass range for the
decay modes $B^{-,0}\rightarrow J/\psi\pi^- K^{0,+}$ and
$B^{-,0}\rightarrow \psi(2S)\pi^- K^{0,+}$ are shown in
Fig.~\ref{Z_all_k_corr}(a) and Fig.~\ref{Z_all_k_corr}(b),
respectively. The points represent the data after correcting for
efficiency and subtracting the events in the \DeltaE\ sideband. The
dashed curves show the reflection from \kpi\ assuming a flat \costhk\
distribution. The solid curves are obtained by weighting each event
according to Eq.~(\ref{eq:weight}). The shaded bands associated with
the solid curves indicate the effect of interpolation using
$\pm1\sigma$ normalized moment values, as described above.

\begin{figure*}[!htbp]
\begin{center}
\includegraphics[width=17.0cm]{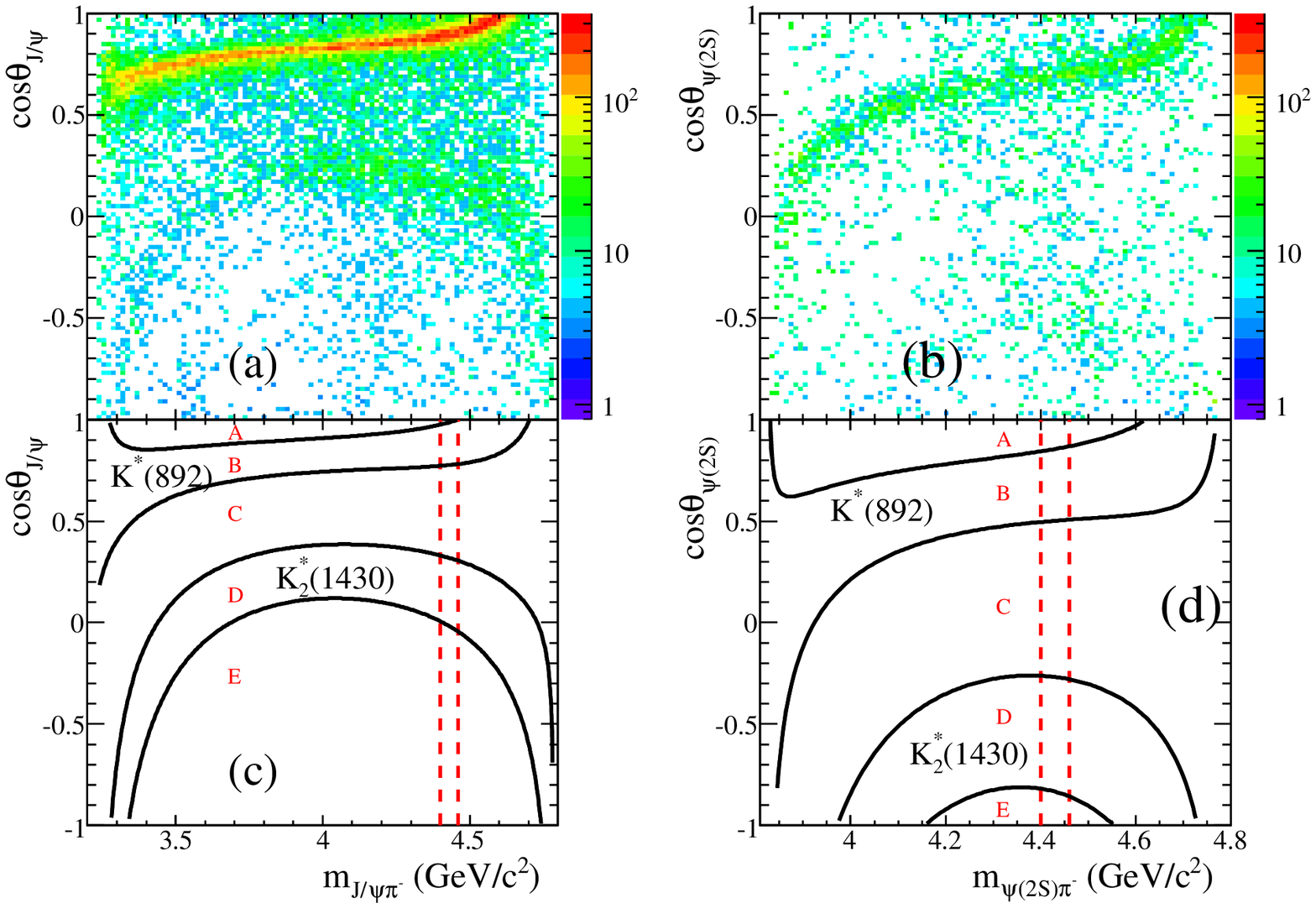}
\caption{The $\cos\theta_{\psi}$ versus $m_{\psi\pi^-}$ rectangular
Dalitz plots for (a) $B^{-,0}\rightarrow J/\psi\pi^- K^{0,+}$, and (b)
$B^{-,0}\rightarrow \psi(2S)\pi^- K^{0,+}$; (c) and (d), the
corresponding plots indicating the loci of the \Ksone\ and \Kstwo\
resonance bands defined in the text; regions A-E, defined by
Eqs.~(\ref{eq:kpiregionsA})-(\ref{eq:kpiregionsE}), are indicated. The
dashed vertical lines show the mass range $4.400<m_{\psi\pi^-}<4.460$
\gevcc\/.}
\label{fig:dp1}
\end{center}
\end{figure*}

We emphasize that the absolute normalization of the curves shown in
Figs.~\ref{Z_all_k_corr}(a) and \ref{Z_all_k_corr}(b) is established
by the scale factor used to normalize each of our ten-million-event MC
samples to the corresponding corrected number of events, as shown in
Figs.~\ref{fig:kpigeneration}(a) and \ref{fig:kpigeneration}(b),
respectively.

Comparison of the dashed and solid curves of Fig.~\ref{Z_all_k_corr}
shows that it is important to modulate the \costhk\ distributions
using the normalized moment weights of Eq.~(\ref{eq:weight}). Since the
individual $P_i(\cos\theta_K)$ functions integrate to zero over
\costhk\/, the incorporation of the $w_j$ weights does not affect
the distributions of Fig.~\ref{fig:kpigeneration}. This also means
that the associated dashed and solid curves of Fig.~\ref{Z_all_k_corr}
integrate to the same total number of events.

The reasons for the enhancement of the solid curves at high \psipi\
mass values, and their suppression at lower mass values are made clear
by the rectangular Dalitz plots of Figs.~\ref{fig:dp1}(a) and
\ref{fig:dp1}(b). We plot $\cos\theta_{\psi}$ against $m_{\psi\pi^-}$,
where $\cos\theta_{\psi}$ is the normalized dot-product of the \psipi\
three-momentum vector in the parent $B$ meson rest frame and the
$\psi$ three-momentum vector in the \psipi\ rest frame. A guide to the
structures observed in these plots is provided by
Figs.~\ref{fig:dp1}(c) and \ref{fig:dp1}(d), where we indicate the
locus of the \Ksone\ band ($0.795-0.995$ \gevcc\/) and that of the
\Kstwo\ band ($1.330-1.530$ \gevcc\/), as chosen in
Ref.~\cite{:2007wga}; in addition, we label regions (A)-(E), defined
in Sec.~\ref{sec:comparison}. The dashed vertical lines indicate the
\z\ region from Ref.~\cite{:2007wga}. The high-$m_{\psi\pi^-}$ region
of each of these bands corresponds to $\cos\theta_K<0$, and in
Fig.~\ref{fig:dp1}(a) this region is clearly populated preferentially
for both $K^{\ast}$ bands, and even for the \kpi\ mass range in
between. This corresponds to the backward-forward asymmetry observed
in Fig.~\ref{fig:dp3}, and to the negative values observed for
$\langle P_1^U\rangle $ in Fig.~\ref{fig:moment1}(a), and to the
positive values of $\langle P_3^U\rangle $ in
Fig.~\ref{fig:moment2}(a). These high-$m_{\psi\pi^-}$ enhancements are
compensated by the low-$m_{\psi\pi^-}$ suppression of the solid curve
relative to the dashed curve in Fig.~\ref{Z_all_k_corr}(a), since the
integral along any \mkpi\ locus is independent of the \costhk\
distribution. Similar behavior is observed for
Fig.~\ref{Z_all_k_corr}(b), but at a reduced statistical
level. Backward-forward asymmetry is observed in
Fig.~\ref{fig:dp3}(b), where $\langle P_1^U\rangle $ is primarily
negative in Fig.~\ref{fig:moment1}(c) and $\langle P_3^U\rangle $ is
positive in Fig.~\ref{fig:moment2}(c), so that the net effect on the
\psipi\ mass distribution of Fig.~\ref{Z_all_k_corr}(b) is much the
same as in Fig.~\ref{Z_all_k_corr}(a).

In Figs.~\ref{Z_all_k_corr}(c) and \ref{Z_all_k_corr}(d), we show the
residuals (data - solid curves) for the distributions in
Figs.~\ref{Z_all_k_corr}(a) and \ref{Z_all_k_corr}(b),
respectively. The dashed vertical lines indicate $m_{\psi\pi^-}=4.433$
\gevcc\/~\cite{:2007wga}. 

There is an excess of events in Fig.~\ref{Z_all_k_corr}(c) for
\mjpsipi\ $\sim 4.61$ \gevcc\/, as well as in
Fig.~\ref{fig:residuals}(b) and Fig.~\ref{fig:Z_k*_corr}(c) below. The
effect is associated with the \Ksone\ region of \kpi\ mass
(Fig.~\ref{fig:ranges}(b)), for which the \mjpsipi\ distribution
decreases steeply in the high-mass region. The mass resolution there
is $\sim 9$ \mevcc\ (Fig.~\ref{fig:res}(a),(c)) and we have not
incorporated this into our calculations. For this reason, the
calculated curve falls systematically below the data, hence yielding a
spurious peak in the residual distribution. For the corresponding
\mpsitwospi\ distributions, the mass resolution is $\sim 4$ \mevcc\
(Fig.~\ref{fig:res}(b),(d)) at $\sim 4.6$ \gevcc\/, the statistical
fluctuations are larger, and no similar effect is observed
(Figs.~\ref{Z_all_k_corr}(d), \ref{fig:residuals}(g), and
\ref{fig:Z_k*_corr}(d)). Apart from this, the distribution of the
residuals shows no evidence of statistically significant departure
from zero at any \jpsipi\ mass value.

In Fig.~\ref{Z_all_k_corr}(b), and correspondingly in
Fig.~\ref{Z_all_k_corr}(d), the small excess of events at $\sim 4.48$
\gevcc\ provides the only indication of a narrow signal. As shown in
Sec.~\ref{sec:zsig}, this yields a $2.7 \sigma$ enhancement with mass
$\sim 4.476$ \gevcc\ and width consistent with that reported in
Ref.~\cite{:2007wga}. The dot-dashed curve in
Fig.~\ref{Z_all_k_corr}(b) was obtained from the dashed curve by
modulating the \kpi\ angular distribution using instead the normalized
\kpi\ moments from the $B^{-,0}\rightarrow J/\psi\pi^- K^{0,+}$ data,
which show no evidence of a \z\ signal. This curve and the solid curve
differ only slightly in the range $\sim 4.2$ \gevcc\ to $\sim 4.55$
\gevcc\/, so that the \kpi\ background function at $\sim 4.48$ \gevcc\
is not very sensitive to the modulation procedure, nor to the presence
of a small, narrow \mpsitwospi\ enhancement (see Sec.~\ref{sec:zsig}
for a quantitative discussion).

 We conclude that the \mpsitwospi\ distribution of
Fig.~\ref{Z_all_k_corr}(b), and the residual distribution of
Fig.~\ref{Z_all_k_corr}(d), do not provide confirmation of the \z\
signal reported in Ref.~\cite{:2007wga}.

\section{Comparison to the Belle results}
\label{sec:comparison}
We now compare our results to those obtained by Belle for
$B\rightarrow\psi(2S)\pi^- K$~\cite{:2007wga}.
\subsection{The \psipi\ mass resolution}
In Sec.~\ref{sec:resolution} we showed (Fig.~\ref{fig:res}) our mass
resolution dependence on $Q$-value, and obtained HWHM $\sim 4$ \mevcc\
for the \psitwospi\ system at the \z\/. In Ref.~\cite{:2007wga}, it is
stated only that the mass resolution is 2.5 \mevcc\/. Since the width
of the \z\ is $\sim 45$ \mev\/~\cite{:2007wga}, mass resolution should
not be an issue for the comparison of similar data samples (see
Sec.~\ref{sec:direct_com}).

\subsection{Efficiency}
\begin{figure}[!htbp]
\begin{center}
\includegraphics[width=8.5cm]{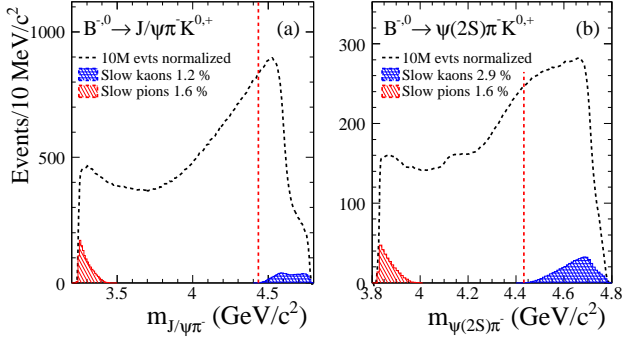}
\caption{(a) The curve of Fig.~\ref{Z_all_k_corr}(a), and (b) that of
Fig.~\ref{Z_all_k_corr}(b), obtained by \costhk\ weighting; the shaded
regions near threshold correspond to $P(\pi^-)<0.1$ \gevc\ in the
laboratory frame, while the cross-hatched regions near maximum \psipi\
mass represent $P(K)<0.25$ \gevc\ in the laboratory frame; the dashed
vertical lines indicate $m_{\psi\pi^-}=4.433$ \gevcc\/.}
\label{fig:mass}
\end{center}
\end{figure}

We have made a detailed study of efficiency over each Dalitz plot for
each \jpsi\ and \psitwos\ decay mode separately
(Sec.~\ref{sec:efficiency}), and have identified efficiency losses
associated with low-momentum pions and kaons in the laboratory frame
(Fig.~\ref{fig:effpik}). We illustrate the effect of such losses on
the $m_{\psi\pi^-}$ distributions using our ten-million-event
$B^{-,0}\rightarrow \psi\pi^- K^{0,+}$ samples weighted to take
account of the \kpi\ angular structure (Sec.~\ref{sec:reflection}). In
Fig.~\ref{fig:mass} we show the \psipi\ distributions obtained as for
Fig.~\ref{Z_all_k_corr} (solid curves). We then require that the
momentum of the $\pi$ be less than 100 \mevc\ in the laboratory frame
(Fig.~\ref{fig:effpik}) and obtain the shaded distributions in the
$m_{\psi\pi^-}$ threshold regions. Similarly, the requirement that the
kaon momentum be less than 250 \mevc\ in the laboratory frame
(Fig.~\ref{fig:effpik}) yields the cross-hatched regions near maximum
\psipi\ mass~\cite{boost}. It follows that the regions of lower
efficiency discussed in Appendix~\ref{append1} should have no
significant effect on the region of the \z\/.

\begin{figure}[!htbp]
\begin{center}
\includegraphics[width=8.5cm]{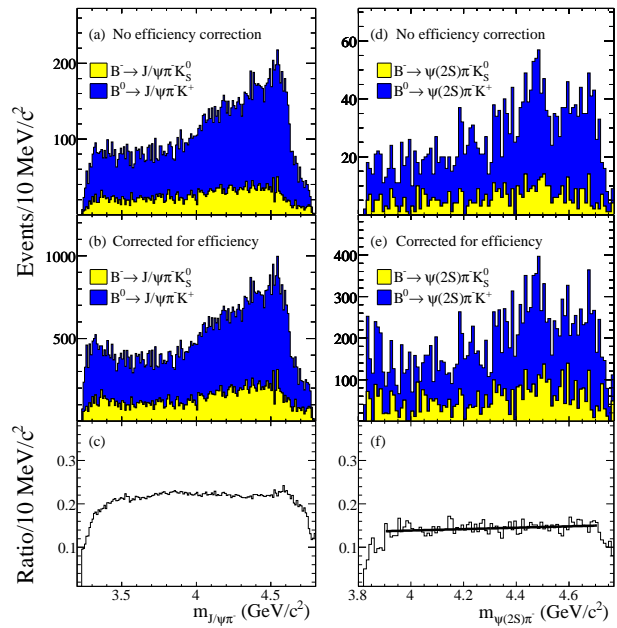}
\caption{The cumulative \psipi\ mass distributions for the $K^0_S$ and
$K^+$ decay modes (a) and (d) before, and (b) and (e) after,
efficiency correction. The ratio of uncorrected to corrected data is
shown in (c) and (f). The fitted line in (f) is described in the
text.}
\label{fig:ratio_corr_uncor}
\end{center}
\end{figure}

As a direct check of the effect of our efficiency-correction
procedure, we show our $m_{\psi\pi^-}$ distributions before and after
correction in Figs.~\ref{fig:ratio_corr_uncor}(a),(b) and
\ref{fig:ratio_corr_uncor}(d),(e) for $B^{-,0}\rightarrow J/\psi\pi^-
K^{0,+}$ and $B^{-,0}\rightarrow \psi(2S)\pi^- K^{0,+}$,
respectively. Then in Fig.~\ref{fig:ratio_corr_uncor}(c) and
\ref{fig:ratio_corr_uncor}(f) we show the ratio of the uncorrected and
corrected distributions as a measure of average efficiency. As
expected, the average value decreases rapidly near threshold, and near
the maximum value for both distributions. Away from these regions, the
efficiency increases slowly with increasing mass. We conclude that our
event reconstruction efficiency should have no effect on any \z\
signal in our data.

In Fig.~\ref{fig:ratio_corr_uncor}(f) we show the result of a linear
fit, excluding the regions $m_{\psi(2S)\pi^-}<3.9$ \gevcc\ and
$m_{\psi(2S)\pi^-}>4.71$ \gevcc\/, which are seriously affected by the
loss of low momentum pions and kaons, respectively. The fitted
efficiency value increases from 13.7 $\%$ to 15.0 $\%$ over the fitted
region. The low-efficiency regions are excluded when we compare our
uncorrected $m_{\psi(2S)\pi^-}$ distribution to that from Belle
(Sec.~\ref{sec:direct_com}). This is due to the fact that for both
experiments the reconstruction efficiency for very low momentum
charged-particle tracks in the laboratory frame decreases rapidly to
zero ({\it cf.} Fig.~\ref{fig:effpik}). There is no detailed
discussion of efficiency in Ref.~\cite{:2007wga}.

\subsection{The \kpi\ mass and \costhk\ structures}
\label{subsec:kpicos}
In Ref.~\cite{:2007wga}, the \Ksone\ and \Kstwo\ regions of the Dalitz
plot are removed, and the remaining non-\z\ contribution to the
\mpsitwospi\ mass distribution is described by a second-order
polynomial in \mpsitwospi\ multiplied by the momentum of the \psitwos\
in the \psitwospi\ rest frame and by the factor
$\sqrt{m_{max}-m_{\psi(2S)\pi^-}}$; for two-out-of-three phase space,
this latter factor should be the momentum of the recoil $K$ in the $B$
rest frame. Also, the \DeltaE\ sideband contributions are not
subtracted prior to their fit. There is no discussion of a possible
description of the background in terms of \kpi\ mass and angular
structures.

In our analysis, we have considered the effect of the \kpi\ mass and
\costhk\ structures (Sec.~\ref{sec:kpi} and Sec.~\ref{sec:legndre}),
and have shown how the observed features affect the \psipi\ mass
distributions in Sec.~\ref{sec:reflection}. The resulting dashed
curves of Fig.~\ref{fig:mass} cannot be described in terms of
second-order polynomials. However, each corresponds to a projection of
the entire Dalitz plot. In order to make a direct comparison to the
Belle data, we investigate the relevant regions of \kpi\ mass
in the following section.

\subsection{Regions of \kpi\ mass}
The $\psi(2S)\pi$ mass distribution shown in Fig.~2 of
Ref.\cite{:2007wga} has a ``$K^{\ast}$ veto'' applied. This means that
events within 100 \mevcc\ of the \Ksone\ or the \Kstwo\ have been
removed, and hence that the \kpi\ mass range has, in effect, been
divided into five regions, as follows:
\begin{eqnarray}
\label{eq:kpiregionsA}
\mathrm{region\, A:  } && m_{K\pi^-} < 0.795 \,  \mathrm{\gevcc\/} \, , \\
\label{eq:kpiregionsB}
\mathrm{region\, B:  } &&0.795<m_{K\pi^-} < 0.995 \,  \mathrm{\gevcc\/} \, , \\
\label{eq:kpiregionsC}
\mathrm{region\, C:  } &&0.995<m_{K\pi^-} < 1.332 \,  \mathrm{\gevcc\/} \, , \\
\label{eq:kpiregionsD}
\mathrm{region\, D:  } &&1.332<m_{K\pi^-} < 1.532 \,  \mathrm{\gevcc\/} \, , \\
\label{eq:kpiregionsE}
\mathrm{region\, E:  } &&  m_{K\pi^-}> 1.532 \,  \mathrm{\gevcc\/} \, . 
\end{eqnarray}
These regions are labeled in Fig.~\ref{fig:dp1}(c) and
Fig.~\ref{fig:dp1}(d).  The $\psi(2S)$ mass distribution of
Ref.~\cite{:2007wga} thus contains events (with sideband contribution)
from regions A, C, and E.

\begin{figure*}[!htbp]
\begin{center}
\includegraphics[width=17.0cm]{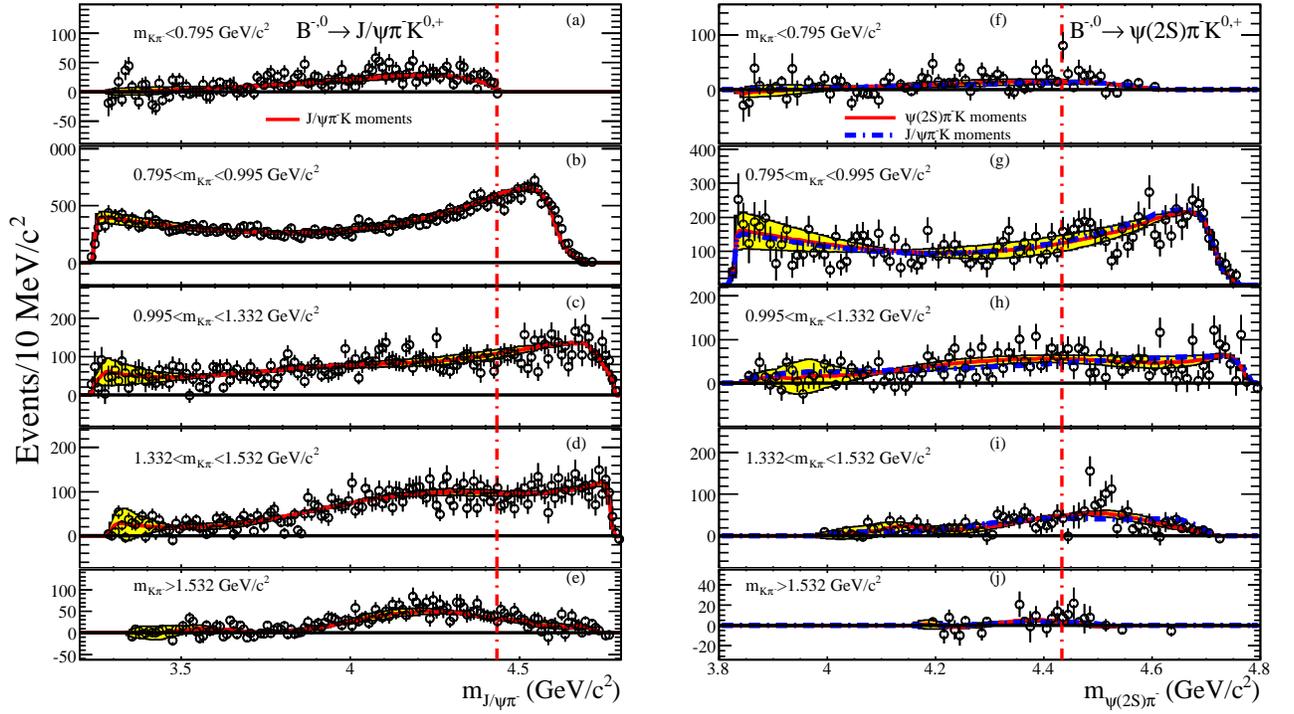}
\caption{The \psipi\ mass distributions in regions A-E of \kpi\ mass
for the combined decay modes (a-e) $B^{-,0}\rightarrow J/\psi\pi^-
K^{0,+}$, and (f-j) $B^{-,0}\rightarrow \psi(2S)\pi^- K^{0,+}$; the
open dots represent the data, and the solid curves and shaded bands
are as in Fig.~\ref{Z_all_k_corr}, but calculated for the relevant
\kpi\ mass region using the same overall normalization constant as for
Fig.~\ref{fig:kpigeneration}. In (f-j), the dot-dashed curves are
obtained using \kpi\ normalized moments for $B^{-,0}\rightarrow
J/\psi\pi^- K^{0,+}$ in Eq.~(\ref{eq:weight}), instead of those from
$B^{-,0}\rightarrow \psi(2S)\pi^- K^{0,+}$; the dot-dashed vertical
lines indicate $m_{\psi\pi^-}=4.433$ \gevcc\/.}
\label{fig:ranges}
\end{center}
\end{figure*}

\begin{figure*}[!htbp]
\begin{center}
\includegraphics[width=17.cm]{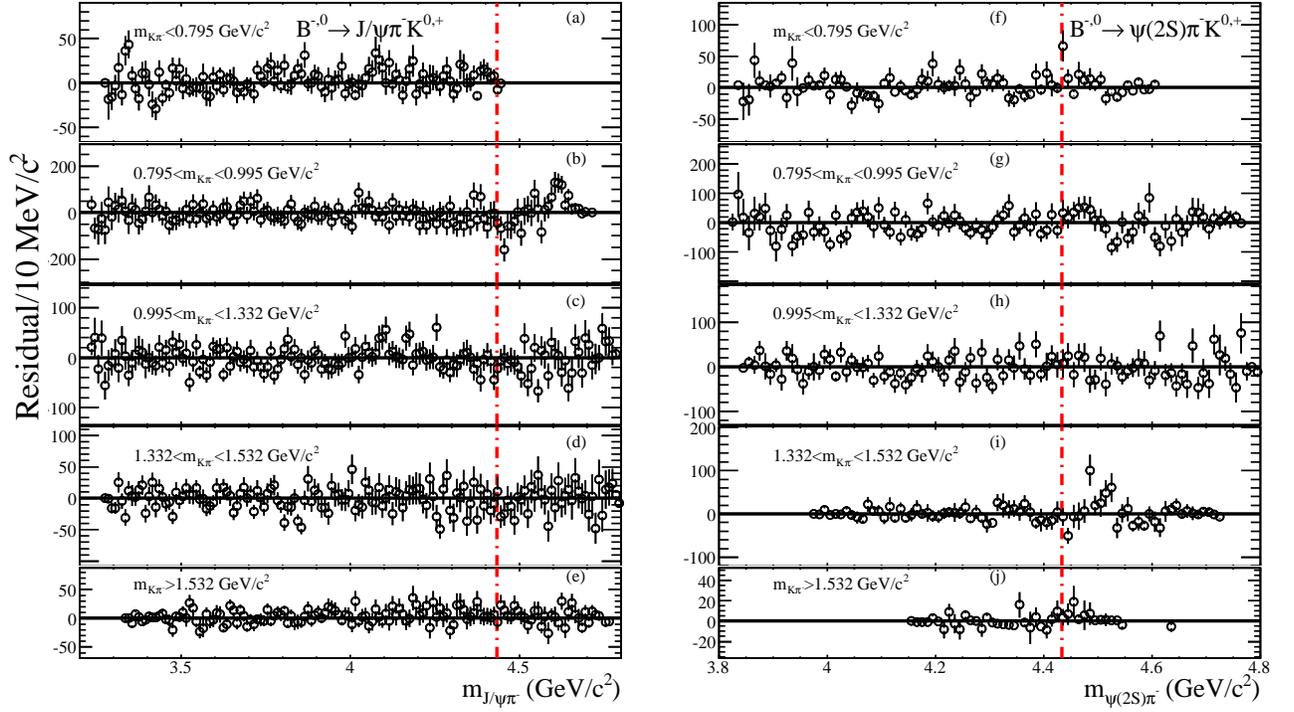}
\caption{The residuals (data - solid curve) corresponding to
Fig.~\ref{fig:ranges}; the dot-dashed vertical lines indicate
$m_{\psi\pi^-}=4.433$ \gevcc\/.}
\label{fig:residuals}
\end{center}
\end{figure*}

\begin{figure*}[!htbp]
\begin{center}
\includegraphics[width=17.0cm]{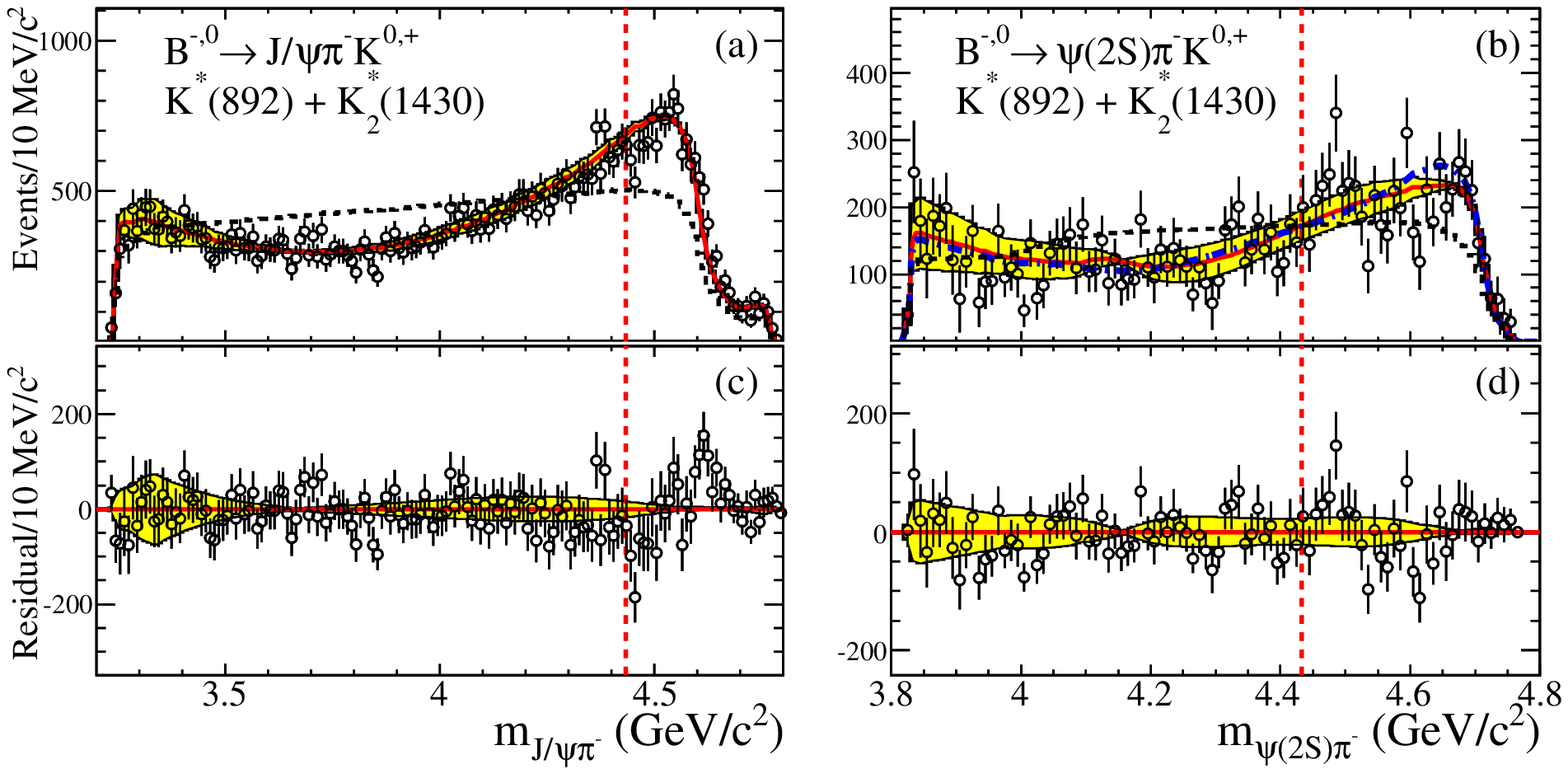}
\caption{The \psipi\ mass distributions for (a) $B^{-,0}\rightarrow
J/\psi\pi^- K^{0,+}$, and (b) $B^{-,0}\rightarrow \psi(2S)\pi^-
K^{0,+}$, for \mkpi\ regions $B$ and $D$ combined; the open dots
represent the data, the solid, dashed, and dot-dashed curves, and the
shaded bands, correspond to those of Fig.~\ref{Z_all_k_corr}(a),(b);
(c) and (d), the corresponding residual distributions. The dashed
vertical lines indicate $m_{\psi\pi^-}=4.433$ \gevcc\/.}
\label{fig:Z_k*_corr}
\includegraphics[width=17.0cm]{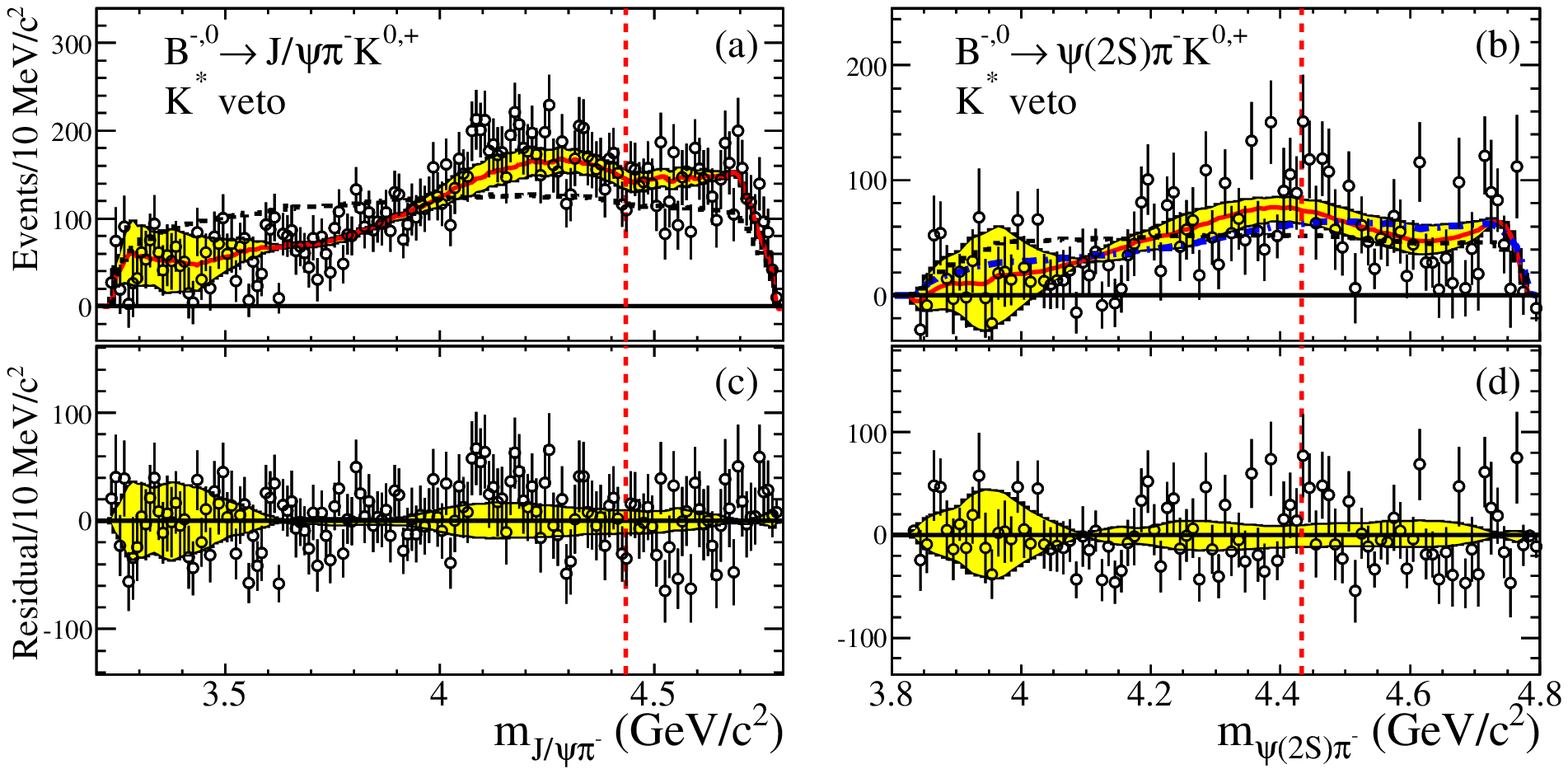}
\caption{The \psipi\ mass distributions for (a) $B^{-,0}\rightarrow
J/\psi\pi^- K^{0,+}$, and (b) $B^{-,0}\rightarrow \psi(2S)\pi^-
K^{0,+}$, for \mkpi\ regions $A$, $C$, and $E$, combined; the open
dots represent the data, the solid, dashed, and dot-dashed curves, and
the shaded bands, correspond to those of
Fig.~\ref{Z_all_k_corr}(a),(b); (c) and (d), the corresponding
residual distributions. The dashed vertical lines indicate
$m_{\psi\pi^-}=4.433$ \gevcc\/.}
\label{fig:Z_nok_corr}
\end{center}
\end{figure*}

In Fig.~\ref{fig:ranges} we show the corrected $m_{\psi\pi^-}$
distributions for regions A-E of \kpi\ mass. The solid curves and
shaded bands correspond to those in Fig.~\ref{Z_all_k_corr}, with the
same overall normalization constants as obtained for
Fig.~\ref{fig:kpigeneration}, {\it{i.e.}}, there is no renormalization
in the separate \kpi\ mass regions. In Figs.~\ref{fig:ranges}(f)-(j),
the dot-dashed curves were obtained using the normalized moments from
$B^{-,0}\rightarrow J/\psi\pi^- K^{0,+}$ in conjunction with
Eq.~(\ref{eq:weight}). For regions A and B, there is almost no
difference between the solid and dot-dashed curves, while in the other
regions the differences are less than, or of the order of, the
statistical fluctuations in the associated data. The residuals
(obtained by subtracting the solid curves from the data) corresponding
to Fig.~\ref{fig:ranges} are shown in Fig.~\ref{fig:residuals}, and
show no evidence of structure. In Figs.~\ref{fig:Z_k*_corr} and
Fig.~\ref{fig:Z_nok_corr} we make similar comparisons for the combined
data in the $K^{\ast}$ regions ($B$ and $D$), and for the
$K^{\ast}$-veto region (A, C, and E). Again the residuals reveal no
significant structure.

\subsection{Direct comparison}
\label{sec:direct_com}
\begin{figure}[!htbp]
\begin{center}
\includegraphics[width=8.5cm]{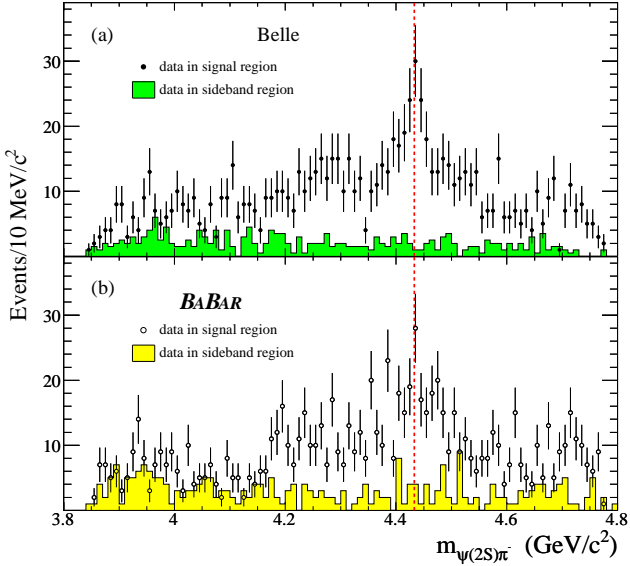}
\caption{a) The \psitwospi\ mass distribution after $K^{\ast}$ veto
from Ref.~\cite{:2007wga}; the data points represent the signal region
(we have assigned $\sqrt{N}$ errors at present), and the shaded
histogram represents the background contribution estimated from the
\DeltaE\ sideband regions; b) shows the corresponding distribution
from the \babar\ analysis. The dashed vertical line indicates
\mpsitwospi\/$=4.433$ \gevcc\/.}
\label{fig:diff}
\end{center}
\end{figure}

The \psitwospi\ mass distribution of Fig.~\ref{fig:diff}(a) is a
reproduction of that in Ref.~\cite{:2007wga}, except for the addition
of error bars~\cite{thankBelle}. In Fig.~\ref{fig:diff}(b) we show the
equivalent distribution for our combined analysis samples for the $B$
meson decay processes of Eqs.~(\ref{eq:modetwo}) and
(\ref{eq:modesix}) for the $K^{\ast}$-veto region (A, C, and E). The
mass intervals are the same as for Fig.~\ref{fig:diff}(a), but no
efficiency-correction has been performed. As mentioned in
Sec.~\ref{subsec:kpicos}, when we make quantitative comparisons
between Fig.~\ref{fig:diff}(a) and Fig.~\ref{fig:diff}(b) we exclude
the low-efficiency regions near threshold and at high mass, and use
only the region $3.9<m_{\psi(2S)\pi^-}<4.71$ \gevcc\/. We make a
global comparison of the data samples in
Table~\ref{table-comparison}. The \babar\ sample contains $\sim 8\%$
more background than does the Belle sample. The net signal ratio is
$1.18\pm0.09$ in favor of Belle, although the corresponding integrated
luminosity ratio is 1.46. It follows that for \babar\/, net signal per
unit luminosity is $\sim 1.34$, while for Belle it is 1.08, and that
this significant increase in signal yield comes at the cost of only a
modest increase in background level. For both experiments the
background distribution increases toward threshold, and differs
markedly in \psitwospi\ mass dependence from the signal.
\begin{figure}[!htbpp]
\begin{center}
\includegraphics[width=8.5cm]{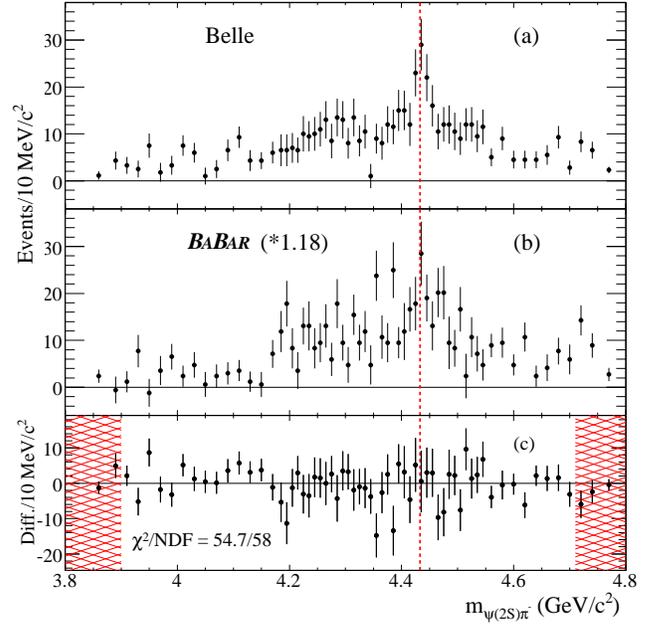}
\caption{a) The distribution of Fig.~\ref{fig:diff}(a) after combining
mass intervals for $m_{\psi(2S)\pi^-}<4.18$ \gevcc\ and
$m_{\psi(2S)\pi^-}>4.55$ \gevcc\/, and carrying out sideband
subtraction. b) The distribution of Fig.~\ref{fig:diff}(b) after
following the same procedure, and in addition scaling by 1.18, as
described in the text. c) The difference, (a)-(b), where the errors
have been combined in quadrature; $\chi^2/NDF=54.7/58$
(Probability=59.9\%) excluding the low-efficiency regions
(cross-hatched).}
\label{fig:diff2}
\end{center}
\end{figure}
\begin{table*}
  \begin{center}
    \caption{Summary of \mpsitwospi\ data from regions A, C, and
    $E$, combined (Fig.~\ref{fig:diff}) for Belle and \babar\ in the
    mass range 3.9-4.71 \gevcc\/.}
    \begin{tabular*}{1.0\textwidth}{@{\extracolsep{\fill}}lcc} \hline\hline \\
      Category \,\,\,\,\,& \,\,\,\,\,Belle [Fig.~\ref{fig:diff}(a)] \,\,\,\,\,& \,\,\,\,\, \babar\ [Fig.~\ref{fig:diff}(b)] \\ \\ \hline\hline \\
      Total signal region events ($N$)  & $824\pm29$ & $786\pm28$ \\ \\
      Sideband contribution ($B$) & $172\pm 9$ & $234\pm 15$ \\ \\
      $B/N$ & $21.5\%$  & $29.8\%$                            \\ \\
      Net signal        & $652\pm 30$           & $552\pm 32$ \\ \\ \hline\hline
   \end{tabular*}
\label{table-comparison}
\end{center}
\end{table*}

We have shown our efficiency-corrected and sideband-subtracted mass
distribution corresponding to Fig.~\ref{fig:diff}(b) in
Fig.~\ref{fig:Z_nok_corr}(b), and should compare the latter to the
equivalent Belle distribution. However, this is not available, and so
we make do with the distributions of Fig.~\ref{fig:diff} instead. In
order to justify the use of $\sqrt{N}$ error assignments, we combine
adjacent mass intervals for $m_{\psi(2S)\pi^-}<4.18$ \gevcc\ and
$m_{\psi(2S)\pi^-}>4.55$ \gevcc\ so that we obtain at least ten events
(signal + sideband) in each mass interval. We then create the
sideband-subtracted distributions of Fig.~\ref{fig:diff2}(a) and
Fig.~\ref{fig:diff2}(b) from the data of Fig.~\ref{fig:diff}(a) and
Fig.~\ref{fig:diff}(b), respectively. In Fig.~\ref{fig:diff2}(b), we
have scaled our data by the factor 1.18 to compensate for the
statistical difference between the experiments. In
Fig.~\ref{fig:diff2}(c) we show the result of subtracting the
distribution of Fig.~\ref{fig:diff2}(b) from that of
Fig.~\ref{fig:diff2}(a), with errors combined in quadrature. There is
no evidence of any statistically significant difference, in particular
near $m_{\psi(2S)\pi^-}=4.433$ \gevcc\/, indicated by the dashed
vertical line. If the low-efficiency
(Fig.~\ref{fig:ratio_corr_uncor}(f)) cross-hatched regions of
Fig.~\ref{fig:diff2}(c) are excluded, the $\chi^2-$value for the
remaining region is found to be 54.7 for 59 mass intervals. There is
one normalization constant (1.18), so that the comparison yields
$\chi^2/NDF=54.7/58$, with corresponding probability $59.9\%$.

We conclude that the Belle and \babar\ distributions of
Fig.~\ref{fig:diff2} are statistically consistent. We have shown in
Fig.~\ref{Z_all_k_corr} and
Figs.~\ref{fig:ranges}-\ref{fig:Z_nok_corr} that all of our corrected
\psipi\ distributions are well-described by reflections of the mass
and angular structures of the \kpi\ system. We refer to this as our
\kpi\ background, and in Sec.~\ref{sec:babar_fits} we quantify the
extent to which an additional \z\ signal is required to describe the
corrected \babar\ \psipi\ mass distributions.

\begin{figure}[!htbpp]
\begin{center}
\includegraphics[width=8.5cm]{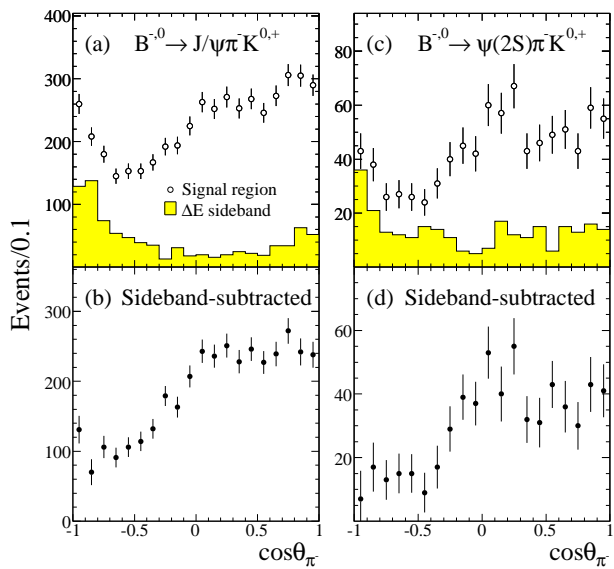}
\caption{The $\cos\theta_{\pi}$ distributions for the combined \kpi\
mass regions A, C, and E, (a), (b) for the decay modes
$B^{-,0}\rightarrow J/\psi\pi^- K^{0,+}$, and (c), (d) for
$B^{-,0}\rightarrow \psi(2S)\pi^- K^{0,+}$. In (a) and (c), the points
represent the data in the signal region, and the shaded histograms
show the \DeltaE\ sideband events. In (b) and (d), the signal region
distributions are shown after sideband subtraction. No efficiency
correction is applied to the data.}
\label{fig:costhnpi}
\end{center}
\end{figure}

We have mentioned previously that it is the backward-forward asymmetry
in the \kpi\ angular distribution as a function of \mkpi\ which yields
the high mass enhancements seen in our \psipi\ mass distributions. We
show this effect explicitly in Fig.~\ref{fig:costhnpi}, where we plot
the distribution of $\cos\theta_{\pi}=-\cos\theta_K$ for regions A, C,
and E of \kpi\ mass. For $\cos\theta_{\pi}\sim 1$, \mpsitwospi\ is
near its maximum value, and $m^2_{\psi\pi}$ is related linearly to
$\cos\theta_{\pi}$, and so it is not surprising that the sideband
subtracted distributions of Fig.~\ref{fig:costhnpi}(b) and
Fig.~\ref{fig:costhnpi}(d) bear a strong shape resemblance to the
corresponding \mpsitwospi\ distributions of
Fig.~\ref{fig:Z_nok_corr}(a) and Fig.~\ref{fig:Z_nok_corr}(b),
respectively. We note also that the increase in the sideband
distribution of Fig.~\ref{fig:costhnpi}(c) towards
$\cos\theta_{\pi}\sim -1$ corresponds to the increase in
Fig.~\ref{fig:costhnpi}(b) toward \psipi\
threshold. Figure~\ref{fig:costhnpi} illustrates how important it is
to take into account the angular structure in the \kpi\ system in
creating the shape of the associated \psipi\ mass distribution, even
after the removal of the $K^{\ast}$ regions.

\section{Fits to the corrected \psipi\ mass distributions}
\label{sec:babar_fits}
\subsection{The fit results}
\begin{figure*}[!htbpp]
\begin{center}
\includegraphics[width=17.0cm]{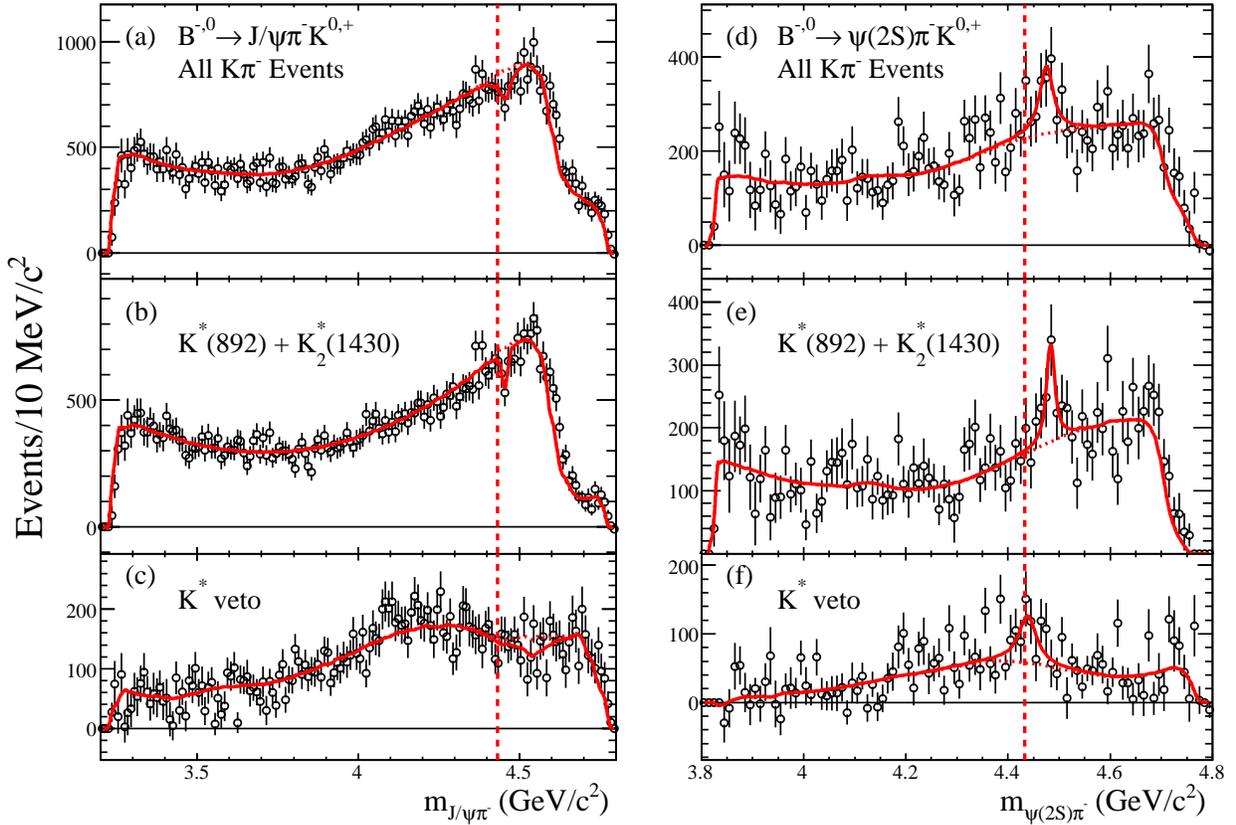}
\caption{The results of the fits to the corrected mass distributions,
(a)-(c) for \jpsipi\/, and (d)-(f) for \psitwospi\/. In each figure,
the open dots represent the data, and the solid curve represents the
fit function, which consists of the \kpi\ background distribution with
normalization free, and a relativistic $S$-wave BW line shape, with
mass, width, and normalization free; the dotted curves indicate the
\kpi\ background functions; the dashed vertical lines indicate
$m_{\psi\pi^-}=4.433$ \gevcc\/.}
\label{fig:babar_cor}
\end{center}
\end{figure*}

In Fig.~\ref{fig:babar_cor} we show the results of the fits to the
corrected \babar\ data of Figs.~\ref{Z_all_k_corr}(a),(b),
Figs.~\ref{fig:Z_k*_corr}(a),(b), and
Figs.~\ref{fig:Z_nok_corr}(a),(b). In each fit, the \kpi\ background
distribution has been multiplied by a free normalization parameter; an
$S$-wave BW line shape, with free normalization, mass and width
parameters, has been added incoherently in order to quantify the
search for a \z\ signal. In each figure, the solid curve represents
the fit result; the parameter values for the corresponding \z\ signal
are summarized in Table~\ref{tab:summ_fits}. Table~\ref{tab:summ_fits}
also contains the results obtained when the mass of the \z\ is fixed
(4.433 \gevcc\/), and when both the mass and width (45 \mev\/) are
fixed~\cite{:2007wga}. For all fits, $\chi^2/NDF$ is acceptable, but
deteriorates slightly, or fails to improve, as first the mass is fixed
and then the mass and width are fixed. We start each fit at the mass
and width values of $m=4.433$ \gevcc\ and $\Gamma=45$ \mev\/.

For the fit to the \jpsipi\ mass distribution using the
$K^{\ast}$-veto sample with mass and width free, the obtained mass is
$\sim 100$ \mevcc\ larger than that of the \z\/, the width is
essentially undetermined, the signal is more than $2\sigma$ negative,
and it remains at least $1\sigma$ negative for the other fits.

For the $K^{\ast}$ region, the fitted mass value is closer to that of
the \z\/, but otherwise the results are very similar to those for the
$K^{\ast}$-veto region in that negative signal values are obtained.

For the total sample, the results are no better. The signal is
negative by $\sim 2.1\sigma$, and remains negative by at least
$2\sigma$ as the constraints are applied.

We conclude that there is no evidence for \z\ production via the decay
sequence $B^{-,0}\rightarrow Z(4430)^-K^{0,+}$, \z\/$\rightarrow
J/\psi\pi^-$.

For the fit to the $\psi(2S)\pi^-$ $K^{\ast}$-veto sample (which is
equivalent to the Belle analysis sample, but sideband-subtracted and
efficiency-corrected), we obtain mass and width values which are
consistent with theirs, but a positive signal which is only $\sim
1.9\sigma$ from zero; fixing mass and width increases this to only
$\sim 3.1\sigma$. Since our efficiency in the \z\ region is almost
constant ({\it{cf}}. Fig.~\ref{fig:ratio_corr_uncor}(f)), our
corrected signal size with mass and width free ($426 \pm 229$ events)
corresponds to $\sim 61 \pm 33$ observed events. This converts to a
signal of $\sim 70$ events in Fig.~\ref{fig:diff2}(b), and we estimate
a similar value for the Belle distribution of
Fig.~\ref{fig:diff2}(a). The signal size reported in
Ref.~\cite{:2007wga} is $\sim 120$ events, obtained on the basis of a
background description which ignores \kpi\ mass and angular
structures.  It is interesting that we find a small positive signal
with mass and width consistent with the Belle values. However, with
the present small data sample, it seems impossible to decide whether
this is due to the production of a real state, or to the imprecision
of the normalized \kpi\ moments, primarily in the region between the
two $K^{\ast}$'s.

For the $K^{\ast}$ region, our fitted mass value is $\sim 50$ \mevcc\
higher than the Belle value, and the signal deviates from zero only by
$\sim 2.5\sigma$; the imposition of Belle mass and width values yields
a signal which is less than $\sim 0.5\sigma$ from zero.

We note that each of these regions corresponds to approximately half
of the $\cos\theta_{\psi(2S)}$ range
({\it{cf}}. Fig.~\ref{fig:dp1}(d)), so that for a flat \z\ angular
distribution, we would expect naively that the signal size for each
region would be the same. The values in Table~\ref{tab:summ_fits} are
consistent with this, however the central mass values differ by $\sim
5$ standard deviations. Although there could be significant
interference effects associated with the $K^{\ast}$ regions, it does
not seem possible that these could cause the signal to be displaced by
approximately one full-width. This, together with the fact that both
signals are in the 2-3 standard deviation range, suggests that a more
likely interpretation is that both are simply statistical
fluctuations.

Finally, for the complete sample, our fitted mass value is $\sim 40$
\mevcc\ higher than the Belle value; the width is consistent with
Belle's, but the signal size is only $\sim 2.7\sigma$ from zero; this
is reduced to $\sim 2.1\sigma$ when the Belle parameters are imposed.

We conclude that there is no convincing evidence for production of the
\z\ state via the decay sequence $B^{-,0}\rightarrow
Z(4430)^-K^{0,+}$, \z\/$\rightarrow$\psitwospi\/, especially since the
fits take no account of the \kpi\ background uncertainty associated
with the normalized \kpi\ moments.
\begin{table*}
  \begin{center}
    \caption{Results of the fits to the corrected \babar\ \psipi\ mass
    distributions; all fits use the relevant \babar\/ \kpi\ background
    shape.}
    \begin{tabular*}{1.0\textwidth}{@{\extracolsep{\fill}}lccc} \hline\hline \\
      Type of fit                    & total                   & $K^{\ast}$ region       &  $K^{\ast}$ veto      \\ \\ \hline\hline \\
      \jpsipi\ sample;            & $m=4455\pm 8$ \mevcc\   & $m=4454\pm 4$ \mevcc\   & $m=4545\pm30$ \mevcc\     \\ \\
     \z\ mass and width free         & $\Gamma=42\pm 27$ \mev\ & $\Gamma=17\pm 12$ \mev\ & $\Gamma=100\pm 96$ \mev\ \\ \\
                                     & $N_Z=-901\pm 420$       & $N_Z=-514\pm223$         & $N_Z=-411\pm181$         \\ \\
                                     & $\chi^2/NDF=131/154$    & $\chi^2/NDF=135/154$    & $\chi^2/NDF=159/154$     \\ \\ \hline \\
       \jpsipi\ sample;           & $\Gamma=82\pm 77$ \mev\ & $\Gamma=79\pm 39$ \mev\ & $\Gamma=10\pm 12$ \mev\  \\ \\
       \z\ fixed mass and free width & $N_Z=-1098\pm 490$      & $N_Z=-810\pm393$        & $N_Z=-86\pm 64$          \\ \\
                                     & $\chi^2/NDF=136/155$    & $\chi^2/NDF=143/155$    & $\chi^2/NDF=162/155$     \\ \\ \hline \\
       \jpsipi\ sample;           & $N_Z=-704\pm 249$       & $N_Z=-540\pm225$        & $N_Z=-147\pm 105$        \\ \\
       \z\ fixed mass and width      & $\chi^2/NDF=137/156$    & $\chi^2/NDF=144/156$    & $\chi^2/NDF=164/156$     \\  \\ \hline \hline \\

      \psitwospi\ sample;          & $m=4476\pm 8$ \mevcc\   & $m=4483\pm 3$ \mevcc\    & $m=4439\pm8$ \mevcc\     \\ \\ 
     \z\ mass and width free         & $\Gamma=32\pm 16$ \mev\ & $\Gamma=17\pm 12$ \mev\  & $\Gamma=41\pm 33$ \mev\ \\ \\
                                     & $N_Z=703\pm 260$        & $N_Z=447\pm177$          & $N_Z=426\pm229$         \\ \\
                                     & $\chi^2/NDF=93/96$      & $\chi^2/NDF=91/96$       & $\chi^2/NDF=106/96$     \\ \\ \hline \\
       \psitwospi\ sample;         & $\Gamma=97\pm 77$ \mev\ & $\Gamma=100\pm 82$ \mev\ & $\Gamma=36\pm 26$ \mev\  \\ \\
       \z\ fixed mass and free width & $N_Z=710\pm 440$        & $N_Z=246\pm247$          & $N_Z=414\pm 194$          \\ \\
                                     & $\chi^2/NDF=101/97$     & $\chi^2/NDF=102/97$      & $\chi^2/NDF=107/97$     \\ \\ \hline \\
       \psitwospi\ sample;         & $N_Z=440\pm 212$        & $N_Z=89\pm162$           & $N_Z=431\pm 137$        \\ \\
       \z\ fixed mass and width      & $\chi^2/NDF=101/98$     & $\chi^2/NDF=101/98$      & $\chi^2/NDF=107/98$     \\ \\ \hline \hline

   \end{tabular*}
\label{tab:summ_fits}
\end{center}
\end{table*}

\subsection{Can the \kpi\ background absorb a \z\ signal?}
\label{sec:zsig}
Since our evidence for the existence of a \z\ signal in the
$B^{-,0}\rightarrow \psi(2S)\pi^- K^{0,+}$ data sample is less than
compelling, it is reasonable to ask whether our use of the normalized
\kpi\ moments to modulate the $\psi(2S)\pi^-$ background shape might
have removed part, or all, of a real \z\ signal.

Firstly, our \kpi\ background curves in Figs.~\ref{Z_all_k_corr},
\ref{fig:Z_k*_corr}, and \ref{fig:Z_nok_corr} show no tendency to peak
in a narrow region around the \z\ mass position, indicated by the
dashed vertical line in each figure. Secondly, for the \psitwospi\
distributions in these figures, the dot-dashed curves obtained using
the normalized \kpi\ moments from the $B^{-,0}\rightarrow J/\psi \pi^-
K^{0,+}$ modes, for which there is no \z\ signal, do not differ
significantly in shape in the signal region from the solid curves from
the $B^{-,0}\rightarrow \psi(2S) \pi^- K^{0,+}$ modes. Finally, in
Ref.~\cite{:2007wga} it is stated that it is not possible to create a
narrow peak from the $S$-,$P$-, and $D$-wave amplitude structure of
the \kpi\ system.  We agree with this statement and provide a
quantitative demonstration below. However, we first show how a \z\
signal would affect the \costhk\ versus \mkpi\ Dalitz plot.

\begin{figure}[!htbpp]
\begin{center}
\includegraphics[width=8.5cm]{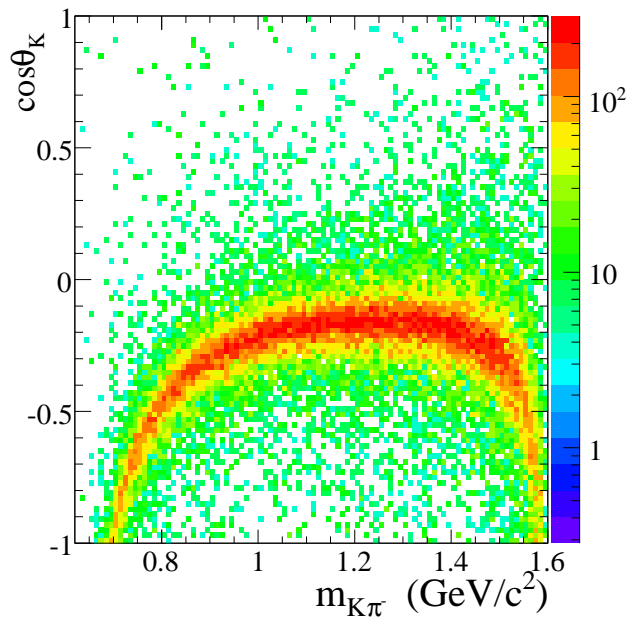}
\caption{The \costhk\/ versus \mkpi\ rectangular Dalitz plot for the
$B^{-,0}\rightarrow Z(4430)^-K^{0,+}$, \z\/$\rightarrow\psi(2S)\pi^-$
MC events generated using $m_Z=4.433$ \gevcc\/, $\Gamma_Z=0.045$
\gev\/, and with isotropic \z\ decay.}
\label{fig:MC_Z4430}
\end{center}
\end{figure}

To this end, we have generated a MC sample of events corresponding to
$B^{-,0}\rightarrow Z(4430)^-K^{0,+}$, $Z(4430)^-\rightarrow
\psi(2S)\pi^-$ where the \z\ has the Belle central mass and width
values, and decays isotropically. Figure~\ref{fig:MC_Z4430} shows the
Dalitz plot which results. The \z\ events yield a narrow locus
confined almost entirely to the region $\cos\theta_K<0$.

\begin{figure}[!htbpp]
\begin{center}
\includegraphics[width=8.5cm]{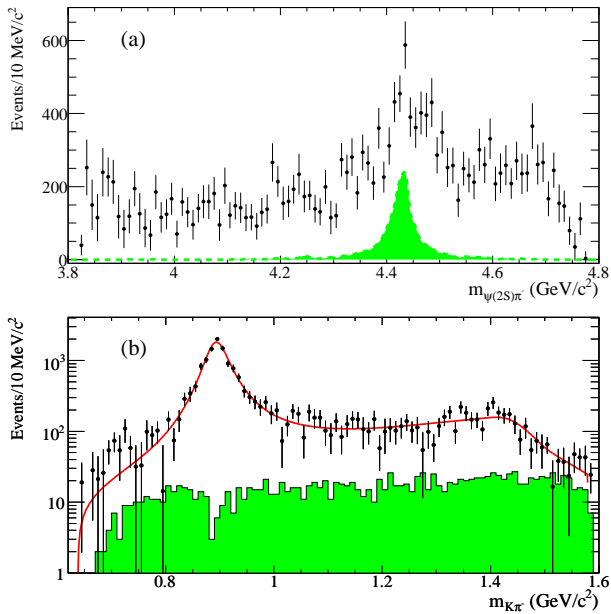}
\caption{(a): The combined \mpsitwospi\ mass distribution for the data
of Fig.~\ref{Z_all_k_corr}(b) and the MC \z\ sample generated as
described in the text; the shaded histogram represents the MC sample;
(b) the \kpi\ mass distribution corresponding to
Fig.~\ref{fig:test1}(a); the shaded histogram represents the
reflection of the MC \z\ events, and the curve shows the result of the
fit.}
\label{fig:test1}
\end{center}
\end{figure}

\begin{figure}[!htbpp]
\begin{center}
\includegraphics[width=8.5cm]{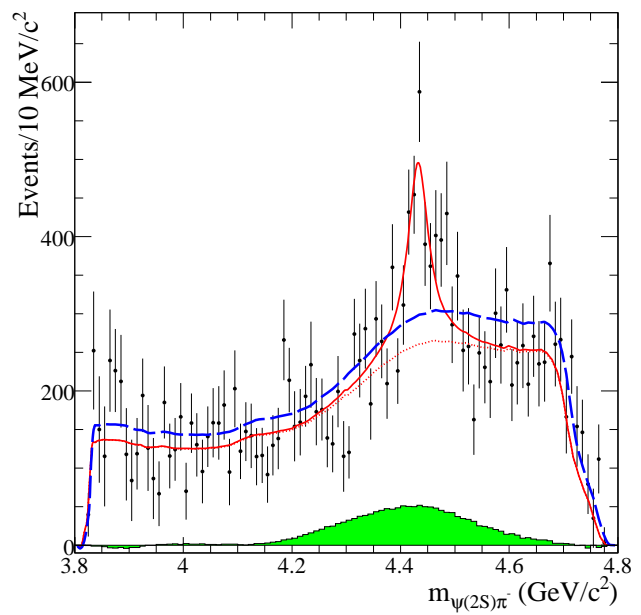}
\caption{The \mpsitwospi\ distribution of Fig.~\ref{fig:test1}(a) for
the data and MC samples combined as described in the text. The dashed
curve represents the \kpi\ background distribution, obtained as
described in the text, and the shaded histogram represents the
difference between this dashed curve and the solid curve of
Fig.~\ref{Z_all_k_corr}(b), {\it {i.e.}} it represents the impact of
the MC $Z(4430)^-$ signal. The solid curve is the result of a fit
using the \kpi\ background function and a relativistic $S$-wave BW,
and the dotted curve shows the resulting renormalized dashed curve.}
\label{fig:test2}
\end{center}
\end{figure}

For the Legendre polynomials which we have used to modulate the \kpi\
angular distribution, such a distribution yields the following
behavior:
\begin{itemize}
\item $P_1(\cos\theta_K)$ is almost always negative;
\item $P_2(\cos\theta_K)$, which is negative for
$|\cos\theta_K|<0.58$, is negative for $\sim 0.75<m_{K\pi}<1.55$
\gevcc\ {\it{i.e.}} almost always;
\item $P_3(\cos\theta_K)$, which is positive for
$-0.78<\cos\theta_K<0$, is positive over almost the entire \mkpi\
range above 1.2 \gevcc\/, which is where we make use of
$P_3(\cos\theta_K)$ in the data;
\item $P_4(\cos\theta_K)$, which is positive for $|\cos\theta_K|<0.33$
and $|\cos\theta_K|>0.88$, is mainly positive for $m_{K\pi^-}>1.2$
\gevcc\/.
\end{itemize}
We estimate the effect of such a \z\ signal on our \kpi\ background by
adding a MC-generated \z\ signal incoherently to the corrected
\psitwospi\ data of Fig.~\ref{Z_all_k_corr}(b), and subjecting this
new sample to the procedure for creating our \kpi\ background
contribution to the \psitwospi\ mass distribution. As for
Fig.~\ref{fig:MC_Z4430}, we use $m_Z=4433$ \mevcc\ and $\Gamma_Z=45$
\mev\ in the simulation, and generate a flat $\cos\theta_\psi$ angular
distribution. On the basis of the Belle result, we would expect an
observed signal of $\sim 200$ events for the full $\cos\theta_\psi$
angular range ({\it{cf.}} Fig.~\ref{fig:dp1}(d)), and this would yield
an efficiency-corrected signal of $\sim 1500$ events ({\it{cf.}}
Fig.~\ref{fig:ratio_corr_uncor}(f)).  Consequently, we generate 1500
events, of which 1493 events are within the (\mes\/, \DeltaE\/) signal
region. The \mpsitwospi\ distribution for these events is shown as the
shaded histogram of Fig.~\ref{fig:test1}(a); the points with error
bars show the result of combining these MC events with our corrected
\psitwospi\ data distribution (Fig.~\ref{Z_all_k_corr}(b)). The \mkpi\
distribution for this combined sample is shown in
Fig.~\ref{fig:test1}(b), where the solid curve shows the result of a
fit to the combined data, and the shaded histogram indicates the
reflection of the \z\ MC events. We use this new fit curve to generate
10 million events corresponding to $B^{-,0}\rightarrow \psi(2S) \pi^-
K^{0,+}$ as before, and normalize this sample to the combined data and
MC-generated \z\ sample. We add the unnormalized \kpi\ moments for the
simulated \z\ events to the corrected \kpi\ data moments, and use the
distribution of Fig.~\ref{fig:test1}(b) to create new normalized \kpi\
moments as in Sec.~\ref{sec:reflection}. Finally, we follow the
weighting procedure described in Sec.~\ref{sec:reflection} to create
the new \kpi\ background description shown by the dashed curve of
Fig.~\ref{fig:test2} in comparison to the combined \mpsitwospi\
distribution of Fig.~\ref{fig:test1}(a).

Clearly, this dashed curve does not describe the \z\ signal region.
However, the shape of the curve has been changed compared to that of
Fig.~\ref{Z_all_k_corr}(b) because of the effect of the \z\ signal
events on the low-order Legendre polynomial \kpi\ moments. This is
shown by the shaded histogram in Fig.~\ref{fig:test2}, which
represents the difference between the dashed curve in this figure and
the solid curve in Fig.~\ref{Z_all_k_corr}(b).

A complete representation of the highly localized \z\ distribution in
Fig.~\ref{fig:MC_Z4430} requires the use of Legendre polynomials to
order more than 30, and so our low-order representation of the \kpi\
angular structure is unable to do this. The shaded histogram does
reach a maximum at about the \z\ mass value, but corresponds to a
width of $\sim 160$ \mev\/, which is almost four times larger than the
input signal value.

A fit to the distribution of Fig.~\ref{fig:test2} with the
normalization of the new \kpi\ background, and the normalization, mass
and width of the \z\ signal, free yields $m_{Z(4430)^-} = 4433 \pm 3$
\mevcc\/, $\Gamma_{Z(4430)^-} = 34 \pm 12$ \mev\/, and $N_{Z(4430)^-}
= 1402 \pm 315$ events so that the width is reduced to $\sim 75\%$,
and the signal to $\sim 94\%$, of the input value, while the mass is
essentially unchanged. The solid curve in Fig.~\ref{fig:test2}
represents the fit result, and the dotted curve shows the reduced
level of \kpi\ background. If the \z\ width is fixed to 45 \mev\/, we
obtain $N_{Z(4430)^-} = 1558 \pm 220$ events. This is consistent with
the input value, and presumably more properly reflects the effect of
the statistical uncertainties in the underlying data distribution. We
therefore use this to estimate the magnitude of the systematic signal
reduction factor, and so obtain a value of $\sim 90\%$.

This is a direct demonstration in support of the statement in the
Belle letter~\cite{:2007wga} that a narrow peak in the \psipi\ mass
distribution cannot be generated by only $S$-, $P$-, and $D$-wave
amplitudes in the \kpi\ system, and indicates that our use of low-order
Legendre polynomials in creating our \kpi\ background could lead to an
approximately $10\%$ systematic reduction of a narrow \z\ signal of
the reported magnitude.

\subsection{Branching fractions}
In Table~\ref{tab:BF}, we summarize the branching fraction values and
their 95$\%$ c.l. upper limits, obtained for the individual $B$ decay
modes studied in the present analysis by repeating the fits of
Figs.~\ref{fig:babar_cor}(a) and \ref{fig:babar_cor}(d), but with \z\
mass and width fixed to the central values obtained by Belle. The
errors quoted are statistical, and the upper limits were obtained
using these values. For the \psitwospi\ modes, the branching-fraction
and upper-limit values have been increased by $10\%$ in order to take
account of possible reduction of \z\ signal size, as described in
Sec.~\ref{sec:zsig}.

The branching fraction for the decay mode $B^0\rightarrow
Z(4430)^-K^+, Z(4430)^-\rightarrow \psi(2S)\pi^-$ from Belle is
$(4.1\pm1.0\pm1.4)\times 10^{-5}$, to be compared to our upper limit
of $3.1\times 10^{-5}$ at $95\%$ c.l.
\begin{table*}
  \begin{center}
    \caption{The \z\ signal size, the branching fraction value, and
    its $95\%$ c.l. upper limit for each decay mode; the errors quoted
    are statistical, and the upper limits were obtained using these
    values; the \z\ mass and width have been fixed to the central
    values obtained by Belle~\cite{:2007wga}.}
    \begin{tabular*}{1.0\textwidth}{@{\extracolsep{\fill}}lccc} \hline\hline\\

      Decay mode                                                            & $Z(4430)^-$ signal & Branching fraction & Upper limit  \\ 
                                                                            &                    &($\times 10^{-5}$)  &($\times 10^{-5}$ at 95$\%$ c.l.) \\ \\ \hline\hline \\
      $B^-\rightarrow Z(4430)^-\bar{K}^0, Z(4430)^-\rightarrow J/\psi\pi^-$ & $-17\pm 140$ & $-0.1 \pm 0.8$ & $1.5$ \\ \\
      $B^0\rightarrow Z(4430)^- K^+, Z(4430)^-\rightarrow J/\psi\pi^-$ & $-670\pm 203$ & $-1.2 \pm 0.4$ & $0.4$ \\ \\
      $B^-\rightarrow Z(4430)^-\bar{K}^0, Z(4430)^-\rightarrow \psi(2S)\pi^-$ & $148\pm 117$ & $2.0 \pm 1.7$ & $4.7$ \\ \\
      $B^0\rightarrow Z(4430)^- K^+, Z(4430)^-\rightarrow \psi(2S)\pi^-$ & $415\pm 170$ & $1.9 \pm 0.8$ & $3.1$ \\ \\ \hline \hline
    \end{tabular*}
\label{tab:BF}
\end{center}
\end{table*}

\section{Summary and Conclusions}
\label{sec:conclusions}
We have searched for evidence supporting the existence of the \z\ in
the \psipi\ mass distributions resulting from the decays
$B^{-,0}\rightarrow \psi \pi^- K^{0,+}$ in a large data sample
recorded by the \babar\ detector at the PEP-II $e^+e^-$ collider at
SLAC. Since the relevant Dalitz plots are dominated by mass and
angular distribution structures in the \kpi\ system, we decided to
investigate the extent to which the reflections of these features
might describe the associated \psipi\ mass distributions. To this end,
we obtained a detailed description of the mass and angular structures
of the \kpi\ system based on the expected underlying $S$-, $P$-, and
$D$-wave \kpi\ amplitude contributions, and in the process even found
evidence of $D-F$ wave interference in the \jpsi\ decay modes. The
fractional $S$-, $P$-, and $D$-wave intensity contributions to the
corrected \kpi\ mass distributions for $B^-\rightarrow J/\psi\pi^-
K^0_S$ and $B^0\rightarrow J/\psi\pi^- K^+$ were found to be the same
within error, as are the branching fraction values after all
corrections, and so we combined these data in our analysis of the
$J/\psi\pi^-$ mass distributions. We observed similar features for the
corresponding \psitwos\ decay modes, even though the analysis sample
is $\sim 6.6$ times smaller than that for \jpsi\
(Table~\ref{table_sig_bkg}), and we combined the charged and neutral
$B$-meson samples for the \psitwos\ analysis also.

We next investigated the \psipi\ mass distributions on the basis of
our detailed analysis of the \kpi\ system. We used a MC generator to
create large event samples for $B^{-,0}\rightarrow \psi \pi^- K^{0,+}$
with \kpi\ mass distribution generated according to the overall fit
function obtained from the corrected data, but with a uniform \costhk\
distribution. The \costhk\ dependence was then modulated using
normalized Legendre polynomial moments whose values were obtained from
our corrected data by linear interpolation.

The total corrected $J/\psi\pi^-$ mass distribution
(Fig.~\ref{Z_all_k_corr}(a)) is well described by this \kpi\
background, whose form can be seen more clearly in
Fig.~\ref{fig:mass}(a). The residuals (Fig.~\ref{Z_all_k_corr}(c))
show no evidence of a \z\ signal, and this is true also for the
various regions of \kpi\ mass shown in Fig.~\ref{fig:ranges}(a)-(e),
Fig.~\ref{fig:Z_k*_corr}(a),(c), and
Fig.~\ref{fig:Z_nok_corr}(a),(c). When we fit the data using a
function which allows the presence of a \z\ signal, we obtain only
negative \z\ signal intensities (Fig.~\ref{fig:babar_cor}(a)-(c),
Table~\ref{tab:summ_fits}). We find this to be the case also for the
$B^{-}$ and $B^{0}$ modes separately, and summarize the corresponding
branching fraction upper limits in Table~\ref{tab:BF} for \z\ mass and
width fixed at the central values obtained by Belle~\cite{:2007wga}.

We conclude that there is no evidence to support the existence of a
narrow resonant structure in the \jpsipi\ mass distributions for our
data on the decay modes \modeFourEight\/.

The corresponding corrected $\psi(2S)\pi^-$ distributions of
Figs.~\ref{Z_all_k_corr}(b),(d), Figs.~\ref{fig:ranges}(f)-(j) and
Figs.~\ref{fig:residuals}(f)-(j), Figs.~\ref{fig:Z_k*_corr}(b) and
(d), and Figs.~\ref{fig:Z_nok_corr}(b) and (d) likewise show no clear
evidence of a narrow signal at the \z\ mass position.

We have directly compared our uncorrected \psitwospi\ data with
$K^{\ast}$ veto, to those from Ref.~\cite{:2007wga}; there is no
evidence of statistically significant difference
(Fig.~\ref{fig:diff2}(c)).

In order to quantify our \z\ production rate estimates, we fit our
total corrected \psipi\ mass distributions of
Figs.~\ref{Z_all_k_corr}(a) and \ref{Z_all_k_corr}(b), using the \kpi\
background shapes shown in these figures, together with a \z\ line
shape. For \jpsipi\ we obtain a negative signal, while for \psitwospi\
we obtain a 2.7 standard deviation signal with fitted width consistent
with the value obtained by Belle, but with central mass value $\sim
43$ \mevcc\ higher than that reported by Belle, which corresponds to a
$+4.7$ standard deviation difference. These fit results are shown in
Figs.~\ref{fig:babar_cor}(a) and \ref{fig:babar_cor}(d), respectively,
and are summarized in Table~\ref{tab:summ_fits}. We repeated these
fits for the individual decay modes of
Eqs.(\ref{eq:modefour})-(\ref{eq:modesix}) with \z\ mass and width
fixed to the central values reported by Belle, and obtained the
branching fraction and upper limit values summarized in
Table~\ref{tab:BF}. In particular, we find a branching fraction upper
limit for the process ${\cal B}(B^0\rightarrow
Z(4430)^-K^+,Z^-\rightarrow \psi(2S)\pi^-)<3.1\times 10^{-5}$ at
95$\%$ c.l., and a corresponding value for the reaction ${\cal
B}(B^-\rightarrow Z(4430)^-\bar{K}^0,Z^-\rightarrow
\psi(2S)\pi^-)<4.7\times 10^{-5}$ at 95$\%$ c.l. We conclude that our
analyses provide no significant evidence for the existence of the
\z\/.

It will be of great interest to see whether or not the \z\ is
confirmed by a future analysis based upon a significantly larger data
sample than is available at present.
\section{Acknowledgments}
\label{sec:acknow}
We are grateful for the 
extraordinary contributions of our \pep2\ colleagues in
achieving the excellent luminosity and machine conditions
that have made this work possible.
The success of this project also relies critically on the 
expertise and dedication of the computing organizations that 
support \babar.
The collaborating institutions wish to thank 
SLAC for its support and the kind hospitality extended to them. 
This work is supported by the
US Department of Energy
and National Science Foundation, the
Natural Sciences and Engineering Research Council (Canada),
the Commissariat \`a l'Energie Atomique and
Institut National de Physique Nucl\'eaire et de Physique des Particules
(France), the
Bundesministerium f\"ur Bildung und Forschung and
Deutsche Forschungsgemeinschaft
(Germany), the
Istituto Nazionale di Fisica Nucleare (Italy),
the Foundation for Fundamental Research on Matter (The Netherlands),
the Research Council of Norway, the
Ministry of Education and Science of the Russian Federation, 
Ministerio de Educaci\'on y Ciencia (Spain), and the
Science and Technology Facilities Council (United Kingdom).
Individuals have received support from 
the Marie-Curie IEF program (European Union) and
the A. P. Sloan Foundation.

\appendix
\addcontentsline{toc}{chapter}{Appendices}
\section{THE DALITZ PLOT EFFICIENCY-CORRECTION PROCEDURE}
\label{append1}
The efficiency is obtained using samples of Monte Carlo events
corresponding to the decay processes of Eqs. (1)-(4) generated uniformly
over the final state Dalitz plot.

In general, the phase space volume element in the Dalitz plot
corresponding to the decay $B\rightarrow\psi\pi K$ is given by:
\begin{eqnarray} 
\label{eq:a1}
d\rho \sim d(m^2_{K\pi})\cdot d(m^2_{\psi\pi}) \, ,
\end{eqnarray}
where \mkpi\ ($m_{\psi\pi^-}$) is the invariant mass of the \kpi\
(\psipi\/) system.

However, when the efficiency is studied in such rectilinear area
elements, those elements at the plot boundary are partially outside
the plot, and this leads to a rather cumbersome efficiency treatment.
The phase space volume element of Eq.~(\ref{eq:a1}) may be transformed
to
\begin{eqnarray} 
d\rho' \sim p\cdot \frac{q}{m_{K\pi}}\cdot m_{K\pi} \, d(m_{K\pi})\, d(\rm cos\theta_{K}) \, ,
\end{eqnarray}
{\it{i.e.}} 
\begin{eqnarray} 
d\rho' \sim p\cdot q\, d(m_{K\pi})
d(\rm cos\theta_{K})\, ,
\end{eqnarray}
where $p$ is the momentum of the $\psi$ daughter of the $B$ in the $B$
rest frame, and $q$ is the momentum of the $K$ in the rest frame of
the \kpi\ system. This expression is such that the phase space density
is uniform in $\cos\theta_{K}$ at a given value of \mkpi\/.

The range of $\cos\theta_{K}$ is [-1,1], and that of \mkpi\ is
from threshold to $m_B-m_{\psi}$, so that the resultant ``Dalitz
Plot'' is rectangular in shape, with the factor $p\cdot q$
representing the Jacobian of the variable transformation.  A plot of
this kind can then be used readily to study efficiency behavior over
the entire phase space region without the problems incurred at the
boundary of a conventional Dalitz plot (see, for example, Appendix B
of Ref.~\cite{ziegler}).

The reconstruction efficiency calculated using the Monte Carlo
simulated events is parametrized as a function of \mkpi\ and
$\cos\theta_{K}$, and then used to correct the data by weighting each
event by the inverse of its parametrized efficiency value.  For a
given mass interval $I=[m_{K\pi^-}, m_{K\pi^-}+dm_{K\pi^-}]$, let $N$ be the
number of generated events, and let $n_{reco}$, represent the number
of reconstructed events. The generated $\cos\theta_{K}$ distribution
is flat, but in general efficiency effects will cause the
reconstructed $\cos\theta_{K}$ distribution to have structure.
Writing the angular distribution in terms of appropriately normalized
Legendre polynomials,
\begin{eqnarray} 
\label{eq:gen} 
\frac{dN}{d\cos\theta_{K}} = N\langle P_0\rangle P_0(\cos\theta_{K}) 
\end{eqnarray}
and,
\begin{eqnarray}
\label{eq:dndcosthk}
\frac{dn_{reco}}{d\cos\theta_{K}}= n_{reco}\sum_{i=0}^{L}\langle P_i\rangle P_i(\cos\theta_K)
\end{eqnarray}
where the normalizations are such that, 
\begin{eqnarray}
\int_{-1}^{1}P_i(\cos\theta_{K})P_{\it j} (\cos\theta_{K}) d(\cos\theta_{K}) =\delta_{ij} \, ,
\end{eqnarray}
where $P_i=\sqrt{2\pi}Y^0_i$, and $Y^0_i$ is a spherical harmonic
function. The value of $L$ is obtained empirically.

Using this orthogonality condition, the coefficients in the expansion
are obtained from
\begin{eqnarray} 
\langle P_j \rangle \, = \frac{1}{n_{reco}}\int_{-1}^{1}{ P_j(\cos\theta_{K})\frac{dn_{reco}}{d\cos\theta_{K}}d(\cos\theta_{K}) }\, ,
\end{eqnarray}
where the integral is given, to a good approximation for a large
enough MC sample, by $\sum_{i=1}^{n_{reco}} { P_{\it
j}(\cos\theta_{K_i}) }$. The index $i$ runs over the reconstructed
events in mass interval $I$, such that $n_{reco}\langle P_{\it j}\rangle \sim
\sum_{i=1}^{n_{reco}} { P_{\it j}(\cos\theta_{K_i}) }$, and the
effect of efficiency loss on the angular distribution is represented
through these coefficients.  The absolute efficiency, calculated as a
function of $\cos\theta_{K}$ and \mkpi\/, in mass interval $I$,
is then given by
\begin{eqnarray} 
\label{eq:EffLP}
E(\cos\theta_{K}, m_{K\pi^-})= \frac{n_{reco}\left (\sum_{i=0}^{L}\langle P_i\rangle P_i(\cos\theta_{K})\right )} {N\langle P_0\rangle P_0(\cos\theta_{K})} \, . \nonumber \\ \, 
\end{eqnarray} 
With
\begin{eqnarray}
\label{eq:calcE0}
E_0=\frac{n_{reco}}{N}
\end{eqnarray}
and
\begin{eqnarray}
\label{eq:calcEj}
E_j=2\frac{n_{reco}\langle P_{\it j}\rangle }{N}=2\frac{\sum_{i=1}^{n_{reco}}{P_{\it j}(\cos\theta_{K_i}) }}{N},
\end{eqnarray}
for a large enough sample (note that the factor 2 enters since
$\langle P_0\rangle P_0(\cos\theta_{K})=1/2$), Eq.~(\ref{eq:EffLP}) becomes,
\begin{eqnarray}
\label{eq:EffParam}
\lefteqn{E(\cos\theta_{K},m_{K\pi^-})=E_0}\nonumber \\ &+E_1 P_1(\cos\theta_{K})+...+E_L P_L(\cos\theta_{K}) \, .
\end{eqnarray}
The mean value of the $P_{\it j}(\cos\theta_{K_i})$, with
$i=1,..., n_{reco}$, corresponding to mass interval $I$, is written as
$\langle P_{\it j}\rangle $.
The r.m.s. deviation of the $P_{\it j}(\cos\theta_{K_i})$ w.r.t.
$\langle P_{\it j}\rangle $, $\sigma$, is given by
\begin{eqnarray} 
\label{eq:sig}
\sigma^2 = \sum_{i=1}^{n_{reco}} { \frac{\left( P_{\it j}(\cos\theta_{K_i})- \langle P_{\it j}\rangle \right )^2 } {n_{reco}-1}} \, .
\end{eqnarray} 
The error on the mean is then $\delta \langle P_{\it j}\rangle =\frac{\sigma}{\sqrt
n_{reco}}$ and from Eq.~(\ref{eq:sig}),
\begin{eqnarray}
\delta \langle P_j\rangle & =& \sqrt{ \frac{\sum_{i=1}^{n_{reco}} {\left(
P_j(\cos\theta_{K_i})- \langle P_j\rangle \right )^2 }}{n_{reco}(n_{reco}-1)} } \nonumber  \\& = &\sqrt{ \frac{\left [
\sum_{i=1}^{n_{reco}} \frac{\left( P_{\it j}(\cos\theta_{K_i})\right
)^2} {n_{reco}}\right ]-\langle P_j\rangle ^2}{n_{reco}-1} }\, . \nonumber \\ \, 
\end{eqnarray}

The uncertainty in the parameter $E_0=\frac{n_{reco}}{N}$ is given by:
\begin{eqnarray}
\label{eq:errE0}
\delta (E_0)=E_0\sqrt{\frac{1}{n_{reco}} +\frac{1}{N}}\, ,
\end{eqnarray}
and the uncertainty in the coefficient $E_j$ is given by:
\begin{eqnarray}
\label{eq:errEj}
\lefteqn{\delta (E_j) = \frac{2}{N} \cdot }\nonumber \\
&\sqrt{\sum_{i=1}^{n_{reco}} {\left(P_j(\cos\theta_{K_{\it i}})\right )^2}
+\frac{\left(\sum_{i=1}^{n_{reco}} {\left( P_j(\cos\theta_{K_{\it i}})\right )}\right )^2} {N}
}\,. \nonumber \\ \, 
\end{eqnarray}

For each of the processes represented by Eqs.(1)-(4), the efficiency
analysis is carried out in 50 \mevcc\ \kpi\ mass intervals from
threshold to the maximum value accessible. As shown in
Sec.~\ref{sec:efficiency}, Fig.~\ref{fig:effi_modes}, the \kpi\
mass dependence of the average efficiency parameter, $E_{0}$, depends
on the decay mode of the $\psi$ involved, and so is obtained by using
the MC sample for that particular decay mode.  The angular dependence
represented by $E_{1}$, $E_{2}$,...{\it{etc.}} does not depend on the
individual $\psi$ decay mode, and so these coefficients are calculated
by combining the MC samples for the individual $\psi$ modes. For the
$B$ meson decay processes of Eqs.~(\ref{eq:modefour})-~(\ref{eq:modesix}),
the main features of the angular dependence of the efficiency are very
similar, and so we present the results only for \modeSix\ by way of
illustration.

\begin{figure*}[!htbp]
\begin{center}
\includegraphics[width=18.0cm]{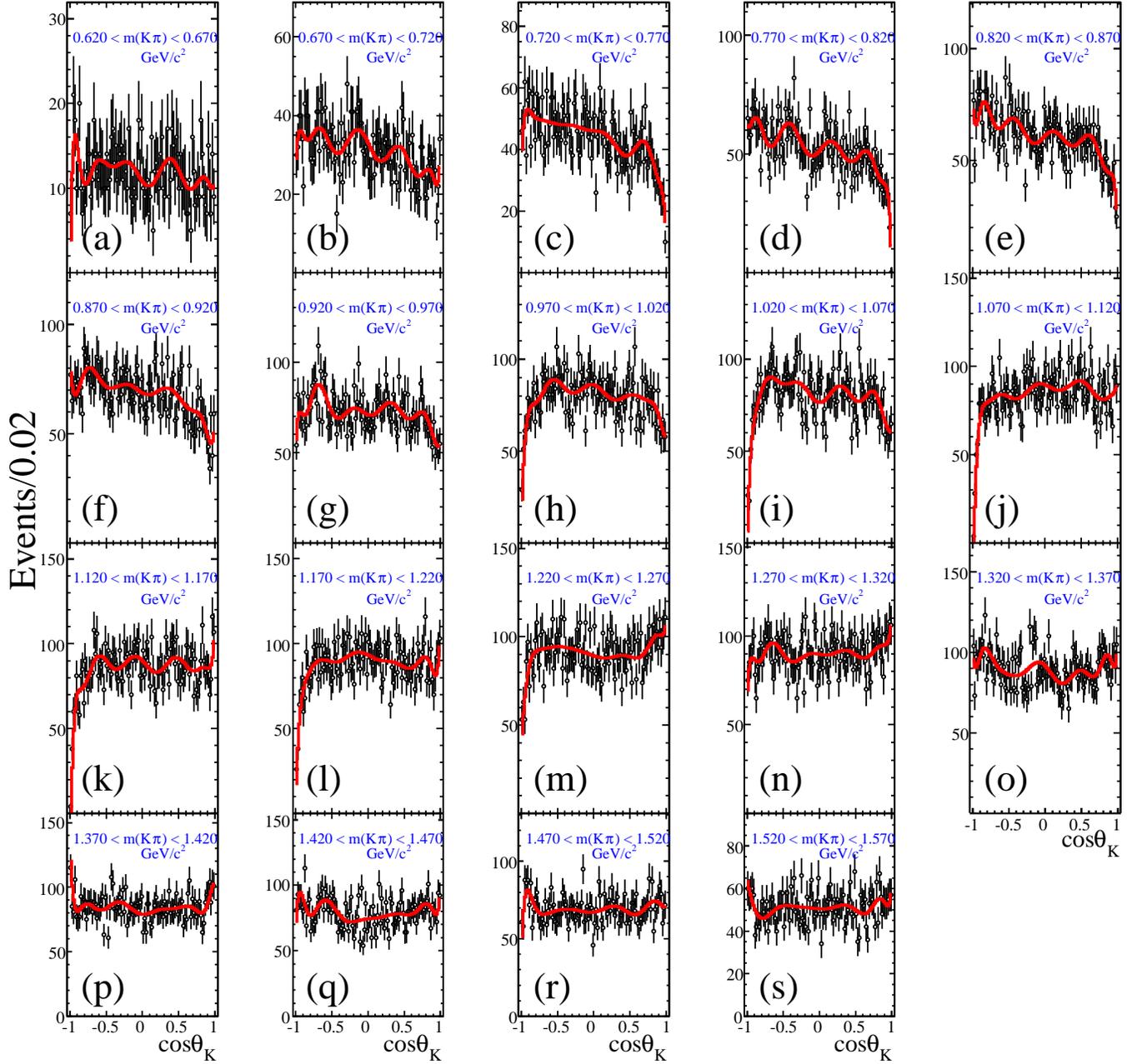}
\caption{The $\cos\theta_K$ distributions in 50 \mevcc\ \kpi\ mass
intervals for the decay mode \modeSix\/. The points represent the
data, and the curves show the functions calculated from the moments.}
\label{fig:cos6}
\end{center}
\end{figure*}

\begin{figure*}[!htbp]
\begin{center}
\includegraphics[width=18.0cm]{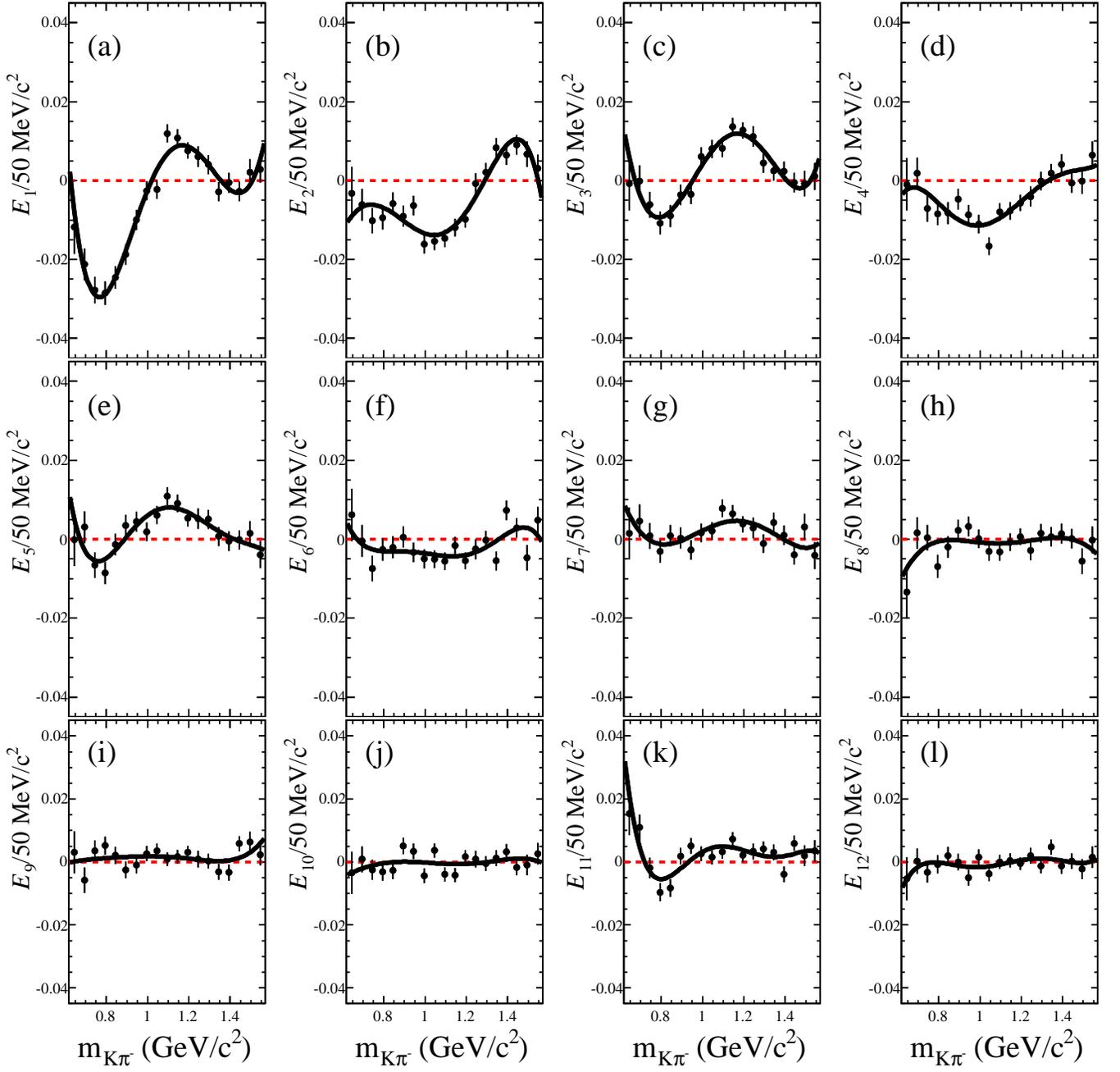}
\caption{The \kpi\ mass dependence of the coefficients $E_1$ through $E_{12}$
for the decay mode \modeSix\/.  The points show the calculated values,
and the curves result from fits to the data using a fifth-order
polynomial.}
\label{fig:Ei6unc}
\end{center}
\end{figure*}

Simulated MC events are subjected to the same reconstruction and
event-selection procedures as those applied to the data. For the
process of Eq.~(\ref{eq:modesix}), the $\cos\theta_{K}$ distributions
for the surviving MC events are shown for each \kpi\ mass interval in
Fig.~\ref{fig:cos6}. We chose a small interval size (0.02) in order to
investigate the significant decrease in efficiency observed for
$\cos\theta_{K}\sim+1$ and $0.720<m_{K\pi^-}<0.920$ \gevcc\
(Figs.~\ref{fig:cos6}(c)-(f)) and for $\cos\theta_{K}\sim-1$ and
$0.970<m_{K\pi^-}<1.270$ \gevcc\
(Figs.~\ref{fig:cos6}(h)-(m)). Representation of such localized losses
requires the use of high-order Legendre polynomials; we find that
$L=12$ yields a satisfactory description, as demonstrated by the
curves in Fig.~\ref{fig:cos6}. The \kpi\ mass dependence of the
resulting values of $E_{1}$ - $E_{12}$ is shown in
Fig.~\ref{fig:Ei6unc}, and is parametrized in each case by the fifth-
order polynomial curve shown. These parameterizations, together with
those describing the \kpi\ mass dependence of $E_{0}$ for the
individual $\psi$ decay modes (Sec.~\ref{sec:efficiency}) enable us to
calculate the efficiency at any point in the relevant rectangular
Dalitz plot, and hence for each event in the corresponding data
sample. We then assign to each event a weight given by the inverse of
this efficiency value, and by using this weight are able to create
efficiency-corrected distributions.

As discussed in Sec.~\ref{sec:efficiency}, the efficiency loss for
$\cos\theta_{K}\sim+1$ is due to the failure to reconstruct low
momentum charged pions in the laboratory frame, while that for
$\cos\theta_{K}\sim-1$ is due to the similar loss of low momentum
kaons (Fig.~\ref{fig:effpik}).


\begin{thebibliography}{10}
\bibitem{CC} The use of charge conjugate reactions is implied
throughout this paper.
\bibitem{psi} We use ``$\psi$'' to represent ``\jpsi\ or \psitwos\/''
unless these are required explicitly.
\bibitem{boost} We have verified that the difference in the Lorentz
boost between the \babar\ and Belle experiments does not significantly
affect these losses of low-momentum particles.
\end{thebibliography}
\end{document}